\documentclass[12pt]{article}
\usepackage{amsfonts,amsmath,hyperref,graphicx,color,tikz}
\makeatletter
\def\amsbb{\use@mathgroup \M@U \symAMSb}
\makeatother
\usepackage{bbold}

\newcommand{\be}{\begin{equation}}
\newcommand{\ee}{\end{equation}}
\newcommand{\bea}{\begin{eqnarray}}
\newcommand{\eea}{\end{eqnarray}}
\newcommand{\beas}{\begin{eqnarray*}}
\newcommand{\eeas}{\end{eqnarray*}}

\newcommand{\Hext}{\overset{\leftrightarrow}{H}\rlap{\phantom{H}}}
\newcommand{\Hextvac}{\overset{\leftrightarrow}{H}\rlap{\phantom{H}}^{(0)}}

\newcommand*{\circled}[1]{\tikz[baseline=(char.base)]{\node[shape=circle, draw, inner sep=2pt] (char) {#1};}}
\def\boxit#1{\setbox0=\hbox{#1} \vbox{\hrule \hbox{\vrule \kern3pt \vbox{\kern 3pt \box0 \kern 3pt} \kern 3pt \vrule} \hrule}}

\newcommand{\Oi}{{\cal O}_1}
\newcommand{\OiL}{{\cal O}_1^{(L)}}
\newcommand{\Oii}{{\cal O}_2}
\newcommand{\Oiipp}{{\cal O}_2^{++}}
\newcommand{\Oiipm}{{\cal O}_2^{+-}}

\newcommand{\Oiimm}{{\cal O}_2^{--}}
\newcommand{\Oiipnot}{{\cal O}_2^{+0}}
\newcommand{\Oiimnot}{{\cal O}_2^{-0}}
\newcommand{\Oiinotp}{{\cal O}_2^{0+}}
\newcommand{\Oiinotm}{{\cal O}_2^{0-}}

\newcommand{\Oiii}{{\cal O}_3}
\newcommand{\OiiiR}{{\cal O}_3^{(R)}}

\newcommand{\identity}{\mathbb{1}}

\begin{document}
\begin{titlepage}

\medskip

\begin{center}

{\Large \bf Modular Hamiltonians for future-perturbed states}

\vspace{10mm}

\renewcommand\thefootnote{\mbox{$\fnsymbol{footnote}$}}
Xiaole Jiang${}^{1,2}$\footnote{xjiang2@gradcenter.cuny.edu},
Daniel Kabat${}^{1,2}$\footnote{daniel.kabat@lehman.cuny.edu},
Aakash Marthandan${}^{1,2}$\footnote{amarthandan@gradcenter.cuny.edu},
Debajyoti Sarkar${}^{3}$\footnote{dsarkar@iiti.ac.in}

\vspace{6mm}

${}^1${\small \sl Department of Physics and Astronomy} \\
{\small \sl Lehman College, City University of New York} \\
{\small \sl 250 Bedford Park Blvd.\ W, Bronx NY 10468, USA}

\vspace{4mm}

${}^2${\small \sl Graduate School and University Center, City University of New York} \\
{\small \sl  365 Fifth Avenue, New York NY 10016, USA}

\vspace{4mm}

${}^3${\small \sl Department of Physics} \\
{\small \sl Indian Institute of Technology Indore} \\
{\small \sl Khandwa Road 453552 Indore, India}

\end{center}

\vspace{12mm}

\noindent
We develop a perturbative understanding of the modular Hamiltonian for a 2D CFT, divided into left
and right half-spaces, with a weak local perturbation inserted in the
future wedge.  A formal perturbation series for the modular
Hamiltonian is available, but must be properly interpreted in quantum
field theory.  
We work inside correlation functions with spectator operators, and introduce a prescription for defining complex modular flow
via analytic continuation to properly resolve singularities.
From the correlators, we extract an operator expression
for the modular Hamiltonian.
It takes the form of a local operator in the future wedge plus contact terms with an unconventional singularity structure.
Thanks to this structure the
KMS conditions are satisfied, which independently establishes the
validity of the results.  Similar techniques apply to perturbations
inserted in the past wedge.  We mention various future directions,
including an all-orders speculation for the excited state modular
Hamiltonian.

\end{titlepage}
\setcounter{footnote}{0}
\renewcommand\thefootnote{\mbox{\arabic{footnote}}}

\hrule
\tableofcontents
\bigskip
\hrule

\addtolength{\parskip}{8pt}

\section{Introduction}\label{sec:intro}

The theory of modular operators is central to our understanding of the
structure of observables in quantum field theory.  It provides a
powerful tool for organizing the way in which operator algebras are
associated with subregions.  As such, it's also a powerful tool for
understanding bulk locality and subregion duality in holography.  For
general references see
\cite{Haag:1992hx,Borchers:2000pv,Jones:2015kbb,Witten:2018lha,Sorce:2023gio},
and for applications to holography, see for example
\cite{Papadodimas:2013jku,Jafferis:2015del}.  Despite their
importance, only a few examples of modular Hamiltonians are known
explicitly.  For the vacuum state, the modular Hamiltonian associated
with a division into half-spaces is a Lorentz boost
\cite{Bisognano:1976za}.  This has been generalized to spherical
regions in conformal field theory \cite{Hislop:1981uh}.  Expressions
for deformed half-spaces \cite{Faulkner:2016mzt}, Virasoro excitations
of the vacuum \cite{Sarosi:2017rsq}, and disjoint intervals
\cite{Eisler:2022rnp} are also available.  For a review of results in
2D see \cite{Cardy:2016fqc}.

In this work we consider the modular operator $\Delta$ for a weakly-perturbed state in a 2D CFT.
The modular operator organizes the operator algebras associated with subregions, but it also captures a great deal of information
about the state.  This is most easily appreciated for finite-dimensional systems, where any pure state can be written in a Schmidt basis as
\be
\vert \psi \rangle = \sum_i p_i \, \vert i \rangle \otimes \vert i \rangle \,.
\ee
Assuming the Schmidt coefficients are all non-zero, the corresponding modular operator can be built from the subsystem density matrices as
$\Delta = \rho_L^{-1} \otimes \rho_R$, leading to
\be
\Delta = \sum_{i,j} {p_j^2 \over p_i^2} \, \vert ij \rangle \langle ij \vert \,.
\ee
Thus the modular operator encodes the Schmidt coefficients, and completely characterizes the entanglement between $L$ and $R$, in a way which
carries over in a well-defined manner to a continuum field theory.  In field theory, note that a modular operator uniquely determines
an associated KMS state \cite{Bratteli:1996xq}.

Our focus will be on the modular Hamiltonian $\Hext$, defined by $\Delta = e^{-\Hext}$.  We separate the observables
into left ($x < 0$) and right ($x > 0$) subalgebras, and we perturb
the vacuum by inserting a local operator in the future Rindler wedge.
A formal perturbation series for the modular Hamiltonian $\Hext$ is available
\cite{Sarosi:2017rsq,Lashkari:2018tjh}, see also
\cite{Rosenhaus:2014woa}, but requires careful interpretation.  The
perturbation series involves modular flow in complex time, which is
not {\em a priori} well-defined in quantum field theory.

Before outlining our approach, let us set the problem in context.  If the vacuum is perturbed
by a unitary operator that factorizes into a product $U_L \otimes U_R$ of operators that act on the left and
right Rindler wedges, it is straightforward to show that the perturbed modular operator $\Delta$ is related to the vacuum modular
operator $\Delta^{(0)}$ by $\Delta = U \Delta^{(0)} U^\dagger$.  However, a generic perturbation cannot be factored into $U_L \otimes U_R$.
Previous work \cite{Kabat:2020oic,Kabat:2021akg} examined a perturbation on a spacelike surface that passes through the origin.
A careful treatment of the perturbation at the origin, where factorization breaks down, showed that it generates an additional
contribution to $\Hext$.  The additional contribution, referred to as an endpoint contribution, takes the form of a light-ray operator through the origin.
In the present work, we examine another perturbation for which there is no simple or obvious factorization, namely, a local perturbation inserted
in the future wedge.  One might expect that, if one evolved such a perturbation back to the $t = 0$ slice, it would become a complicated operator
that is smeared across the origin.  In fact, we will find that a local perturbation in the future wedge does not generate an endpoint contribution.
However, we will derive a non-trivial set of rules which are required to resolve singularities in correlators involving $\Hext$.

The approach we take is to insert the first-order change in the
modular Hamiltonian $\delta\Hext$ inside a correlation function with
spectator operators.  This makes the analytic structure explicit, and
allows us to develop a prescription for defining complex modular flow.
We find that we have to start with real modular flow, then
analytically continue off the real axis so that modular time $s$
approaches the boundaries of a strip $- \pi < {\rm Im} \, s < \pi$ in
the complex plane.  This approach builds on the previous work
\cite{Kabat:2020oic,Kabat:2021akg}, which studied weak perturbations
on a spacelike surface.  We study in turn both two-point and
three-point correlators, and show that this prescription makes them
well-defined, with a definite analytic structure and a set of rules for resolving
singularities.  After obtaining results for correlators, we strip off
the spectator operators and obtain an expression for $\delta\Hext$
itself.  We find that it takes the form
\be
\delta\Hext = - i \lambda \left[ E,\Hextvac\right]
\ee
where $\lambda$ is the strength of the
perturbation, $E$ is an operator which we determine, and $\Hextvac$ is
the vacuum modular Hamiltonian.  Although the general expression for
$E$ is somewhat involved, in many situations there is a dramatic
simplification, and $E$ reduces to the local operator which is used to
perturb the state.  We go on to show that the KMS conditions are
satisfied, which independently establishes that the results for
$\delta\Hext$ are correct.

\section{Preliminaries}\label{sec:prelim}

We work in a Lorentzian 2D CFT with light-front coordinates
\be
\xi^\pm = x \pm t \,.
\ee
This divides the spacetime into four Rindler wedges $L$, $R$, $F$, $P$, separated by horizons at $\xi^+ = 0$ and $\xi^- = 0$.  We will also refer to the (Rindler) subregions $R$ and $L$ as $A $ and $\bar{A}$, respectively.

\centerline{\includegraphics{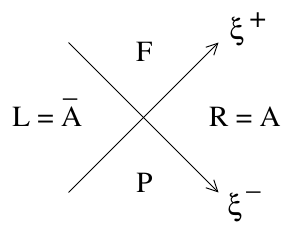}}

\noindent
We imagine slightly perturbing the vacuum to make an excited state
\be
\vert \psi \rangle = e^{-i \lambda G} \vert 0 \rangle \approx \left(\identity - i \lambda G\right) \vert 0 \rangle
\ee
where $\lambda$ is a small parameter and $G$ is a Hermitian operator.  We'd like to find the change in the modular Hamiltonian to first order in $\lambda$.
To do this we assume that $G$ can be factored as $G = G_A \otimes G_{\bar{A}}$, and use the Sarosi--Ugajin formulas \cite{Sarosi:2017rsq,Lashkari:2018tjh}\footnote{The expressions were put in this form in appendix B of \cite{Kabat:2020oic}.}
\bea
\label{deltaHA}
&& \delta H_A = {i \lambda \over 2} \int_{-\infty}^\infty {ds \over 1 + \cosh s} \left(G_A \big\vert_{s - i \pi} \widetilde{G}_{\bar{A}} \big\vert_s - \widetilde{G}_{\bar{A}} \big\vert_s G_A \big\vert_{s + i \pi} \right) \\
\label{deltaHAbar}
&& \delta H_{\bar{A}} = {i \lambda \over 2} \int_{-\infty}^\infty {ds \over 1 + \cosh s} \left(G_{\bar{A}} \big \vert_{s + i \pi}  \widetilde{G}_A \big\vert_s - \widetilde{G}_A \big\vert_s G_{\bar{A}} \big\vert_{s - i \pi} \right) \\
\label{deltaHext}
&& \delta \Hext = \delta H_A - \delta H_{\bar A}
\eea
Here $H_A$ and $H_{\bar{A}}$ are the subregion or one-sided modular Hamiltonians that generate a flow forward in time in $A$ and $\bar{A}$, respectively, and $\Hext$ is the full or two-sided modular Hamiltonian.\footnote{$H_A$ and $H_{\bar{A}}$ are not well-defined as operators, but they are well-defined as sesquilinear forms in the sense that they have well-defined matrix elements between suitable states. See footnote 27 in \cite{Witten:2018lha} and appendix I in \cite{Kudler-Flam:2023hkl}.  Although we will not be particularly careful about
this in what follows, it may justify working with subregion modular Hamiltonians inside correlators.}
Flow with the vacuum modular Hamiltonian $\Hextvac$ is denoted by\footnote{The vacuum modular Hamiltonian for a division into Rindler wedges is $\Hextvac = 2 \pi K$, where $K$ is the boost generator.}
\be
{\cal O}(\xi^+,\xi^-)\big\vert_s = e^{i \Hextvac s / 2 \pi} {\cal O}(\xi^+,\xi^-) \, e^{-i \Hextvac s / 2 \pi}
\ee
and a CPT transformation is denoted by $\widetilde{\cal O} = J {\cal O} J$ where $J$ is vacuum modular conjugation.

The formulas above are well-defined in systems with a finite-dimensional Hilbert space, but in carrying them over to field theory there are several challenges.
\begin{enumerate}
\item
As mentioned previously, the subregion modular Hamiltonians are only defined through their matrix elements.  We will deal with this by working inside correlation functions.
\item
The Sarosi--Ugajin formulas rely on factoring the perturbation as $G = G_A \otimes G_{\bar{A}}$.  At the operator level, it is not clear how to find such a factorization. Fortunately we will not have to address this issue, since in a 2D CFT correlators naturally factorize.
\item
The Sarosi--Ugajin formulas rely on complex modular time.  Vacuum modular time evolution is produced by
\be
e^{-i \Hextvac s / 2 \pi} = \Delta_0^{i s / 2 \pi} 
\ee
where $\Delta_0$ is the vacuum modular operator.  This yields a well-defined unitary operator for $s \in {\mathbb R}$, but whether it can be extended to complex $s$ depends on the state which is being evolved.\footnote{See for example section 4.2 in \cite{Witten:2018lha}.}  We will deal with this by working inside correlation functions and continuing from real to complex $s$, while paying attention to any singularities we encounter.
\end{enumerate}

To construct the perturbation $G$, we use an operator ${\cal O}(\xi^+,\xi^-)$ of dimension $\Delta$ and modular weight $n$.
This means that under vacuum modular flow, generated by the vacuum modular Hamiltonian $\Hextvac$, we have
\bea
\nonumber
{\cal O}(\xi^+,\xi^-)\big\vert_s & = & e^{i \Hextvac s / 2 \pi} {\cal O}(\xi^+,\xi^-) \, e^{-i \Hextvac s / 2 \pi} \\
\label{VacuumFlow}
& = & e^{ns} {\cal O}(e^s \xi^+,e^{-s} \xi^-) \,.
\eea
For example, the stress tensor $T_{++}(\xi^+)$ is an operator with $\Delta = n = 2$.  We will also need the CPT transformation of ${\cal O}$ which is
\be
\label{CPT}
\widetilde{\cal O}(\xi^+,\xi^-) = (-1)^n {\cal O}(-\xi^+,-\xi^-) \,.
\ee
For simplicity we will restrict to the case where $\Delta$ and $n$ are integers, either both even or both odd.\footnote{We do this to ensure that the left- and right-moving
conformal dimensions $h = {\Delta + n \over 2}$, $\bar{h} = {\Delta - n \over 2}$ are both integers.  This avoids branch cuts in correlators, which simplifies the
analysis that follows.  It would be interesting to explore what happens when this condition is relaxed.}  Strictly speaking, to make a normalizeable state,
we should suitably smear ${\cal O}$ and take
\be
G = \int d\xi^+ d\xi^- \, f(\xi^+,\xi^-) \, {\cal O}(\xi^+,\xi^-) \,.
\ee
We discuss this briefly in section \ref{subsec:multiple}.  However in most of what follows we take $G = {\cal O}(\xi^+,\xi^-)$ to be a local operator and see how the analysis goes.  Since the case of a local perturbation inserted in $L$ or $R$ is easily understood \cite{Kabat:2020oic,Kabat:2021akg}, we will focus on local perturbations inserted in $F$. A similar analysis goes through for operators inserted in the $P$ region.

\section{Two-point correlators involving $\delta\Hext$}\label{sec:2pc}

As mentioned in the introduction, we start with an analysis of $\delta\Hext$ inside a two-point correlator. We consider $\delta H_A$ first, and later perform a similar study for $\delta H_{\bar{A}}$. Finally we combine the results to determine a correlator involving $\delta\Hext$.

\subsection{Two-point correlator of $\delta H_A$ at non-singular points\label{sect:2ptNonSing}}

In this section we consider perturbing the state by a local operator
\be
G = \Oii = {\cal O}(\xi_2^+,\xi_2^-)
\ee
which is inserted in the future wedge, meaning $\xi_2^+ > 0$ and $\xi_2^- < 0$.  Our goal is to compute correlators of the form
$\langle \Oi \delta H_A \rangle$, where $\Oi = {\cal O}(\xi_1^+, \xi_1^-)$ is a spectator operator that could be inserted in any Rindler wedge.
For now we'll restrict our attention to operator insertions which keep the correlator $\langle \Oi \delta H_A \rangle$ non-singular.  In section \ref{sect:resolve} we'll see how to
deal with singularities.

First, some preliminaries on CFT correlators.  For an operator of dimension $\Delta$ and modular weight $n$, the 2-point function at non-singular points is
\be
\label{OOcorrelator}
\langle {\cal O}(\xi_1^+,\xi_1^-) {\cal O}(\xi_2^+,\xi_2^-) \rangle = {1 \over \big(\xi_1^+ - \xi_2^+\big)^{\Delta + n} \big(\xi_1^- - \xi_2^-\big)^{\Delta - n}} \,.
\ee
We resolve singularities using the Wightman prescription, $t_i \rightarrow t_i - i \epsilon_i$.  Operator ordering is fixed by the requirement that the value of $\epsilon_i$ decreases monotonically
as one goes from left to right in the correlator.\footnote{One can think of this prescription as inserting a convergence factor $e^{-\epsilon H}$ with $\epsilon \rightarrow 0^+$ between successive operators in the correlator.}
It's convenient to set $\xi_{ij}^\pm = \xi_i^\pm - \xi_j^\pm$, $\epsilon_{ij} = \epsilon_i - \epsilon_j$.  We'll want to consider various operator orderings, so we introduce the notation
\be\label{eq:shorthand}
\langle \left\lbrace {\cal O}(\xi_1^+,\xi_1^-), {\cal O}(\xi_2^+,\xi_2^-) \right\rbrace \rangle = {1 \over \big(\xi_{12}^+ - i \epsilon_{12}\big)^{\Delta + n} \big(\xi_{12}^- + i \epsilon_{12}\big)^{\Delta - n}} \,.
\ee
The notation $\langle \left\lbrace \cdot, \cdot \right\rbrace \rangle$ means that the ordering of operators in the correlator is determined by the sign of $\epsilon_{12}$.  If $\epsilon_{12} > 0$ it stands for
$\langle \Oi \Oii \rangle$, while if $\epsilon_{12} < 0$ it stands for $\langle \Oii \Oi \rangle$.

\subsubsection{Spectator operator in left wedge}

We begin with the following concrete situation.  We insert the perturbing operator in $F$, at position $(\xi_2^+,\xi_2^-)$ with
\be
\xi_2^+ > 0 \qquad \xi_2^- < 0 \,.
\ee
We insert the spectator operator in $L$, at position $(\xi_1^+,\xi_1^-)$ with
\be
\xi_1^+ < 0 \qquad \xi_1^- < 0 \,.
\ee
This is a nice starting point since any left-wedge operator should have a non-singular correlator with $\delta H_A$, which is a right-wedge operator. We can move the spectator operator to other wedges by taking care of the $i\epsilon$ prescriptions appropriately. Also, to begin with, we keep the spectator operator on the left in the correlator, meaning we evaluate $\langle \Oi \delta H_A \rangle$.  In other words we take $\epsilon_{12} > 0$. This will be generalized later.

To use the Sarosi--Ugajin formula, the first step is to factor the perturbation into $G_A \otimes G_{\bar{A}}$.  As a simple case,
suppose the perturbing operator is a tensor product of left-moving and right-moving chiral operators.
\be
\label{factor}
G = {\cal O}(\xi_2^+,\xi_2^-) = {\cal O}_L(\xi_2^+) {\cal O}_R(\xi_2^-)
\ee
When $G$ is inserted in $F$, the left-moving operator ${\cal O}_L(\xi_2^+)$ can be thought of as acting on region $A$, while the right-moving operator ${\cal O}_R(\xi_2^-)$ can be thought of as acting on region $\bar{A}$.\footnote{So the left-moving ${\cal O}_L$ acts on the right region $A = R$, while ${\cal O}_R$ acts on region $\bar{A} = L$.  We apologize for the notation, which we won't use after (\ref{last}).  Note that ${\cal O}_L$, ${\cal O}_R$ refers to the perturbing operator.  Starting with (\ref{FJdeltaHA}) we'll use ${\cal O}^{(L)}$, ${\cal O}^{(R)}$ to denote general (non-chiral) spectator operators that act on the left, right regions.}

\medskip
\centerline{\includegraphics{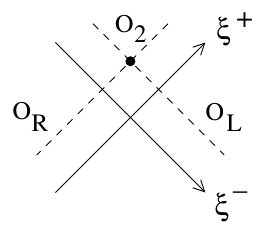}}

\noindent
Thus we identify
\bea
\nonumber
&& G_A = {\cal O}_L(\xi_2^+) \qquad \hbox{\rm operator with $\Delta_L = n_L = {\Delta + n \over 2}$} \\[5pt]
&& G_{\bar{A}} = {\cal O}_R(\xi_2^-) \qquad \hbox{\rm operator with $\Delta_R = {\Delta - n \over 2}$, $n_R = - {\Delta - n \over 2}$}
\eea
and the Sarosi--Ugajin formula (\ref{deltaHA}) for $\delta H_A$ becomes
\be
\label{FdeltaHA}
\delta H_A = {i \lambda \over 2} \int_{-\infty}^\infty {ds \over 1 + \cosh s} \left({\cal O}_L(\xi_2^+)\big\vert_{s - i \pi} \widetilde{{\cal O}}_R(\xi_2^-)\big\vert_s - \widetilde{{\cal O}}_R(\xi_2^-)\big\vert_s {\cal O}_L(\xi_2^+)\big\vert_{s+i\pi}\right) \,.
\ee
We'll consider this simple case of a factorized perturbation first, then argue that the result is more general.

Now let's work on our expression for $\delta H_A$.  From the CPT transformation (\ref{CPT}) we have
\be
\delta H_A = {i \lambda \over 2} \int_{-\infty}^\infty {ds \over 1 + \cosh s} \left({\cal O}_L(\xi_2^+)\big\vert_{s - i \pi} (-1)^{n_R} {\cal O}_R(-\xi_2^-)\big\vert_s
- (-1)^{n_R} {\cal O}_R(-\xi_2^-)\big\vert_s {\cal O}_L(\xi_2^+)\big\vert_{s+i\pi}\right) \,.
\ee
This involves complex modular flow, which we will define by analytic continuation inside correlators.  To do this we introduce a parameter $r$, which is to be analytically continued $r \, : \, 0 \rightarrow \pi$,
and set
\be
\delta H_A = {i \lambda \over 2} \int_{-\infty}^\infty {ds \over 1 + \cosh s} \left({\cal O}_L(\xi_2^+)\big\vert_{s - i r} (-1)^{n_R} {\cal O}_R(-\xi_2^-)\big\vert_s
- (-1)^{n_R} {\cal O}_R(-\xi_2^-)\big\vert_s {\cal O}_L(\xi_2^+)\big\vert_{s+ir}\right) \,.
\ee
At this stage we can use the vacuum modular flow (\ref{VacuumFlow}) to obtain (putting $r = \pi$ in the overall phase)
\be
\label{last}
\delta H_A = {i \lambda \over 2} \int_{-\infty}^\infty {ds \over 1 + \cosh s} (-1)^n e^{ns} \left({\cal O}_L(e^{s-ir} \xi_2^+) {\cal O}_R(-e^{-s} \xi_2^-)
- {\cal O}_R(-e^{-s} \xi_2^-) {\cal O}_L(e^{s+ir} \xi_2^+)\right) \,.
\ee
We can re-write this in terms of the original perturbing operator as
\be
\delta H_A = {i \lambda \over 2} \int_{-\infty}^\infty {ds \over 1 + \cosh s} (-1)^n e^{ns} \left({\cal O}(e^{s-ir} \xi_2^+,-e^{-s} \xi_2^-)
- {\cal O}(e^{s+ir} \xi_2^+, -e^{-s} \xi_2^-) \right) \,.
\ee
Going forward, it will be convenient to change variables to $w = e^{-s}$ from the modular time $s$ and write
\be
\label{operatordeltaHA}
\delta H_A = i \lambda \int_0^\infty {dw \over (w + 1)^2} \left(-{1 \over w}\right)^n \left({\cal O}\big(e^{-ir} {1 \over w} \xi_2^+,-w \xi_2^-\big)
- {\cal O}\big(e^{ir} {1 \over w} \xi_2^+, -w \xi_2^-\big) \right) \,.
\ee
This is the expression for $\delta H_A$ that we will use.
Once again, we obtained this result assuming that the perturbation has the factorized form (\ref{factor}).  However the correlator (\ref{OOcorrelator}) holomorphically factorizes, even if
the operator ${\cal O}$ itself doesn't have any simple or obvious factorization.  So it seems plausible that (\ref{operatordeltaHA}) is correct in general.  We
will proceed assuming this is the case.

Although it might appear that (\ref{operatordeltaHA}) provides an operator expression for $\delta H_A$, the fact is that $\delta H_A$ can only be
defined through its matrix elements, rather than as an operator.  Also we have to be careful about the analytic continuation $r \, : \, 0 \rightarrow \pi$
that we are using to define complex modular flow (these subtleties will be further clarified in section \ref{sec:exop}).  To address these issues we insert the expression for $\delta H_A$ in a correlation function with a
spectator operator.  As mentioned above, we take the spectator to be inserted at a point $(\xi_1^+,\xi_1^-)$ in the left Rindler wedge, and to be positioned on the left in the correlator,
returning to the general case in section \ref{subsubsec:generic2p}.

Denoting an operator in the left wedge by ${\cal O}^{(L)}$, from the correlator (\ref{OOcorrelator}) we have
\bea
\nonumber
&& \langle {\cal O}^{(L)}(\xi_1^+,\xi_1^-) \, \delta H_A \rangle = i \lambda \int_0^\infty {dw \over (w + 1)^2} \left(-{1 \over w}\right)^n \bigg[
{1 \over (\xi_1^+ - e^{-ir} {1 \over w} \xi_2^+ - i \epsilon)^{\Delta + n}} \, {1 \over (\xi_1^- + w \xi_2^- + i \epsilon)^{\Delta - n}} \\
\label{FJdeltaHA}
& & \hspace{5cm} - {1 \over (\xi_1^+ - e^{ir} {1 \over w} \xi_2^+ - i \epsilon)^{\Delta + n}} \, {1 \over (\xi_1^- + w \xi_2^- + i \epsilon)^{\Delta - n}} \bigg] \,.
\eea
The correlator is defined with a Wightman prescription, with $\epsilon_{12} = \epsilon \rightarrow 0^+$.  As a function of complex $w$, the first term in square
brackets has poles at
\be
\label{poles1}
w = e^{-ir} \left({\xi_2^+ \over \xi_1^+} + i\epsilon \right) \qquad\text{and}\qquad
w = - {\xi_1^- \over \xi_2^-} + i\epsilon
\ee
whereas the second term has poles at
\be
\label{poles2}
w = e^{ir} \left({\xi_2^+ \over \xi_1^+} + i\epsilon \right) \qquad\text{and}\qquad
w = - {\xi_1^- \over \xi_2^-} + i\epsilon \,.
\ee
There is also a pole at $w = -1$ coming from the integration measure.\footnote{Since $\xi_2^+$ and $\xi_2^-$ are non-zero, the integrand
has good behavior $\sim w^\Delta$ as $w \rightarrow 0$ and $\sim {1 \over w^{\Delta + 2}}$ as $w \rightarrow \infty$.  So we have the complete list of poles,
and there are no additional singularities.  The behavior as $w \rightarrow 0$ changes if $\xi_2^+ = 0$, and the behavior as $w \rightarrow \infty$ changes if $\xi_2^- = 0$.
This can produce singularities that lead to the endpoint contributions to $\delta \Hext$
studied in \cite{Kabat:2020oic,Kabat:2021akg}.\label{footnote:nosing}}  Recall that the
spectator operator is inserted in the left Rindler wedge, with $\xi_1^+ < 0$ and $\xi_1^- < 0$, while the perturbing operator is inserted in the future Rindler wedge,
with $\xi_2^+ > 0$ and $\xi_2^- < 0$.  Then all of the poles begin in the left half-plane, with ${\rm Re} \, w < 0$.  As we continue $r \, : \, 0 \rightarrow \pi$,
one of the poles in the first term rotates clockwise and hits the integration contour from above.
In the second term, one of the poles rotates counter-clockwise and approaches but does not hit the integration contour from below.
The two lines can be combined into a single contour integral that encircles the rotated pole.

\centerline{\includegraphics{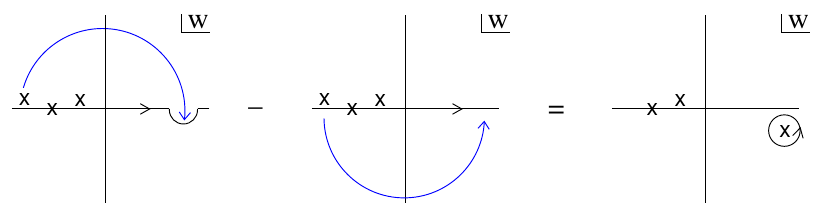}}

\noindent
The resulting expression for the correlator is
\be
\label{FJdeltaHA2}
\langle {\cal O}^{(L)}(\xi_1^+,\xi_1^-) \, \delta H_A \rangle = i \lambda \hspace{-5mm} \oint\limits_{\,\,\,w = - {\xi_2^+ \over \xi_1^+} - i \epsilon} \hspace{-2mm} {dw \over (w + 1)^2} \left(-{1 \over w}\right)^n
{1 \over (\xi_1^+ + {1 \over w} \xi_2^+ - i \epsilon)^{\Delta + n}} \, {1 \over (\xi_1^- + w \xi_2^- + i \epsilon)^{\Delta - n}} \,.
\ee
We have kept track of the $i \epsilon$ which is inherited from the CFT correlator, however this expression has a smooth $\epsilon \rightarrow 0$ limit.  So we might
as well set $\epsilon = 0$ and take
\be
\label{FJdeltaHA3}
\langle {\cal O}^{(L)}(\xi_1^+,\xi_1^-) \, \delta H_A \rangle = i \lambda \hspace{-5mm} \oint\limits_{\,\,\,w = - {\xi_2^+ \over \xi_1^+}} \hspace{-2mm} {dw \over (w + 1)^2} \left(-{1 \over w}\right)^n
{1 \over (\xi_1^+ + {1 \over w} \xi_2^+)^{\Delta + n}} \, {1 \over (\xi_1^- + w \xi_2^-)^{\Delta - n}} \,.
\ee
As a concrete example, suppose the perturbing operator ${\cal O}$ has $\Delta = 2$ and $n = 0$.  Then the contour integral yields
\be
\label{Delta2n0}
\Delta = 2, \, n = 0 \, : \enskip \langle {\cal O}^{(L)}(\xi_1^+,\xi_1^-) \, \delta H_A \rangle = - 4 \pi \lambda {\xi_1^+ \xi_2^+ \left((\xi_1^+)^2 \xi_1^- - (\xi_2^+)^2 \xi_2^-\right) \over
\big(\xi_1^+ - \xi_2^+\big)^3 \big(\xi_2^+ \xi_2^- - \xi_1^+ \xi_1^-\big)^3} \,.
\ee

\subsubsection{Spectator operator at generic points}\label{subsubsec:generic2p}

We obtained the contour integral expression for $\langle {\cal O}(\xi_1^+,\xi_1^-) \, \delta H_A \rangle$ given in (\ref{FJdeltaHA3}) by assuming that the spectator operator
is inserted in the left Rindler wedge, and is located on the left in the correlator.  There are quite a few other possibilities.  As we now show, for generic spectator
positions $(\xi_1^+,\xi_1^-)$ all possibilities lead to exactly the same result.  As a warning, at non-generic spectator positions the correlator
can be singular.  We leave these singularities aside for now and return to analyze them in the next section.

The essential point is to keep track of how the poles in (\ref{poles1}), (\ref{poles2}) move as $r \, : \, 0 \rightarrow \pi$.  If $\xi_1^+ < 0$, meaning that the spectator
operator is inserted in the $L$ or $P$ Rindler wedge, then the mobile poles begin near the negative real $w$ axis.  If $\epsilon_{12} > 0$, meaning that the spectator
operator is on the left in the correlator, then the pole in the first term rotates clockwise and hits the integration contour from above.  This is the situation we
analyzed in the previous section, and it leads to the outcome (\ref{FJdeltaHA3}).

Another possibility is to keep the spectator operator in the $L$ or $P$ Rindler wedge, but to place it on the right in the correlator, meaning that $\epsilon_{12} < 0$.
In this case the mobile poles begin just below the negative real axis.  Now it is the pole in the second term which rotates counter-clockwise and hits the integration
contour from below.  Fortunately, thanks to the $-$ sign in front of the second term in the Sarosi--Ugajin formula, the outcome is the same.

\centerline{\includegraphics{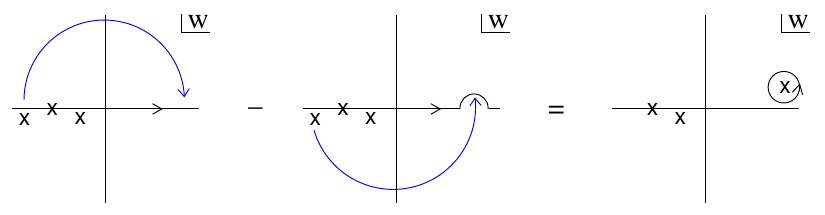}}

\noindent
Another possibility is to put the spectator operator in the $R$ or $F$ Rindler wedge, with $\xi_1^+ > 0$.  If we place the spectator operator on the left in the correlator,
meaning that $\epsilon_{12} > 0$, the mobile poles begin just above the positive real axis.  In the first term the pole rotates clockwise and drags a small loop of
integration contour with it.  In the second term the pole rotates counter-clockwise.  The straight parts of the contours cancel between the two terms, and we are left
with a single contour integral that encircles the rotated pole.

\centerline{\includegraphics{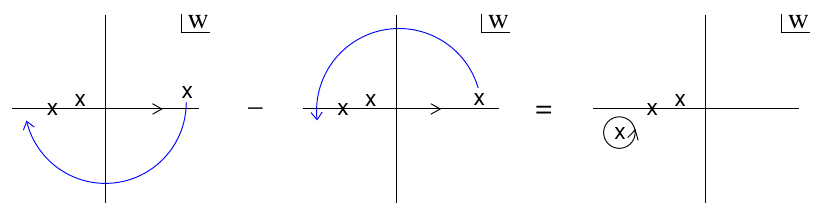}}

\noindent
A final possibility is to have the spectator operator in the $R$ or $F$ Rindler wedge, with $\xi_1^+ > 0$, but to place it on the right in the correlator,
so that $\epsilon_{12} < 0$.  In this case the mobile poles begin just below the positive real axis.  Now in the second term the pole rotates counter-clockwise
and drags a small loop of integration contour with it.  Again the straight parts of the contours cancel between the two terms, and we are left
with a single contour integral that encircles the rotated pole.  Thanks to the $-$ sign in front of the second term in the Sarosi--Ugajin formula,
the outcome is the same

\centerline{\includegraphics{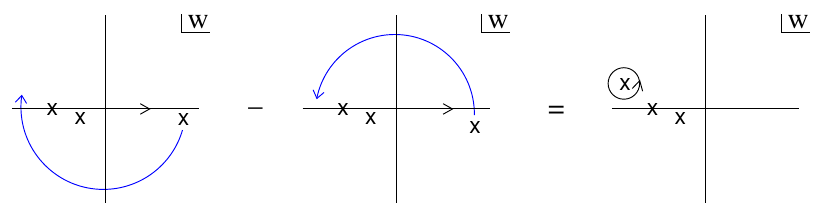}}

To summarize, at generic operator positions, the spectator operator can be located in any Rindler wedge, and can be inserted at any position in the correlator.
We always get the same outcome, that the two-point correlator is given by a counter-clockwise contour integral that encircles the rotated pole.  The result is given in (\ref{FJdeltaHA3}), which we repeat here for clarity.
\be
\label{FJdeltaHA4}
\langle {\cal O}(\xi_1^+,\xi_1^-) \, \delta H_A \rangle = i \lambda \hspace{-5mm} \oint\limits_{\substack{{\rm counter-clockwise} \\ w = - {\xi_2^+ \over \xi_1^+}}} \hspace{-2mm} {dw \over (w + 1)^2} \left(-{1 \over w}\right)^n
{1 \over (\xi_1^+ + {1 \over w} \xi_2^+)^{\Delta + n}} \, {1 \over (\xi_1^- + w \xi_2^-)^{\Delta - n}}
\ee

Let us make a few comments on this result.
\begin{enumerate}
\item
Although we derived this for a two-point correlator, the argument goes through in general.  For an arbitrary number of spectator operators, the generic correlator with
$\delta H_A$ is given by a contour integral that surrounds the rotated poles.  We present explicit results for three-point correlators in section \ref{sect:threepoint}.
\item
At special spectator operator positions, the correlator with $\delta H_A$ is singular, as can be seen in the $\Delta = 2$, $n = 0$ example (\ref{Delta2n0}).
We will study these singularities in detail in the next section.  They arise when an integration contour gets pinched between two poles.
\item
One might worry that (for example) a mobile pole could start out just above the positive real axis, rotate clockwise, and immediately pinch a fixed pole just below the
positive real axis.  Would this obstruct the analytic continuation $r \, : \, 0 \rightarrow \pi$?  The answer is no, because one can slightly change the starting point.
That is, one can slightly displace the initial position of the mobile pole so that it passes to the left or right of the fixed pole rather than hitting it.
\end{enumerate}
The last point is related to a feature of the construction that may appear surprising.  Vacuum modular flow is produced by
\be
e^{-i \Hextvac s / 2 \pi} = \Delta_0^{i s / 2 \pi} \,.
\ee
This is a unitary operator for $s \in {\mathbb R}$, but whether it can be extended to complex $s$ depends on the state that is being evolved.\footnote{As discussed in section 4.2 of \cite{Witten:2018lha},
it is holomorphic in the strip $-\pi < {\rm Im} \, s < 0$ when acting on states produced by bounded operators
in the right Rindler wedge, and it is holomorphic in the strip $0 < {\rm Im} \, s < \pi$ when acting on states produced by bounded operators in the left Rindler
wedge.  For certain states it can be continued to arbitrary complex $s$ and for other states it can't be continued at all.}  To apply the Sarosi--Ugajin formula we must be able to work inside a general correlator and continue into the strip $-\pi < {\rm Im} \, s < \pi$.  Although this is not a priori guaranteed to be possible, the discussion in point 3 shows that there is no obstruction in the correlators we are considering.

\subsection{Resolving singularities in $\delta H_A$\label{sect:resolve}}

We've seen that the correlator of $\delta H_A$ with a spectator operator is given by a contour integral that encircles the rotating pole.
The structure with any number of spectator operators is similar.  The integrand has mobile or rotating poles, located at
\be
\label{rotating}
w = w_m = e^{-ir} \left({\xi_2^+ \over \xi_k^+} - i\epsilon_{2k} \right) \quad \hbox{\rm and} \quad w = w_m = e^{ir} \left({\xi_2^+ \over \xi_k^+} - i\epsilon_{2k} \right) \,,
\ee
coming from the first and second term of the Sarosi--Ugajin formula for $\delta H_A$.  Here $\Oii$ is the perturbing operator and ${\cal O}_k$ denotes any spectator operator.
These mobile poles rotate clockwise or counter-clockwise as we continue $r \, : \, 0 \rightarrow \pi$.  The integrand also has fixed or non-rotating poles, present in both terms
of the Sarosi-Ugajin formula, which are located at
\be
\label{fixed}
w = w_f = - {\xi_k^- \over \xi_2^-} - i\epsilon_{2k} \,.
\ee
There is also a fixed pole at
\be
w = w_0 = -1
\ee
coming from the integration measure.  One can see this structure for a single spectator operator in \eqref{FJdeltaHA}.  In section \ref{sect:threepoint} we treat two spectator operators,
and the structure can be seen explicitly in (\ref{3ptContour}).

By following the same contour manipulations that led to (\ref{FJdeltaHA3}), the correlator of $\delta H_A$ with any string of spectator operators is given by a contour integral around the mobile poles after they have been rotated $180^\circ$. We either get a counter-clockwise contour from the first term in Sarosi--Ugajin, or we get a clockwise contour from the second term but with an overall minus sign.  In either case, the final expression is the same.

For generic positions of the spectator operators, this is the end of the story.  However there are singularities at special operator positions, as can be seen in (\ref{Delta2n0}).  Our goal here is to understand and resolve these singularities.

In general, singularities arise when two poles collide and pinch an integration contour, and the way in which the singularity is resolved depends on how the poles
approach each other.  As a prototype example, consider
\be
I(a,b) = \int_{-\infty}^\infty dx \, {1 \over (x-a)(x-b)} \,.
\ee
We can define $I(a,b)$ by analytic continuation starting from ${\rm Im} \, a > 0$ and ${\rm Im} \, b < 0$, which leads to
\be
I(a,b) = {2 \pi i \over a - b} \,.
\ee
The singularity at $a = b$ is due to the poles pinching the integration contour, and the behavior as $a \rightarrow b$ depends on how $a$ approaches
$b$ in the complex plane: it diverges, but with a phase that depends on how $a$ approaches $b$.

In our case, recall that the integration contour encircles the mobile poles (\ref{rotating}).  Two of these mobile poles can collide and produce a singularity when
\be
\label{mobile-mobile}
{\xi_2^+ \over \xi_j^+} - i \epsilon_{2j} = {\xi_2^+ \over \xi_k^+} - i \epsilon_{2k}
\ee
or equivalently (since $\xi_2^+ > 0$, and recalling that $\epsilon_{2j} = \epsilon_2 - \epsilon_j$, $\epsilon_{2k} = \epsilon_2 - \epsilon_k$) when
\be
\label{mobile-mobile2}
\xi_j^+ - i \epsilon_j = \xi_k^+ - i \epsilon_k \,.
\ee
Thus the prescription for resolving a singularity at $\xi_j^+ = \xi_k^+$ is inherited from the CFT.  We simply impose the Wightman prescription, $t_i \rightarrow
t_i - i \epsilon_i$.

Another possibility is that a mobile pole could rotate and collide with a fixed pole -- either one of the poles listed in (\ref{fixed}), or the pole at $w = -1$ coming from the integration measure.  A collision with one of the poles listed in (\ref{fixed}) produces a singularity when
\be
\label{mobile-fixed}
\xi_2^+ \xi_2^- = \xi_j^+ \xi_k^- \qquad \hbox{\rm (including the case $j = k$)}
\ee
while a collision with the pole at $w = -1$ produces a singularity when
\be
\label{mobile-measure}
\xi_2^+ = \xi_j^+ \,.
\ee
The prescription for resolving these singularities depends on whether the pole is rotating clockwise or counter-clockwise, in other words,
whether the singularity comes from the first or second term in the Sarosi--Ugajin formula.  To see this, we examine the behavior as $r$ approaches $\pi$ more closely.
Setting $r = \pi - \delta$, with $\delta$ a small positive quantity, the Sarosi--Ugajin formula becomes  (since $\xi_2^+ > 0$)
\be
\delta H_A = i \lambda \int_0^\infty {dw \over (w+1)^2} \left(-{1 \over w}\right)^n \left[{\cal O}\left(-{1 \over w} (\xi_2^+ + i \delta),-w \xi_2^-\right)
- {\cal O}\left(-{1 \over w} (\xi_2^+ - i \delta), -w \xi_2^-\right) \right] \,.
\ee
This means a singularity that comes from the first term in Sarosi--Ugajin, in which the pole rotates clockwise, is resolved by the prescription
$\xi_2^+ \rightarrow \xi_2^+ + i \delta$.  A singularity that comes from the second term, in which the pole rotates counter-clockwise, is resolved by
$\xi_2^+ \rightarrow \xi_2^+ - i \delta$.

In applying this prescription, an important and somewhat subtle order of limits must be taken.  The original CFT correlator of local operators is defined by a Wightman prescription $t_i \rightarrow t_i - i \epsilon_i$.  This yields infinitesimal parameters $\epsilon_{ij}$ that specify how singularities are resolved when a pair of local operators are null separated.  Correlators involving $\delta H_A$ pick up additional singularities when a mobile pole collides with a fixed pole.  These additional singularities are resolved by
$\xi_2^+ \rightarrow \xi_2^+ \pm i \delta$, depending on whether the mobile pole has rotated clockwise or counter-clockwise.  The prescription is that one should first
calculate at $r = 0$, using a Wightman CFT correlator, and then analytically continue $r \, : \, 0 \rightarrow \pi$.  This means the parameters $\epsilon_{ij}$ approach
zero first, before the parameter $\delta$ is sent to zero.  So the resolution of the singularity is entirely controlled by whether the mobile pole has rotated clockwise
or counter-clockwise.\footnote{To see that this prescription is necessary, consider working in the opposite order of limits and sending
$\delta \rightarrow 0$ first.  In some cases the Wightman prescription inherited from the CFT, $t_i \rightarrow t_i - i \epsilon_i$, is sufficient to resolve singularities in
correlators involving $\delta H_A$.  But in other cases the Wightman prescription is not sufficient, in particular for correlators $\langle {\cal O} \delta H_A {\cal O} \rangle$
in which $\delta H_A$ is sandwiched between local operators in the left and right Rindler wedges.  In this case applying the Wightman prescription leaves certain
singularities ambiguous, in a manner curiously similar to the behavior found in \cite{Kabat:2018pbj}.  For further discussion see appendix \ref{appendix:contact}.\label{footnote:ambiguous}}

At this stage we've seen that the resolution of certain singularities depends on whether the mobile pole responsible for the singularity has rotated clockwise or counter-clockwise.  Although accurate, this is not a convenient characterization.  As we now show, the resolution can be read off from properties of the spectator operator, namely,
the Rindler wedge in which it is inserted and its position in the correlator relative to $\delta H_A$.

To show this, we go through the various possibilities that can lead to a singularity from a mobile pole colliding with a fixed pole.
\begin{enumerate}
\item
Suppose a spectator operator ${\cal O}_j$ is inserted in the $R$ or $F$ Rindler wedge, so that $\xi_j^+ > 0$.  Then there are mobile poles that start near the positive
real axis and rotate clockwise or counter-clockwise according to
\be
w_m = e^{\pm i r} \left({\xi_2^+ \over \xi_j^+} - i \epsilon_{2j}\right) \,.
\ee
If $\epsilon_{2j} > 0$, so that the pole starts below the positive real axis, it must rotate counter-clockwise to generate a pinch singularity with a pole near the negative
real axis, which means $\xi_2^+ \rightarrow \xi_2^+ - i \delta$.  On the other hand if $\epsilon_{2j} < 0$, so that the pole starts above the positive real axis, then it must rotate clockwise to generate a pinch singularity, which means $\xi_2^+ \rightarrow \xi_2^+ + i \delta$.

\centerline{\includegraphics{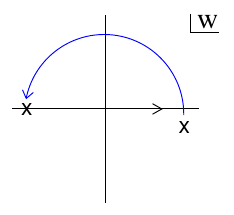} \hspace{2cm} \includegraphics{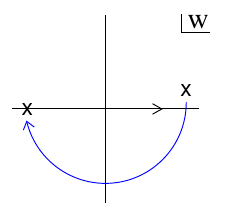}}

Since the sign in front of $i\delta$ is correlated with the sign of
$\epsilon_{2j}$, we can summarize the outcome as
\be\label{eq:pres}
\xi_2^+ \rightarrow \xi_2^+ - i \epsilon_{2j} \,.
\ee
\item
Suppose a spectator operator ${\cal O}_j$ is inserted in the $R$ or $P$ Rindler wedge, so that $\xi_j^- > 0$.  Then there is a fixed pole near the positive real axis at
\be
w_f = - {\xi_j^- \over \xi_2^-} - i \epsilon_{2j} \,.
\ee
If $\epsilon_{2j} > 0$, so that the fixed pole is located below the positive real axis, it could be pinched against a mobile pole that starts near the negative real axis
and rotates clockwise, which means $\xi_2^+ \rightarrow \xi_2^+ + i \delta$.  On the other hand if $\epsilon_{2j} < 0$, so that the fixed pole is located above the
positive real axis, then it could be pinched against a mobile pole that starts near the negative real axis and rotates counter-clockwise, which means
$\xi_2^+ \rightarrow \xi_2^+ - i \delta$.  Note that whether the mobile pole starts above or below the negative real axis doesn't matter.

\centerline{\includegraphics{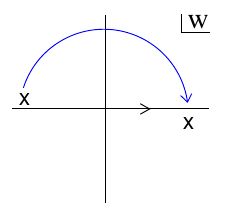} \hspace{2cm} \includegraphics{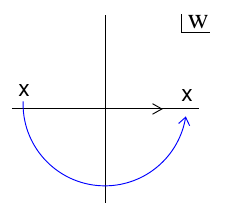}}

Again the sign in front of $i\delta$ is correlated with the sign of $\epsilon_{2j}$, so we can summarize the outcome as
\be
\xi_2^+ \rightarrow \xi_2^+ + i \epsilon_{2j} \,.
\ee
\end{enumerate}

Now we can list how the singularities in (\ref{mobile-fixed}), (\ref{mobile-measure}) are resolved.
\begin{itemize}
\item
The singularities at $\xi_2^+ \xi_2^- = \xi_j^+ \xi_k^-$ are resolved by\footnote{Since $\xi_2^+ \xi_2^- < 0$, $\xi_j^+$ and $\xi_k^-$ must have opposite signs to encounter the singularity.  The one that is positive controls how the singularity is resolved, while the other one just comes along for the ride.}
\bea
\nonumber
&& \xi_2^+ \rightarrow \xi_2^+ - i \epsilon_{2j} \quad \hbox{\rm if $\xi_j^+ > 0$ and $\xi_k^- < 0$} \\[5pt]
&& \xi_2^+ \rightarrow \xi_2^+ + i \epsilon_{2k} \quad \hbox{\rm if $\xi_k^- > 0$ and $\xi_j^+ < 0$}
\eea
If $\xi_j^+$ and $\xi_k^-$ have the same sign, we don't need to specify a prescription because there is no singularity to resolve.
\item
The singularities at
$\xi_2^+ = \xi_j^+$ are resolved by\footnote{Since $\xi_2^+$ is positive, $\xi_j^+$ is also positive and therefore controls how the singularity is resolved.}
\be
\label{xi2p-prescription}
\xi_2^+ \rightarrow \xi_2^+ - i \epsilon_{2j} \,.
\ee
This is equivalent to saying the singularity is slightly displaced, from $\xi_2^+ = \xi_j^+$ to
\be
\xi_2^+ - i \epsilon_2 = \xi_j^+ - i \epsilon_j \,.
\ee
This is exactly the Wightman prescription that one would inherit from the CFT, so for these singularities no special treatment is needed.
\end{itemize}

We now summarize the rules for resolving singularities in correlators with $\delta H_A$.  We assume the perturbing operator $\Oii$ is inserted in the
$F$ Rindler wedge, and we include the case of colliding mobile poles (\ref{mobile-mobile2}) for completeness.

\bigskip

\boxit{\vbox{
\noindent
Singularities at $\xi_I^+ = \xi_J^+$, where $I,J$ include the perturbing operator $\Oii$, are resolved by the Wightman prescription inherited from the CFT.
\be
\label{inherited}
\xi_I ^+ \rightarrow \xi_I^+ - i \epsilon_I
\ee
Singularities at $\xi_2^+ \xi_2^- = \xi_j^+ \xi_k^-$, where $j,k$ are spectator operators including the case $j = k$, are resolved by
\bea
\nonumber
&& \xi_2^+ \rightarrow \xi_2^+ - i \epsilon_{2j} \quad \hbox{\rm if $\xi_j^+ > 0$ and $\xi_k^- < 0$} \\[5pt]
\label{prescription}
&& \xi_2^+ \rightarrow \xi_2^+ + i \epsilon_{2k} \quad \hbox{\rm if $\xi_k^- > 0$ and $\xi_j^+ < 0$}
\eea}}

\noindent
It is worth commenting on the resolution (\ref{prescription}).  From the CFT point of view, a singularity at $\xi_2^+ \xi_2^- = \xi_j^+ \xi_k^-$ would seem to
involve three distinct infinitesimal quantities $\epsilon_2$, $\epsilon_j$, $\epsilon_k$, although one would expect them to only enter in the combinations $\epsilon_{2j}$
and $\epsilon_{2k}$.  We see that in fact, due to the analytic continuation $r \, : \, 0 \rightarrow \pi$, only one of the two combinations plays a role.  Moreover the
combination that appears, and the sign in front of the combination, is determined by which of the spectator operators contributes a positive light-front coordinate to
producing the singularity.

\subsection{General two-point correlator with $\delta H_A$}

We now have a recipe for determining general correlators involving $\delta H_A$.  The first step is to determine the correlator at generic (non-singular) points,
given by a contour integral that surrounds the rotated poles.  The next step is to resolve singularities, following the prescription developed in section \ref{sect:resolve}.
Here we give examples of two-point correlators built from operators with $\Delta = 2$ and $n = 0$.  We adopt the notation $\langle \left\lbrace \Oi, \delta H_A \right\rbrace \rangle$ to indicate that the order of operators in the correlator is
determined by the Wightman prescription.

In this case the generic correlator was already obtained in (\ref{Delta2n0}).
\be\label{eq:2pb4iep}
\langle \left\lbrace {\cal O}(\xi_1^+,\xi_1^-), \, \delta H_A \right\rbrace \rangle = - 4 \pi \lambda {\xi_1^+ \xi_2^+ \left((\xi_1^+)^2 \xi_1^- - (\xi_2^+)^2 \xi_2^-\right) \over
\big(\xi_1^+ - \xi_2^+\big)^3 \big(\xi_2^+ \xi_2^- - \xi_1^+ \xi_1^-\big)^3} \,.
\ee
The singularity at $\xi_1^+ = \xi_2^+$ is to be resolved by the Wightman prescription (\ref{inherited}).  For the singularity at $\xi_2^+ \xi_2^- = \xi_1^+ \xi_1^-$
there are two possibilities.
\begin{enumerate}
\item
If $\Oi$ is in the $F$ wedge, so that $\xi_1^+ > 0$ and $\xi_1^- < 0$, it is resolved by $\xi_2^+ \rightarrow \xi_2^+ + i \epsilon_{12}$ or equivalently (since
$t_1$ is positive in $F$) by $\xi_2^+ \rightarrow \xi_2^+ + i \epsilon_{12}t_1$.
\item
If $\Oi$ is in the $P$ wedge, so that $\xi_1^+ < 0$ and $\xi_1^- > 0$, it is resolved by $\xi_2^+ \rightarrow \xi_2^+ - i \epsilon_{12}$ or equivalently (since
$t_1$ is negative in $P$) by $\xi_2^+ \rightarrow \xi_2^+ + i \epsilon_{12}t_1$.
\end{enumerate}
The resulting correlator can be written in a unified way as (remember that $\xi_2^-$ is negative)
\be
\label{resolved2pt}
\langle \left\lbrace {\cal O}(\xi_1^+,\xi_1^-), \, \delta H_A \right\rbrace \rangle = - 4 \pi \lambda {\xi_1^+ \xi_2^+ \left((\xi_1^+)^2 \xi_1^- - (\xi_2^+)^2 \xi_2^-\right) \over
\big(\xi_1^+ - \xi_2^+ - i \epsilon_{12}\big)^3 \big(\xi_2^+ \xi_2^- - \xi_1^+ \xi_1^- - i \epsilon_{12} t_1\big)^3} \,.
\ee

\subsection{General two-point correlator with $\delta H_{\bar{A}}$\label{subsec:iepdeltahabar}}

The same game can be played with a two-point correlator involving $\delta H_{\bar{A}}$. The analysis is similar to what we just encountered for $\delta H_A$, so we will be rather brief and only point out some necessary equations and details. 

Starting from the Sarosi-Ugajin formula \eqref{deltaHAbar}, the expression of $\delta H_{\bar{A}}$ analogous to (\ref{operatordeltaHA}) is
\begin{equation}\label{eq:deltahabarSU}
	\delta H_{\bar{A}}=i\lambda\int_{0}^{\infty}\frac{dw}{(w+1)^2}\left(-\frac{1}{w}\right)^n\left[\mathcal{O}(-\frac{1}{w}\xi_2^+,e^{-ir}w\xi_2^-)-\mathcal{O}(-\frac{1}{w}\xi_2^+,e^{ir}w\xi_2^-)\right]\,.
\end{equation}
We expect $\delta H_{\bar{A}}$ to be an operator in the left wedge.  To avoid singularities, it's convenient to introduce a spectator operator $\Oiii=\mathcal{O}(\xi_3^+,\xi_3^-)$ in the right wedge, with
\be
\xi_3^+ > 0 \quad {\rm and} \quad \xi_3^- > 0 \,.
\ee
Using the two-point function \eqref{OOcorrelator}, we find that the poles in $\langle \delta H_{\bar{A}}\,\mathcal{O}(\xi_3^+,\xi_3^-)\rangle$ are located at (in addition to the pole at $w=w_0=-1$)
\bea
\nonumber
&& w=w_f=-\frac{\xi_2^+}{\xi_3^+}+i\epsilon_{23} \qquad\quad \text{(fixed pole)} \\
\label{eq:2pfpoles}
&& w=w_m = e^{\pm ir}\left(\frac{\xi_3^-}{\xi_2^-}+i\epsilon_{23}\right)\qquad \text{(mobile pole)}\,.
\eea
When inserted in a correlator, the fixed pole comes from both terms of \eqref{eq:deltahabarSU}, while the $\pm$ sign indicates
a mobile pole coming from the first and second term of \eqref{eq:deltahabarSU}, respectively.

Continuing $r \, : \, 0 \rightarrow \pi$, while performing the appropriate contour gymnastics for various locations of $\Oiii$ in the correlator, and also for $\Oiii$ inserted in various wedges, we find that -- as long as there are no pinch singularities -- the two-point function is given by
\begin{equation}\label{FJdeltaHAbar4}
	\langle\{\delta H_{\bar{A}},\mathcal{O}(\xi_3^+,\xi_3^-)\}\rangle = 
	i \lambda \hspace{-5mm} \oint\limits_{\substack{{\rm clockwise} \\ w = - {\xi_3^- \over \xi_2^-}}} \hspace{-2mm} {dw \over (w + 1)^2} \left(-{1 \over w}\right)^n
{1 \over (\frac{1}{w}\xi_2^+ + \xi_3^+)^{\Delta + n} (w\xi_2^- + \xi_3^-)^{\Delta - n}} \,.
\end{equation}
This equation is the analog of (\ref{FJdeltaHA4}), but for $\delta H_{\bar{A}}$.
The integration contour runs clockwise (opposite to (\ref{FJdeltaHA4})!), encircling the rotated pole at $w=-\frac{\xi_3^-}{\xi_2^-}$.  Since we have assumed there are no
singularities (generic operator positions), the integral has a smooth $\epsilon_{23} \to 0$ limit, and we have dropped all the $i\epsilon_{23}$'s.  As a concrete example, for $\Delta=2$ and $n=0$, and re-labeling $\Oiii$ as $\Oi$ to facilitate comparison
with \eqref{eq:2pb4iep}, we find
\begin{equation}\label{eq:habar2pf}
	\langle\{\delta H_{\bar{A}},\mathcal{O}(\xi_1^+,\xi_1^-)\}\rangle=-\frac{4\pi \lambda\, \xi_2^-\xi_1^-\left[(\xi_2^-)^2\xi_2^+-(\xi_1^-)^2\xi_1^+\right]}{(\xi_{21}^-)^3(\xi_2^+\xi_2^--\xi_1^+\xi_1^-)^3}\,.
\end{equation}

Now that we know what the generic two-point function looks like (as in section \ref{sect:2ptNonSing}) we can examine the way in which singularities
are resolved (as in section \ref{sect:resolve}). In a general $n$-point correlator involving $\delta H_{\bar{A}}$, there are mobile poles at
\begin{equation}\label{eq:m-pole-generic}
	w=w_m=e^{\pm ir}\left(\frac{\xi_k^-}{\xi_2^-}+i\epsilon_{2k}\right)\,,
\end{equation}
and there are fixed poles at 
\begin{equation}\label{eq:f-pole-generic}
	w=w_0=-1\qquad\text{and}\qquad w=w_f=-\frac{\xi_2^+}{\xi_k^+}+i\epsilon_{2k}\,.
\end{equation}
This generalizes the two-point result \eqref{eq:2pfpoles}.
Once again, the collision between the mobile poles is handled by the standard Wightman prescription, namely 
\begin{equation}
	\xi_k^- \rightarrow \xi_k^- + i \epsilon_k \,.
\end{equation}
A collision between $w_m$ and $w_f$ requires more care. It gives rise to singularities of the form $\xi_k^-\xi_j^+=\xi_2^+\xi_2^-$. Following a similar analysis as in section \ref{sect:resolve}, we realize that the first term of \eqref{eq:deltahabarSU}, in which the mobile pole rotates counter-clockwise, picks up the prescription
$\xi_2^-\to \xi_2^--i\delta$. The second term gives rise to a clockwise rotation of the mobile poles and picks up the prescription $\xi_2^-\to \xi_2^-+i\delta$.
Looking at the various possibilities, we find that
\begin{equation}
	\xi_2^-\to \xi_2^-+i\epsilon_{2k} \qquad\text{for}\qquad \xi_k^-<0 \quad \text{and}\quad \xi_j^+>0
\end{equation}
and 
\begin{equation}
	\xi_2^-\to \xi_2^--i\epsilon_{2j} \qquad\text{for}\qquad \xi_k^->0 \quad \text{and}\quad \xi_j^+<0\,.
\end{equation}
The above also helps us evaluate what happens when $w_m$ hits $w_0$, which turns out to be the standard Wightman prescription. To summarize,

\bigskip

\boxit{\vbox{
\noindent
Singularities at $\xi_I^- = \xi_J^-$, where $I,J$ include the perturbing operator $\Oii$, are resolved by the Wightman prescription inherited from the CFT.
\be
\label{inherited2}
\xi_I ^- \rightarrow \xi_I^- + i \epsilon_I
\ee
Singularities at $\xi_2^+ \xi_2^- = \xi_j^+ \xi_k^-$, where $j,k$ are spectator operators including the case $j = k$, are resolved by
\bea
\nonumber
&& \xi_2^- \rightarrow \xi_2^- + i \epsilon_{2k} \quad \hbox{\rm if $\xi_k^- < 0$ and $\xi_j^+ > 0$} \\[5pt]
\label{prescription2}
&& \xi_2^- \rightarrow \xi_2^- - i \epsilon_{2j} \quad \hbox{\rm if $\xi_k^- > 0$ and $\xi_j^+ < 0$}
\eea}}

\noindent
This prescription tells us how to resolve the singularities of any correlator involving $\delta H_{\bar{A}}$.  It's the analog of the prescription \eqref{inherited}, \eqref{prescription} that we obtained for $\delta H_A$.  For example, for $\Delta = 2$ and $n = 0$,
applying the prescription to the generic two-point correlator \eqref{eq:habar2pf} leads to
\begin{equation}\label{eq:habar2pfcorr}
	\langle\{\delta H_{\bar{A}},\mathcal{O}(\xi_1^+,\xi_1^-)\}\rangle=-\frac{4\pi \lambda\, \xi_2^-\xi_1^-\left[(\xi_2^-)^2\xi_2^+-(\xi_1^-)^2\xi_1^+\right]}{(\xi_{21}^-+i\epsilon_{21})^3(\xi_2^+\xi_2^--\xi_1^+\xi_1^--i\epsilon_{12}t_1)^3}\,.
\end{equation}

\subsection{General two-point correlator with $\delta \Hext$}

It's straightforward to combine our results to obtain the general two-point correlator of $\delta \Hext = \delta H_A - \delta H_{\bar{A}}$.
For example, for $\Delta = 2$ and $n = 0$, we can subtract (\ref{eq:habar2pfcorr}) from (\ref{resolved2pt}) to obtain
\begin{eqnarray}\label{eq:hext2pf}
	\langle\{\delta \Hext,\mathcal{O}(\xi_1^+,\xi_1^-)\}\rangle=\frac{4\pi \lambda(\xi_2^+\xi_1^--\xi_1^+\xi_2^-)}{(\xi_{12}^+ - i \epsilon_{12})^3 (\xi_{12}^- + i \epsilon_{12})^3}\,.
\end{eqnarray}
There is a significant cancellation in the $\delta \Hext$ two-point correlator: unlike correlators of $\delta H_A$ or $\delta H_{\bar{A}}$, it only has singularities when the spectator operator is null separated from the perturbation.  We return to this simplification, and explain its origins, in section \ref{sect:simplifying} (see (\ref{simple2pt})).  We will find that such a simplification is common but not universal in correlators involving
$\delta \Hext$.

\section{Three-point correlators involving $\delta\Hext$}\label{sec:3pc}

We now move to the study of three-point correlators involving $\delta\Hext$ and two spectator operators.
We begin with the CFT 3-point correlator of operators ${\cal O}_i$ with dimensions $\Delta_i$ and modular weights $n_i$.  These are related to the left- and right-moving
conformal dimensions by $h_i = {\Delta_i + n_i \over 2}$, $\bar{h}_i = {\Delta_i - n_i \over 2}$.  The Wightman correlator is
\bea
\nonumber
\langle \left\lbrace \Oi,\, \Oii,\, \Oiii \right\rbrace \rangle & = & {1 \over (\xi_{12}^+ - i \epsilon_{12})^{h_1 + h_2 - h_3} (\xi_{13}^+ - i \epsilon_{13})^{h_1 + h_3 - h_2} (\xi_{23}^+ - i \epsilon_{23})^{h_2 + h_3 - h_1}} \\
\label{eq:3pfarbitloc}
& & {1 \over (\xi_{12}^- + i \epsilon_{12})^{\bar{h}_1 + \bar{h}_2 - \bar{h}_3} (\xi_{13}^- + i \epsilon_{13})^{\bar{h}_1 + \bar{h}_3 - \bar{h}_2} (\xi_{23}^- + i \epsilon_{23})^{\bar{h}_2 + \bar{h}_3 - \bar{h}_1}} \,.
\eea
We are using the notation $\lbrace \cdot,\cdot,\cdot \rbrace$ introduced in \eqref{eq:shorthand}, in which operator ordering is fixed by the prescription $t_i \rightarrow t_i - i \epsilon_i$
with the requirement that $\epsilon_i$ decreases monotonically from
left to right in the correlator.

As before, our perturbing operator $\Oii$ will always be inserted in the future wedge, although a similar analysis can be performed when the perturbation is in the past wedge. We will once again analyze correlators involving $\delta H_A$ and $\delta H_{\bar{A}}$ separately, then combine the results to obtain three-point correlators involving $\delta \Hext$. 

\subsection {Three-point correlators with $\delta H_A$\label{sect:threepoint}}

We place $\Oii$ in the future Rindler wedge and use it to perturb the state.  Then at generic (meaning non-singular) points the three-point correlator with $\delta H_A$ can
be obtained by repeating the steps in section \ref{sect:2ptNonSing}, which lead to a contour integral encircling the mobile poles.\footnote{Note that, just as in footnote \ref{footnote:nosing}, the integrand in (\ref{3ptContour}) has good behavior $\sim w^{\Delta_2}$ as $w \rightarrow 0$ and $\sim {1 \over w^{\Delta_2 + 2}}$ as $w \rightarrow \infty$.  This makes the contour manipulations of section \ref{sect:2ptNonSing} possible.\label{footnote:nosing2}}  Additional details and the explicit locations of the poles may be found in appendix \ref{appendix:contact}.
As $r \rightarrow \pi$ we find
\bea
\nonumber
&& \langle \left\lbrace \Oi,\, \delta H_A,\, \Oiii \right\rbrace \rangle = {i \lambda \over (\xi_{13}^+)^{h_1+h_3-h_2} (\xi_{13}^-)^{\bar{h}_1 + \bar{h}_3 - \bar{h}_2}}
\oint\limits_{\substack{{\rm counter-clockwise} \\ w = -\frac{\xi_2^+}{\xi_1^+},\,-\frac{\xi_2^+}{\xi_3^+}}} {dw \over (w + 1)^2} \, \left(-{1 \over w}\right)^{n_2} \\
\nonumber
& & \hspace{5cm} {1 \over (\xi_1^+ + {1 \over w} \xi_2^+)^{h_1 + h_2 - h_3} (- {1 \over w} \xi_2^+ - \xi_3^+)^{h_2 + h_3 - h_1}} \\
\label{3ptContour}
& & \hspace{5cm} {1 \over (\xi_1^- + w \xi_2^-)^{\bar{h}_1 + \bar{h}_2 - \bar{h}_3} (-w \xi_2^- - \xi_3^-)^{\bar{h}_2 + \bar{h}_3 - \bar{h}_1}} \,.
\eea
Here the integration contour encircles the rotated poles at $w = -\xi_2^+/\xi_1^+$ and $w = - \xi_2^+ / \xi_3^+$ with a counter-clockwise orientation.
For the simple case of $\Delta_i = 2$ and $n_i = 0$ the integral straightforwardly gives
\bea\label{eq:deltaHa3pf}
\nonumber
\langle \left\lbrace \Oi,\, \delta H_A,\, \Oiii \right\rbrace \rangle & = & - {2 \pi \lambda \xi_2^+ \over (\xi_{13}^+)^2 \xi_{13}^-} \bigg[
{(\xi_1^+)^2 \over (\xi_{12}^+)^2 (\xi_2^+ \xi_2^- - \xi_1^+ \xi_1^-) (\xi_2^+\xi_2^- - \xi_1^+\xi_3^-)} \\
& & \hspace{2cm} - {(\xi_3^+)^2 \over (\xi_{23}^+)^2(\xi_2^+ \xi_2^- - \xi_3^+ \xi_3^-)(\xi_2^+\xi_2^- - \xi_3^+\xi_1^-)} \bigg] \,.
\eea

\begin{figure}
\centerline{\includegraphics[height=5cm]{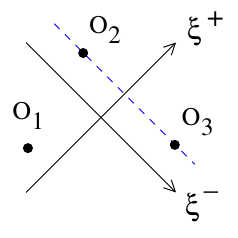} \hspace{2.5cm} \includegraphics[height=5cm]{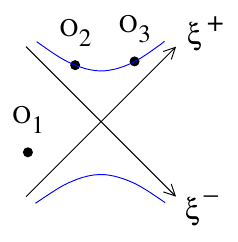}}
\vspace{1.5cm}
\centerline{\includegraphics[height=5cm]{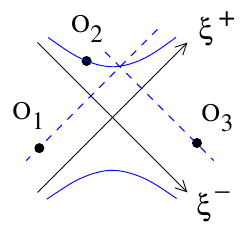} \hspace{2.5cm} \includegraphics[height=5cm]{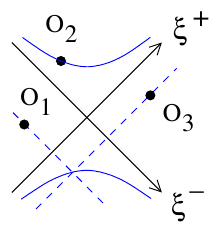}}
\caption{The correlator is singular (i) when $\Oiii$ is null separated from the perturbing operator $\Oii$, (ii) when $\Oiii$ lies on a space-like hyperbola that passes through $\Oii$
or its CPT conjugate $\widetilde\Oii$, (iii) when $\Oi$ and $\Oiii$ are connected by a null ray that bounces off the space-like hyperbola in $F$, (iv) when $\Oi$ and $\Oiii$ are connected by a null ray that bounces off the space-like hyperbola in $P$.\label{fig:singularities}}
\end{figure}

Now let's see how the singularities in (\ref{eq:deltaHa3pf}) are resolved.
In the analysis of three-point functions, it is easier to keep one of the spectator operators fixed in a given wedge while we vary the location of the other spectator operator. Since $\delta H_A$ is an operator in the right wedge, it's simplest to restrict attention to the case where $\Oi$ is inserted in the left wedge, with $\xi_1^+ < 0$ and $\xi_1^- < 0$.  We will indicate this by $\OiL$.
However we leave
the position of $\Oiii$ arbitrary.  Then, following the prescription for resolving singularities (\eqref{inherited} and \eqref{prescription}), we obtain
\bea\label{eq:deltaHa3pfiep}
\nonumber
\langle \big\lbrace \OiL,\, \delta H_A,\, \Oiii \big\rbrace \rangle & = & - {2 \pi \lambda \xi_2^+ \over (\xi_{13}^+ - i \epsilon_{13})^2 (\xi_{13}^- + i \epsilon_{13})} \bigg[
{(\xi_1^+)^2 \over (\xi_{12}^+)^2 (\xi_2^+ \xi_2^- - \xi_1^+ \xi_1^-) (\xi_2^+\xi_2^- - \xi_1^+\xi_3^- - i \epsilon_{23})} \\
\nonumber
& & \hspace{2cm} - {(\xi_3^+)^2 \over (\xi_{23}^+ - i \epsilon_{23})^2(\xi_2^+ \xi_2^- - \xi_3^+ \xi_3^- + i \epsilon_{23} t_3)(\xi_2^+\xi_2^- - \xi_3^+\xi_1^- + i \epsilon_{23})} \bigg] \,. \\
\eea
This result can be understood as follows.  It's important that $\OiL$ is in the $L$ wedge, with $\xi_1^+ < 0$ and $\xi_1^- < 0$, and that $\Oii$ is in the
$F$ wedge, with $\xi_2^+ \xi_2^- < 0$.
\begin{itemize}
\item
Singularities when two operators are null separated are resolved following the CFT Wightman prescription.
\item
The singularity at $\xi_2^+ \xi_2^- = \xi_3^+ \xi_3^-$ is resolved as in (\ref{resolved2pt}).
\item
Since $\xi_1^+ < 0$, the singularity at $\xi_2^+\xi_2^- = \xi_1^+\xi_3^-$ occurs when $\xi_3^- > 0$.  Then $\xi_2^+ \rightarrow \xi_2^+ + i \epsilon_{23}$, but multiplying
by $\xi_2^-$ flips the sign.
\item
Since $\xi_1^- < 0$, the singularity at $\xi_2^+\xi_2^- = \xi_3^+\xi_1^-$ occurs when $\xi_3^+ > 0$.  Then $\xi_2^+ \rightarrow \xi_2^+ - i \epsilon_{23}$, but again multiplying
by $\xi_2^-$ flips the sign.
\end{itemize}
The three-point correlator has the expected light-cone singularities when $\OiL$ and $\Oiii$ are null separated.  There are additional singularities, involving $\delta H_A$, that have
a simple geometric interpretation given in Fig.\ \ref{fig:singularities}.

\subsection{Three-point correlators with $\delta H_{\bar{A}}$}\label{subsec:3pfdeltahabarnonsing}

We now treat three-point functions involving $\delta H_{\bar{A}}$. Using the formula for $\delta H_{\bar{A}}$ \eqref{eq:deltahabarSU} and the 3-point correlator \eqref{eq:3pfarbitloc}, we find that for generic (non-singular) operator locations for $\mathcal{O}(\xi_1^+,\xi_1^-)$ and $\mathcal{O}(\xi_3^+,\xi_3^-)$, one ends up with an integration contour that encircles the rotated poles at $w=-\frac{\xi_1^-}{\xi_2^-}$ and $w=-\frac{\xi_3^-}{\xi_2^-}$ in a clockwise direction.
\bea
\nonumber
&& \langle \left\lbrace \Oi,\, \delta H_{\bar{A}},\, \Oiii \right\rbrace \rangle = {i \lambda \over (\xi_{13}^+)^{h_1+h_3-h_2} (\xi_{13}^-)^{\bar{h}_1 + \bar{h}_3 - \bar{h}_2}}
\oint\limits_{\substack{{\rm clockwise} \\ w = -\frac{\xi_1^-}{\xi_2^-},\,-\frac{\xi_3^-}{\xi_2^-}}} {dw \over (w + 1)^2} \, \left(-{1 \over w}\right)^{n_2} \\
\nonumber
& & \hspace{5cm} {1 \over (\xi_1^+ + {1 \over w} \xi_2^+)^{h_1 + h_2 - h_3} (- {1 \over w} \xi_2^+ - \xi_3^+)^{h_2 + h_3 - h_1}} \\
\label{3ptContour2}
& & \hspace{5cm} {1 \over (\xi_1^- + w \xi_2^-)^{\bar{h}_1 + \bar{h}_2 - \bar{h}_3} (-w \xi_2^- - \xi_3^-)^{\bar{h}_2 + \bar{h}_3 - \bar{h}_1}}
\eea
For example, for $\Delta_i = 2$ and $n_i = 0$, evaluating the residues gives
\begin{eqnarray}\label{eq:deltahabar3p}
	\langle\{\mathcal{O}(\xi_1^+,\xi_1^-),\delta H_{\bar{A}},\mathcal{O}(\xi_3^+,\xi_3^-)\}\rangle=-\frac{2\pi\lambda\,\xi_2^-}{(\xi_{13}^-)^2\xi_{13}^+}\Big[\frac{(\xi_1^-)^2}{(\xi_{12}^-)^2(\xi_2^+\xi_2^--\xi_1^+\xi_1^-)(\xi_2^+\xi_2^--\xi_1^-\xi_3^+)}\nonumber\\
	-\frac{(\xi_3^-)^2}{(\xi_{23}^-)^2(\xi_2^+\xi_2^--\xi_3^+\xi_3^-)(\xi_2^+\xi_2^--\xi_1^+\xi_3^-)}\Big]\,.
\end{eqnarray}

Next we consider how the singularities in (\ref{eq:deltahabar3p}) are resolved, using the prescriptions discussed in section \ref{subsec:iepdeltahabar}. But much like the analysis of $\delta H_A$ in section \ref{sect:threepoint}, we will first fix one of the spectator operators, namely $\mathcal{O}(\xi_3^+,\xi_3^-)$, in the right wedge, denoting it $\OiiiR$. This simplifies matters because $\delta H_{\bar{A}}$ is an operator in the left wedge. However, we will keep the location of $\mathcal{O}(\xi_1^+,\xi_1^-)$ arbitrary.
Then the prescriptions in section \ref{subsec:iepdeltahabar} lead to
\begin{align}\label{eq:habar3pfcorrotherfixing}
	&\langle\{\mathcal{O}(\xi_1^+,\xi_1^-),\delta H_{\bar{A}},\OiiiR(\xi_3^+,\xi_3^-)\}\rangle=-\frac{2\pi\lambda\,\xi_2^-}{(\xi_{13}^-+i\epsilon_{13})^2(\xi_{13}^+-i\epsilon_{13})}\nonumber\\
	&\Big[\frac{(\xi_1^-)^2}{(\xi_{12}^-+i\epsilon_{12})^2(\xi_2^+\xi_2^--\xi_1^+\xi_1^--i\epsilon_{12}t_1)(\xi_2^+\xi_2^--\xi_1^-\xi_3^+-i\epsilon_{12})}
	-\frac{(\xi_3^-)^2}{(\xi_{23}^-)^2(\xi_2^+\xi_2^--\xi_3^+\xi_3^-)(\xi_2^+\xi_2^--\xi_1^+\xi_3^-+i\epsilon_{12})}\Big]\,.
\end{align}
This result can be understood as follows.  It's important that $\OiiiR$ is in the $R$ wedge, with $\xi_3^+ > 0$ and $\xi_3^- > 0$, and that $\Oii$ is in the
$F$ wedge, with $\xi_2^+ \xi_2^- < 0$.
\begin{itemize}
\item
Singularities when two operators are null separated are resolved following the CFT Wightman prescription.
\item
The singularity at $\xi_2^+ \xi_2^- = \xi_1^+ \xi_1^-$ is resolved as in (\ref{resolved2pt}).
\item
Since $\xi_3^+ > 0$, the singularity at $\xi_2^+\xi_2^- = \xi_3^+\xi_1^-$ occurs when $\xi_1^- < 0$.  Then $\xi_2^- \rightarrow \xi_2^- + i \epsilon_{21} = \xi_2^- - i \epsilon_{12}$, and multiplying
by $\xi_2^+$ preserves the sign.
\item
Since $\xi_3^- > 0$, the singularity at $\xi_2^+\xi_2^- = \xi_1^+\xi_3^-$ occurs when $\xi_1^+ < 0$.  Then $\xi_2^- \rightarrow \xi_2^- - i \epsilon_{21}= \xi_2^- + i \epsilon_{12}$, and again multiplying
by $\xi_2^+$ preserves the sign.
\end{itemize}

\subsection{Three-point correlators with $\delta \Hext$}

We proceed to determine the correlator $\langle\{\mathcal{O}(\xi_1^+,\xi_1^-),\delta \Hext,\mathcal{O}(\xi_3^+,\xi_3^-)\}\rangle$ for $\Delta_i = 2$ and $n_i = 0$.
We take $\mathcal{O}(\xi_1^+,\xi_1^-)$ to lie in the left wedge and ${\cal O}(\xi_3^+,\xi_3^-)$ to lie in the right wedge, so that the correlator with $\delta H_{\bar{A}}$ is
\begin{align}\label{eq:habar3pfcorrotherfixingboth}
	&\langle\{\mathcal{O}^{(L)}(\xi_1^+,\xi_1^-),\,\delta H_{\bar{A}},\,\mathcal{O}^{(R)}(\xi_3^+,\xi_3^-)\}\rangle=-\frac{2\pi\lambda\,\xi_2^-}{(\xi_{13}^-)^2(\xi_{13}^+)}\nonumber\\
	&\Big[\frac{(\xi_1^-)^2}{(\xi_{12}^-+i\epsilon_{12})^2(\xi_2^+\xi_2^--\xi_1^+\xi_1^-)(\xi_2^+\xi_2^--\xi_1^-\xi_3^+-i\epsilon_{12})}
	-\frac{(\xi_3^-)^2}{(\xi_{23}^-)^2(\xi_2^+\xi_2^--\xi_3^+\xi_3^-)(\xi_2^+\xi_2^--\xi_1^+\xi_3^-+i\epsilon_{12})}\Big]\,.
\end{align}
The aim now is to subtract this from \eqref{eq:deltaHa3pfiep}, with $O(\xi_3^+,\xi_3^-)$ in the latter equation located in the right wedge. We rewrite that equation here for the reader's convenience. 
\begin{align}\label{eq:habar3pfcorrotherfixingbothDan}
	&\langle\{\mathcal{O}^{(L)}(\xi_1^+,\xi_1^-),\,\delta H_{A},\,\mathcal{O}^{(R)}(\xi_3^+,\xi_3^-)\}\rangle=-\frac{2\pi\lambda\,\xi_2^+}{(\xi_{13}^-)(\xi_{13}^+)^2}\nonumber\\
	&\Big[\frac{(\xi_1^+)^2}{(\xi_{12}^+)^2(\xi_2^+\xi_2^--\xi_1^+\xi_1^-)(\xi_2^+\xi_2^--\xi_1^+\xi_3^--i\epsilon_{23})}
	-\frac{(\xi_3^+)^2}{(\xi_{23}^+-i\epsilon_{23})^2(\xi_2^+\xi_2^--\xi_3^+\xi_3^-)(\xi_2^+\xi_2^--\xi_1^-\xi_3^++i\epsilon_{23})}\Big]
\end{align}
The subtraction is simplest at non-singular points, where $\epsilon_{12}$ and $\epsilon_{23}$ can be neglected.  At non-singular points, the result (valid for spectator operators
in any wedge) is simply
\be
\label{3ptcommutator}
\langle\{\mathcal{O}(\xi_1^+,\xi_1^-),\delta \Hext,\mathcal{O}(\xi_3^+,\xi_3^-)\}\rangle = 2 \pi \lambda \Big(\xi_2^+ {\partial \over \partial \xi_2^+} - \xi_2^- {\partial \over \partial \xi_2^-}\Big) \,
{1 \over \xi_{12}^+ \, \xi_{13}^+ \, \xi_{23}^+ \, \xi_{12}^- \, \xi_{13}^- \, \xi_{23}^-} \,.
\ee
We return to this simplification, which amounts to the statement that $\delta \Hext$ can be replaced by $2 \pi \lambda$ times an infinitesimal Lorentz boost of the perturbation, in section \ref{sect:simplifying}.  Note, however,
that the simplification relies on being able to neglect $\epsilon_{12}$ and $\epsilon_{23}$.  At singular points, where this is not possible, the result is more complicated.  We discuss this further in section \ref{sect:contact}.

\section{Operator expression for $\delta \Hext$}\label{sec:exop}

So far we've developed a prescription for calculating correlators involving $\delta H_A$ and $\delta H_{\bar{A}}$.  The prescription is
based on an analytic continuation $r \, : \, 0 \rightarrow \pi$.  In studying the behavior as $r \rightarrow \pi^-$, a particular order of limits
must be taken.  The rule is to set $r = \pi - \delta$, and to send $\delta \rightarrow 0^+$ more slowly than any other infinitesimal
parameter in the problem.  In particular $\delta$ is taken to approach zero more slowly than the infinitesimal parameters $\epsilon_{ij}$ that
define the Wightman correlator.  This prescription is necessary to obtain well-defined correlators, as discussed in footnote \ref{footnote:ambiguous} and appendix \ref{appendix:contact}.

Having understood how to compute correlators involving $\delta H_A$, we'd like to understand how to strip off the spectator operators in the correlator
and obtain an operator expression for $\delta H_A$ itself.  (We begin by considering $\delta H_A$.  Corresponding expressions for $\delta H_{\bar{A}}$ and
$\delta \Hext$ are given below.)

To do this, it's convenient to return to the original Sarosi--Ugajin formula, which we write as (here $\Oii(\xi_2^+,\xi_2^-)$ is the perturbing operator)
\bea
\nonumber
&& \delta H_A = \delta H_A^{(1)} - \delta H_A^{(2)} \\
&& \delta H_A^{(1)} = i \lambda \int_0^\infty {dw \over (w+1)^2} \left(-{1 \over w}\right)^{n_2} \Oii\big(e^{-ir} {1 \over w} \xi_2^+,-w \xi_2^-\big) \\
\nonumber
&& \delta H_A^{(2)} = i \lambda \int_0^\infty {dw \over (w+1)^2} \left(-{1 \over w}\right)^{n_2} \Oii\big(e^{ir} {1 \over w} \xi_2^+,-w \xi_2^-\big)
\eea
The advantage of writing the Sarosi--Ugajin formula in this way is that inside a correlator the poles rotate in a definite direction in the two terms.

It's instructive to approach the problem in a series of attempts.  The first and most naive way to obtain an operator expression for $\delta H_A^{(1)}$
is simply to set $r = \pi$ in Sarosi--Ugajin.  This leads to
\bea
\hbox{\rm first attempt:} \quad \delta H_A^{(1)} = i \lambda \int_0^\infty {dw \over (w+1)^2} \left(-{1 \over w}\right)^{n_2} \Oii\Big(-{1 \over w} \xi_2^+,-w \xi_2^-\Big) \,.
\eea
This is a perfectly well-defined operator.  It's the integral of $\Oii$ over a spacelike hyperbola that passes through the point $(-\xi_2^+,-\xi_2^-)$.  (This point
is CPT conjugate to the location of the perturbation.)  However, as a candidate operator expression for $\delta H_A^{(1)}$, it is a failure.  It involves a local operator
smeared over a region in the past Rindler wedge.  As such, there is no reason for it to commute with operators in the left Rindler wedge, or for it to be regarded as
an element of the right wedge operator subalgebra.

\begin{figure}
\centerline{\includegraphics[height=4.5cm]{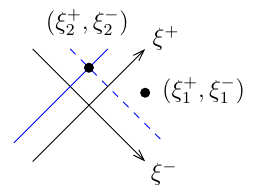}}
\caption{Light cone structure for the correlator $\langle \lbrace {\cal O}(\xi_1^+,\xi_1^-), {\cal O}^{-0}(\xi_2^+,\xi_2^-) \rbrace \rangle$.  The operators commute
on the dashed null ray but not on the solid null ray.\label{fig:half}}
\end{figure}

This is exactly the problem that the continuation $r \, : \, 0 \rightarrow \pi^-$ is supposed to solve.  When $r = 0$ the expression for $\delta H_A^{(1)}$ involves
a local operator integrated over a timelike hyperbola in the right Rindler wedge, and one might hope that as $r \rightarrow \pi^-$ it remains within the right subalgebra.
For this reason we set $r = \pi - \delta$ and consider the operator
\bea
\nonumber
\Oii\big(e^{-ir} {1 \over w} \xi_2^+,-w \xi_2^-\big) & = & \Oii(-e^{i\delta} {1 \over w} \xi_2^+,-w\xi_2^-\big) \\
& \approx &  \Oii(-{1 \over w} \xi_2^+ - i \delta,-w\xi_2^-\big)
\eea
(the approximate equality is for infinitesimal $\delta$, with $w$ and $\xi_2^+$ positive).  This motivates defining an operator
\be
\Oiimnot(\xi_2^+,\xi_2^-) = \Oii(\xi_2^+ - i \delta,\xi_2^-)
\ee
and writing a second attempt at an operator expression for $\delta H_A^{(1)}$ as
\bea
\label{second}
\hbox{\rm second attempt:} \quad \delta H_A^{(1)} = i \lambda \int_0^\infty {dw \over (w+1)^2} \left(-{1 \over w}\right)^{n_2} \Oiimnot\Big(-{1 \over w} \xi_2^+,-w \xi_2^-\Big) \,.
\eea
Is this any better than the first attempt?  To study this question, it is useful to introduce a collection of operators
\bea
{\cal O}_2^{ab}(\xi_2^+,\xi_2^-) = \Oii(\xi_2^+ + a i \delta,\xi_2^- + b i \delta)
\eea
where the labels $a,b$ run over the values $+1$, $-1$, $0$.  These operators have well-defined correlators, for example
\be
\langle \lbrace {\cal O}(\xi_1^+,\xi_1^-), {\cal O}^{ab}(\xi_2^+,\xi_2^-) \rbrace \rangle = {1 \over (\xi_1^+ - \xi_2^+ - a i \delta - i \epsilon_{12})^{\Delta +n}
(\xi_1^- - \xi_2^- - b i \delta + i \epsilon_{12})^{\Delta - n}} \,.
\ee
The case $a = b = 0$ recovers the original local operator of the CFT.  To interpret the other operators, note that due to the order of limits
we have $\delta \gg \vert \epsilon_{12} \vert$.  This leads to
\be
\langle \lbrace {\cal O}(\xi_1^+,\xi_1^-), {\cal O}^{-0}(\xi_2^+,\xi_2^-) \rbrace \rangle = {1 \over (\xi_1^+ - \xi_2^+ + i \delta)^{\Delta + n}
(\xi_1^- - \xi_2^- + i \epsilon_{12})^{\Delta - n}}
\ee
with analogous expressions for the other operators.
At non-null separation, we see that ${\cal O}^{-0}$ is indistinguishable from a local operator whose correlators are defined by a Wightman prescription.
At null separation $\xi_1^- = \xi_2^-$ the correlator is singular, but the singularity is resolved following the standard Wightman prescription.  At null separation
$\xi_1^+ = \xi_2^+$ the correlator again is singular, but now the resolution depends only on $\delta$ and is non-standard.  In particular the resolution is independent of
$\epsilon_{12}$, which means {\em operator ordering does not matter}.  That is, at equal values of $\xi^+$, the operators ${\cal O}$ and ${\cal O}^{-0}$ commute.
We can depict this graphically, as a dashed null ray emanating from $(\xi_2^+,\xi_2^-)$.  See Figure \ref{fig:half}.

We will refer to the operators ${\cal O}^{ab}$ as ``pseudo-local.''  A few comments about these operators are in order.  First, as we've already mentioned, ${\cal O}^{00}$ is
the original local operator of the CFT.  The operators ${\cal O}^{+0}$, ${\cal O}^{-0}$, ${\cal O}^{0+}$, ${\cal O}^{0-}$ could be referred to as ``half-local.''
They commute at null separation in one direction but not the other.  The operators ${\cal O}^{++}$, ${\cal O}^{+-}$, ${\cal O}^{-+}$, ${\cal O}^{--}$ commute
with all local operators in the CFT.  The operators ${\cal O}^{-+}$ and ${\cal O}^{+-}$ can be thought of as Wightman operators, with $\pm \delta$ playing the role
of $\epsilon$.  So ${\cal O}^{-+}$ can be identified with a local operator inserted on the far left in a correlator, and ${\cal O}^{+-}$ can be identified with a local operator inserted on the far right.

\begin{figure}
\centerline{\includegraphics[height=5cm]{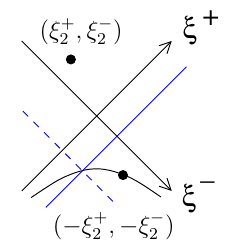}}
\caption{Light cone structure associated with the second attempt at $\delta H_A^{(1)}$.  Since it involves an integral of $\Oiimnot$ over a spacelike
hyperbola in the past Rindler wedge, obtained by boosting the CPT conjugate point $(-\xi_2^+,-\xi_2^-)$, it is guaranteed to commute with local operators in the left Rindler wedge.\label{fig:second}}
\end{figure}

Now let's return to our second attempt at an operator expression for $\delta H_A^{(1)}$, given in (\ref{second}).  Since it involves the pseudo-local
operator ${\cal O}^{-0}$ integrated over a spacelike hyperbola in the past Rindler wedge, then as shown in Figure \ref{fig:second} it commutes with
all local operators in the left wedge.  That is, it behaves like an element of the right operator algebra.

\begin{figure}

\leftline{$r = 0 \, :$ \hspace{4.73mm} \raisebox{-1.3cm}{\includegraphics[height=3cm]{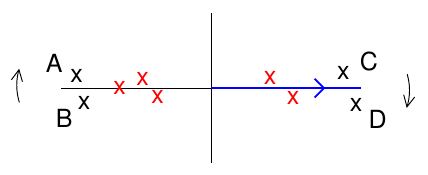}}}

\bigskip

\leftline{$r = \pi - \delta \, :$ \quad \raisebox{-1.4cm}{\includegraphics[height=3cm]{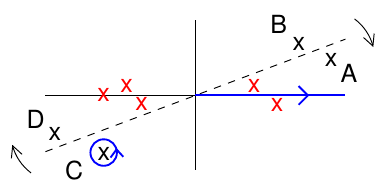}} \quad $=$ \quad \raisebox{-1.43cm}{\includegraphics[height=3cm]{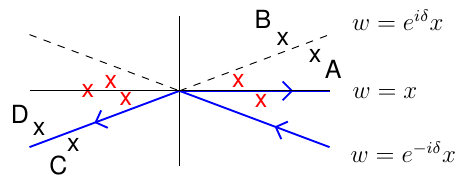}}}

\bigskip

\leftline{\hspace{2.8cm}\includegraphics[height=3cm]{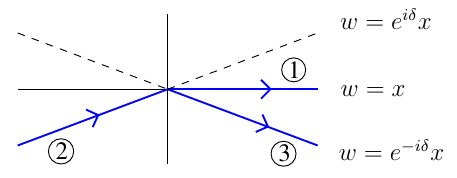}}

\caption{Top panel: the starting point $r = 0$, showing some generic fixed and mobile poles as well as the pole at $w = -1$.  The fixed poles and the pole at $w = -1$
are shown in red.  The mobile poles, shown in black, are labeled {\sf A}, {\sf B}, {\sf C}, {\sf D}.  The mobile poles rotate clockwise in $\delta H_A^{(1)}$.
Middle panel: continuation to $r = \pi - \delta$.  Mobile poles of type {\sf C} get wrapped by the integration contour.  The contour around
{\sf C} can be deformed to a $\wedge$-shaped form in the lower half plane, parametrized by $w = e^{\pm i \delta} x$ for $x \in {\mathbb R}$.  The $\wedge$-shaped contour has the advantage that (after closing the contour at infinity) it only encircles mobile poles of type $C$,
while avoiding all other  fixed and mobile poles.  Moreover it does this in a way that is universal, meaning the contour is independent of the positions of the spectator operators.  Bottom panel: we can strip off the spectator operators to obtain a three-part contour integral expression for $\delta H_A^{(1)}$.\label{fig:deform}}
\end{figure}

Does this mean (\ref{second}) is a good candidate for $\delta H_A^{(1)}$?  It does not, for the following reason.  In continuing $r \, : \, 0 \rightarrow \pi^-$,
it is not enough to analytically continue the integrand.  As we saw by working inside correlators, poles can rotate and collide with the integration contour.  As we analytically continue $r \, : \, 0 \rightarrow \pi^-$, we need to deform the integration contour to avoid crossing any poles.  We need to do this in a way that's universal, independent of the positions of the spectator operators.  It is possible to do this as follows.
A correlator involving $\delta H_A^{(1)}$ generically has mobile poles, located at
\be
\label{op-mobile}
w_m = e^{-ir} \left({\xi_2^+ \over \xi_i^+} + i \epsilon_{i2}\right)\,,
\ee
that rotate clockwise.  It also has fixed poles located at
\be
\label{op-fixed}
w_f = - {\xi_i^- \over \xi_2^-} + i \epsilon_{i2}\,,
\ee
as well as a pole at $w_0 = -1$ from the integration measure.  For generic positions of several mobile and fixed poles, a suitable analytic continuation and
contour deformation is shown in Figure \ref{fig:deform}.

This leads to our third and final attempt at an operator expression.
\bea
\label{1minus2minus3}
&& \delta H_A^{(1)} = \circled{1} - \circled{2} - \circled{3} \\
\nonumber
&& \circled{1} = i \lambda \int_0^\infty {dw \over (w+1)^2} \left(-{1 \over w}\right)^{n_2} \Oii\Big(- e^{i \delta} {1 \over w} \xi_2^+, -w \xi_2^-\Big) \\
\nonumber
&& \circled{2} = i \lambda \int_{-\infty}^0 {e^{i \delta} dx \over \Big(e^{i\delta}x+1\Big)^2} \left(-{1 \over e^{i\delta}x}\right)^{n_2} \Oii\Big(-{1 \over x} \xi_2^+, -e^{i\delta}x \xi_2^-\Big) \\
\nonumber
&& \circled{3} = i \lambda \int_0^\infty {e^{-i \delta} dx \over \Big(e^{-i\delta}x+1)^2} \left(-{1 \over e^{-i\delta}x}\right)^{n_2} \Oii\Big(- e^{2i \delta} {1 \over x} \xi_2^+, -e^{-i\delta}x \xi_2^-\Big)
\eea
Making some approximations appropriate to small $\delta$, using the fact that $\xi_2^\pm$ and $x$ have definite signs when the perturbation is in the future wedge,
and relabeling $x$ as $\pm w$, we obtain the final expression
\bea
\nonumber
\delta H_A^{(1)} &=& i \lambda \int_0^\infty {dw \over (w + 1)^2} \left(-{1\over w}\right)^{n_2} \left.\left(\Oiimnot - \Oiimm\right)\right\vert_{\big(-{1 \over w} \xi_2^+,-w\xi_2^-\big)} \\
&-& i \lambda \int_0^\infty {dw \over (w - 1 + i \delta)^2} \left({1 \over w}\right)^{n_2} \left.\Oiinotm\right\vert_{\big({1\over w} \xi_2^+,w \xi_2^-\big)} \,.
\eea
Note that $\delta H_A^{(1)}$ is built from pseudo-local operators integrated over spacelike hyperbolas in both the past and future wedges, in such a way that it
is guaranteed to commute with local operators in the left wedge.

An identical series of steps leads to an expression for $\delta H_A^{(2)} = - \Big(\delta H_A^{(1)}\Big)^\dagger$.
\bea
\nonumber
\delta H_A^{(2)} &=& i \lambda \int_0^\infty {dw \over (w + 1)^2} \left(-{1\over w}\right)^{n_2} \left.\left(\Oiipnot - \Oiipp\right)\right\vert_{\big(-{1 \over w} \xi_2^+,-w\xi_2^-\big)} \\
&-& i \lambda \int_0^\infty {dw \over (w - 1 - i \delta)^2} \left({1 \over w}\right)^{n_2} \left.\Oiinotp\right\vert_{\big({1\over w} \xi_2^+,w \xi_2^-\big)}
\eea
Assembling the ingredients, we obtain $\delta H_A = \delta H_A^{(1)} - \delta H_A^{(2)}$ as a manifestly Hermitian operator.\footnote{We take Hermitian conjugation to act by, for example, $\left(\Oiimm(\xi^+,\xi^-)\right)^\dagger = \Oiipp(\xi^+,\xi^-)$.}
\bea
\nonumber
\delta H_A &=& + i \lambda \int_0^\infty {dw \over (w + 1)^2} \left(-{1\over w}\right)^{n_2} \left.\left(\Oiimnot - \Oiimm - \Oiipnot + \Oiipp\right)\right\vert_{\big(-{1 \over w} \xi_2^+,-w\xi_2^-\big)} \\
\nonumber
&&- i \lambda \int_0^\infty {dw \over (w - 1 + i \delta)^2} \left({1 \over w}\right)^{n_2} \left.\Oiinotm\right\vert_{\big({1\over w} \xi_2^+,w \xi_2^-\big)} \\
&&+ i \lambda \int_0^\infty {dw \over (w - 1 - i \delta)^2} \left({1 \over w}\right)^{n_2} \left.\Oiinotp\right\vert_{\big({1\over w} \xi_2^+,w \xi_2^-\big)}
\eea
We likewise find (relative to $\delta H_A$, switch the subscripts and insert an overall $-$ sign)
\bea
\nonumber
\delta H_{\bar{A}} &=& - i \lambda \int_0^\infty {dw \over (w + 1)^2} \left(-{1\over w}\right)^{n_2} \left.\left(\Oiinotm - \Oiimm - \Oiinotp + \Oiipp\right)\right\vert_{\big(-{1 \over w} \xi_2^+,-w\xi_2^-\big)} \\
\nonumber
& & + i \lambda \int_0^\infty {dw \over (w - 1 + i \delta)^2} \left({1 \over w}\right)^{n_2} \left.\Oiimnot\right\vert_{\big({1\over w} \xi_2^+,w \xi_2^-\big)} \\
& & - i \lambda \int_0^\infty {dw \over (w - 1 - i \delta)^2} \left({1 \over w}\right)^{n_2} \left.\Oiipnot\right\vert_{\big({1\over w} \xi_2^+,w \xi_2^-\big)} \,.
\eea
Finally, the first-order correction to the extended modular Hamiltonian can be presented as
\bea
\nonumber
\delta \Hext &=& i \lambda \int_0^\infty {dw \over (w + 1)^2} \left(-{1\over w}\right)^{n_2} \left.\left(\Oiimnot + \Oiinotm - 2 \Oiimm\right)\right\vert_{\big(-{1 \over w} \xi_2^+,-w\xi_2^-\big)} \\
\label{op-deltaH}
&-& i \lambda \int_0^\infty {dw \over (w - 1 + i \delta)^2} \left({1 \over w}\right)^{n_2} \left.\left(\Oiimnot + \Oiinotm\right)\right\vert_{\big({1\over w} \xi_2^+,w \xi_2^-\big)} \\
\nonumber
&+& {\rm h.c.}
\eea
One can check that correlators of these operators reproduce (\ref{eq:habar3pfcorrotherfixingboth}), (\ref{eq:habar3pfcorrotherfixingbothDan}).

\section{$\delta\Hext$ as a commutator}

Having obtained an operator expression for $\delta \Hext$, we proceed to point out that $\delta \Hext$ can be written as a commutator, $\delta \Hext = -i \lambda \left[E, \, \Hextvac \right]$.  We will give an expression for $E$ shortly.  In general $E$ takes a rather involved form, much like $\delta \Hext$ itself.  However in many cases of practical
significance $E$ can be replaced by the perturbing operator  $\Oii$.  We detail these cases in section \ref{sect:simplifying}.  The fact that $E$ cannot always be
replaced by $\Oii$ is due to certain singularities in correlators which we refer to as contact terms.  We discuss contact terms briefly
in section \ref{sect:contact} and give expressions in appendix \ref{appendix:contact}.

\subsection{General case: $\delta\Hext$ as a commutator with $E$}

To show that $\delta \Hext$ can be written as a commutator, we return to the starting point.  Consider a perturbed state
\be
\vert \psi \rangle = e^{-i \lambda G} \vert 0 \rangle \,.
\ee
We make the assumption that $G = G_A \otimes G_{\bar{A}}$, and introduce the convenient combinations
\bea
\nonumber
\widehat{G}_A & = &  G_A \big\vert_{- i r} \widetilde{G}_{\bar{A}} - \widetilde{G}_{\bar{A}} G_A \big\vert_{i r} \\[3pt]
\widehat{G}_{\bar{A}} & = &  G_{\bar{A}} \big \vert_{-i r}  \widetilde{G}_A - \widetilde{G}_A G_{\bar{A}} \big\vert_{i r} \,.
\eea
The Sarosi--Ugajin formulas give the first-order change in the subregion modular Hamiltonians.
\bea
\label{commutator_deltaHA}
&& \delta H_A = {i \lambda \over 2} \int_{-\infty}^\infty {ds \over 1 + \cosh s} \, \widehat{G}_A \big\vert_s \\[3pt]
\label{commutator_deltaHAbar}
&& \delta H_{\bar{A}} = - {i \lambda \over 2} \int_{-\infty}^\infty {ds \over 1 + \cosh s} \, \widehat{G}_{\bar{A}} \big\vert_s
\eea
The first-order change in the extended modular Hamiltonian is then
\bea
\nonumber
\delta \Hext & = & \delta H_A - \delta H_{\bar A} \\[5pt]
\label{commutator_deltaHext}
& = & {i \lambda \over 2} \int_{-\infty}^\infty {ds \over 1 + \cosh s} \left( \widehat{G}_A + \widehat{G}_{\bar{A}} \right) \Big\vert_s \,.
\eea

This expression for $\delta \Hext$ can be recast in the form of a commutator acting on $\Hextvac$.  The argument goes as follows.  Recall that vacuum modular flow is defined by
\be
{\cal O}\big\vert_s = e^{i \Hextvac s / 2 \pi} \, {\cal O} \, e^{-i \Hextvac s / 2 \pi}
\ee
which means that
\be
{d \over ds} {\cal O} \big\vert_s = - {i \over 2\pi} \big[ {\cal O}\big\vert_s \, , \Hextvac \big] \,.
\ee
We can re-write (\ref{commutator_deltaHext}) as
\be
\label{commutator_deltaHext2}
\delta \Hext = {i \lambda \over 2} \int_{-\infty}^\infty {ds \over 1 + \cosh s} \left({d \over ds}\right)^{-1} {d \over ds} \left( \widehat{G}_A + \widehat{G}_{\bar{A}} \right) \Big\vert_s \,.
\ee
The derivative produces a commutator, so we have
\be
\label{commutator_deltaHext3}
\delta \Hext = {\lambda \over 4\pi} \left[\int_{-\infty}^\infty {ds \over 1 + \cosh s} \left({d \over ds}\right)^{-1} \left( \widehat{G}_A + \widehat{G}_{\bar{A}} \right) \Big\vert_s, \, \Hextvac \right] \,.
\ee
For reasons that will become clear below, we write this in the form
\be
\label{commutator_deltaHext4}
\delta \Hext = -i \lambda \left[E, \, \Hextvac \right]
\ee
where
\be
\label{commutator_Eext}
E = {i \over 4 \pi} \int_{-\infty}^\infty {ds \over 1 + \cosh s} \left({d \over ds}\right)^{-1} \left( \widehat{G}_A + \widehat{G}_{\bar{A}} \right) \Big\vert_s \,.
\ee
We interpret $\left({d \over ds}\right)^{-1}$ as a Green's function,
\be
\left({d \over ds}\right)^{-1} \left( \widehat{G}_A + \widehat{G}_{\bar{A}} \right) \Big\vert_s = \int_{-\infty}^\infty ds' \, \epsilon(s - s') \left( \widehat{G}_A + \widehat{G}_{\bar{A}} \right) \Big\vert_{s'}
\ee
where
\be
\epsilon(s - s') = \left\lbrace\begin{array}{ll} 1/2 & s > s' \\[3pt] -1/2 & s < s' \end{array}\right.
\ee
Performing the $s$ integral in (\ref{commutator_Eext}) first, then relabeling $s'$ as $s$, we have
\be
E = - {i \over 4 \pi} \int_{-\infty}^\infty ds \, \tanh\big({s \over 2}\big) \left( \widehat{G}_A + \widehat{G}_{\bar{A}} \right) \Big\vert_{s} \,.
\ee
In terms of $w = e^{-s}$ this becomes
\be
E = {i \over 4 \pi} \int_0^\infty {dw \, (w-1) \over w(w + 1)} \left( \widehat{G}_A + \widehat{G}_{\bar{A}} \right) \Big\vert_{s} \,.
\ee
Let us compare this expression for $E$ to the expression for $\delta\Hext$ given in (\ref{commutator_deltaHext}), which in terms of $w$ is
\be
\delta\Hext = {i \lambda} \int_0^\infty {dw \over (w + 1)^2} \left( \widehat{G}_A + \widehat{G}_{\bar{A}} \right) \Big\vert_{s} \,.
\ee
Comparing the expressions we see that for $E$ the factor in front changes,
\be
\label{commutator_change1}
i\lambda \rightarrow {i \over 4\pi}
\ee
and the measure changes,
\be
\label{commutator_change2}
{dw \over (w+1)^2} \rightarrow {dw \, (w - 1) \over w(w+1)} \,.
\ee
In $E$ the pole at $w = -1$ is first order.  One might worry that there is a new pole at $w = 0$.  Fortunately, thanks to the good behavior mentioned in
footnotes \ref{footnote:nosing} and \ref{footnote:nosing2}, provided $\Delta_2 \geq 1$ there are no new poles and the integrand still falls off at infinity.

To proceed, note that the contour manipulations that led to an operator expression for $\delta \Hext$ in section \ref{sec:exop} only
depended on the locations of the poles (see figure \ref{fig:deform}).  So exactly the same manipulations can be carried out for $E$, and we can
obtain an operator expression for $E$, just by making the substitutions (\ref{commutator_change1}), (\ref{commutator_change2}) in our results for $\delta\Hext$.  For example, for a local perturbation in $F$, from \eqref{op-deltaH} we have
\bea
\nonumber
E &=& {i \over 4 \pi} \int_0^\infty {dw \, (w-1) \over w(w + 1)} \left(-{1\over w}\right)^{n_2} \left.\left(\Oiimnot + \Oiinotm - 2 \Oiimm\right)\right\vert_{\big(-{1 \over w} \xi_2^+,-w\xi_2^-\big)} \\
\label{commutator_op-deltaH}
&-& {i \over 4 \pi} \int_0^\infty {dw \, (w-1) \over w(w - 1 + i \delta)} \left({1 \over w}\right)^{n_2} \left.\left(\Oiimnot + \Oiinotm\right)\right\vert_{\big({1\over w} \xi_2^+,w \xi_2^-\big)} \\
\nonumber
&+& {\rm h.c.}
\eea
We then have $\delta \Hext = -i \lambda \left[E, \, \Hextvac \right]$ as in (\ref{commutator_deltaHext4}).

\subsection{Special case: $\delta \Hext$ as a commutator with $\Oii$\label{sect:simplifying}}

In general, $\delta \Hext$ is given by a commutator, $\delta \Hext = -i \lambda \left[E, \, \Hextvac \right]$.  However there are several situations in which $E$ can be
replaced with the perturbing operator $\Oii$, and in these situations we can get away with setting $\delta \Hext = -i \lambda \left[\Oii, \, \Hextvac \right]$.  In fact,
this is responsible for the simplifications we observed in (\ref{eq:hext2pf}) and (\ref{3ptcommutator}), so we've already encountered a few of these situations.  In this section we systematically list the conditions we're aware of under which the replacement $E \rightarrow \Oii$
is justified.

As motivation, we first mention an elementary situation, not directly relevant to this paper, in which $\delta\Hext$ is given by a commutator with the perturbing operator.
Suppose the perturbation $\Oii$ is localized in the right Rindler wedge.  As shown in \cite{Kabat:2020oic}, this immediately leads to $\delta \Hext = -i \lambda \left[\Oii, \, \Hextvac \right]$.

\subsubsection{$\delta \Hext$ acting on the vacuum\label{ActionOnVacuumSection}}

Having obtained an operator expression for $\delta \Hext$ in section \ref{sec:exop}, we can ask what happens if $\delta \Hext$ is inserted on the far left or
far right in a correlator.  In other words, we can ask how $\delta \Hext$ acts on the unperturbed vacuum state $\vert 0 \rangle$.  We will see that the
operator expression for $\delta \Hext$ can be simplified considerably if one is only interested in how it acts on the vacuum.

The starting point for this discussion is the assumption that $\delta \Hext$ stands on the far right in any correlator, or in other words, that the infinitesimal
Wightman parameters $\epsilon_{i2}$ are positive for all $i$.  Referring to (\ref{op-mobile}) and (\ref{op-fixed}), this means that for $\delta H_A^{(1)}$ the
fixed poles are all located in the upper half plane, while we only have mobile poles of types {\sf A} and {\sf C}.  Thus at $r = \pi - \delta$ the picture looks like

\centerline{$\delta H_A^{(1)} \, :$ \qquad \raisebox{-1.5cm}{\includegraphics[height=4cm]{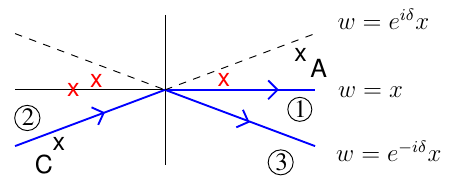}}}

\noindent
Contours \circled{1} and \circled{3} cancel in $\delta H_A^{(1)} = \circled{1} - \circled{2} - \circled{3}$, so we are left with
\be
\label{op-HA1}
\delta H_A^{(1)} = - i \lambda \int_0^\infty {dw \over (w - 1 + i \delta)^2} \left({1 \over w}\right)^{n_2} \left.\Oiinotm\right\vert_{\big({1\over w} \xi_2^+,w \xi_2^-\big)} \,.
\ee
A similar cancellation takes place in $\delta H_A^{(2)}$, where the picture looks like

\smallskip

\centerline{$\delta H_A^{(2)} \, :$ \qquad \raisebox{-1.5cm}{\includegraphics[height=4cm]{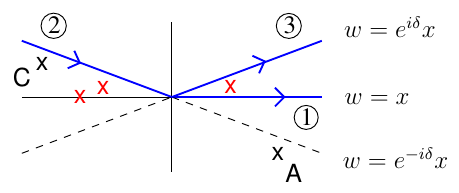}}}

\noindent
In this case contours \circled{2} and \circled{3} cancel, since they can be closed in the upper half plane, and we are left with
\be
\label{op-HA2}
\delta H_A^{(2)} =i \lambda \int_0^\infty {dw \over (w + 1)^2} \left(-{1\over w}\right)^{n_2} \left.\Oiipnot\right\vert_{\big(-{1 \over w} \xi_2^+,-w\xi_2^-\big)} \,.
\ee
Similar cancellations in $\delta H_{\bar{A}}$ lead to
\bea
\label{op-HAbar1}
\delta H_{\bar{A}}^{(1)} & = & - i \lambda \int_0^\infty {dw \over (w - 1 - i \delta)^2} \left({1 \over w}\right)^{n_2} \left.\Oiipnot\right\vert_{\big({1\over w} \xi_2^+,w \xi_2^-\big)} \\
\label{op-HAbar2}
\delta H_{\bar{A}}^{(2)} & = & i \lambda \int_0^\infty {dw \over (w + 1)^2} \left(-{1\over w}\right)^{n_2} \left.\Oiinotm\right\vert_{\big(-{1 \over w} \xi_2^+,-w\xi_2^-\big)}
\eea
For later reference, note that the cancellations we have encountered, between \circled{1} and \circled{3} in $\delta H_A^{(1)}$ and between \circled{2} and \circled{3} in $\delta H_A^{(2)}$,
can be summarized as a set of identities that hold when inserted on the far right in a correlator.
\bea
\label{identity1}
&& \int_0^\infty {dw \over (w + 1)^2} \left(-{1 \over w}\right)^{n_2}\left.\left(\Oiimnot - \Oiimm\right)\right\vert_{\big(-{1 \over w} \xi_2^+,-w\xi_2^-\big)} = 0 \\
\nonumber
&& \int_{-\infty}^0 {dw \over (w + 1 + i \delta)^2} \left(-{1 \over w}\right)^{n_2} \left.\Oiinotp\right\vert_{\big(-{1 \over w} \xi_2^+,-w\xi_2^-\big)} + \int_0^\infty {dw \over (w + 1)^2} \left(-{1 \over w}\right)^{n_2}
\left.\Oiipp\right\vert_{\big(-{1 \over w} \xi_2^+,-w\xi_2^-\big)} = 0 \\ \label{identity2}
\eea

To proceed, note that when a local operator $\Oii(\xi_2^+,\xi_2^-)$ is on the far right in a correlator, the Wightman prescription requires that $\epsilon_{i2}$ is
positive for all $i$.  This can be achieved by making $\epsilon_2$ ``large and negative.''  In some cases including an additional $\delta$ has no effect, which means there are
families of pseudo-local operators which behave identically when acting on the vacuum.  For example, consider the correlators
\be
\langle 0 \vert {\cal O}(\xi_1^+,\xi_1^-) {\cal O}(\xi_2^+,\xi_2^-) \vert 0 \rangle \quad {\rm and} \quad \langle 0 \vert {\cal O}(\xi_1^+,\xi_1^-) {\cal O}^{0-}(\xi_2^+,\xi_2^-) \vert 0 \rangle\,.
\ee
In these correlators all singularities are resolved in the same way, which means that
\be
{\cal O}^{00}(\xi_2^+,\xi_2^-) \vert 0 \rangle = {\cal O}^{0-}(\xi_2^+,\xi_2^-) \vert 0 \rangle \,.
\ee
We denote this sort of relation, of producing the same state when acting on the vacuum, by ${\cal O}^{00} \approx {\cal O}^{0-}$.  There are a number of similar relations, in which $0$ can be replaced by $+$ in the
left subscript, and $0$ can be replaced by $-$ in the right subscript.  The complete list of relations is
\bea
\nonumber
&& {\cal O} = {\cal O}^{00} \approx {\cal O}^{0-} \approx{\cal O}^{+0} \approx {\cal O}^{+-} \\
\label{op-identify}
&& {\cal O}^{0+} \approx {\cal O}^{++} \\
\nonumber
&& {\cal O}^{-0} \approx {\cal O}^{--}
\eea

When $\delta \Hext$ acts on the vacuum, we can use these relations to replace the operators in (\ref{op-HA1}), (\ref{op-HA2}), (\ref{op-HAbar1}), (\ref{op-HAbar2})
with ${\cal O}$.  Then we can assemble
\be
\delta \Hext = \left(\delta H_A^{(1)} - \delta H_A^{(2)}\right) - \left(\delta H_{\bar{A}}^{(1)} - \delta H_{\bar{A}}^{(2)}\right) \,.
\ee
Noting that $\delta H_A^{(2)}$ and $ \delta H_{\bar{A}}^{(2)}$ cancel, we're left with
\bea
\label{op-Hext}
\delta \Hext & \approx & i \lambda \oint_{w = 1} {dw \over (w - 1)^2} \left({1 \over w}\right)^{n_2} \left.\Oii\right\vert_{\big({1\over w} \xi_2^+,w \xi_2^-\big)} \\
\nonumber
& = & - 2 \pi \lambda \left. {d \over dw} \right\vert_{w = 1} \left({1 \over w}\right)^{n_2} \left.\Oii\right\vert_{\big({1\over w} \xi_2^+,w \xi_2^-\big)} \\
\nonumber
& = & i \lambda [\Hextvac, \Oii(\xi_2^+,\xi_2^-)] \,.
\eea
In other words, when placed on the far right in a correlator, $\delta \Hext$ has the same effect on the vacuum as $i \lambda [\Hextvac, \Oii(\xi_2^+,\xi_2^-)]$.
It reduces to a Lorentz boost of the perturbation.
\be
\label{boost}
\delta \Hext \approx 2 \pi \lambda \left(\xi^+ \partial_+  - \xi^- \partial_- + n_2 \right) \left.\Oii\right\vert_{(\xi_2^+,\xi_2^-)}
\ee
In a 2-point function, $\delta \Hext$ is always on either the far left or far right.  This is responsible for the simplification we observed in (\ref{eq:hext2pf}),
which can also be written in the form
\begin{eqnarray}\label{simple2pt}
\langle\{\delta \Hext,\mathcal{O}(\xi_1^+,\xi_1^-)\}\rangle=
2 \pi \lambda \Big(\xi_2^+ {\partial \over \xi_2^+} - \xi_2^- {\partial \over \partial \xi_2^-}\Big) \,
{1 \over (\xi_{12}^+ - i \epsilon_{12})^2 (\xi_{12}^- + i \epsilon_{12})^2}\,.
\end{eqnarray}

We understood this by repeating the original derivation and noting some simplifications that happen when $\delta \Hext$ acts on the vacuum.
Another approach is to start from the final operator result (\ref{op-deltaH}).  When acting on the vacuum, we can use ${\cal O}^{-0} \approx {\cal O}^{--}$, which
leads to a cancellation in the first line.  Moreover the cancellations between \circled{2} and \circled{3} in
$\delta H_A^{(2)}$, summarized in the identity (\ref{identity2}), can be further simplified to\footnote{This further simplification is not required for the argument, but it
lets us present (\ref{identity2}) in a nicer form.  There is also the identity (\ref{identity1}), but ${\cal O}^{-0} \approx {\cal O}^{--}$ makes it trivial.}
\be
\int_{-\infty}^\infty {dw \over (w + 1 + i\delta)^2} \left(-{1 \over w}\right)^{n_2} \Oiipp{\big(-{1 \over w} \xi_2^+,-w\xi_2^-\big)} \approx 0 \,.
\ee
An analogous identity, which follows from a cancellation in $\delta H_{\bar{A}}^{(2)}$, is
\be
\int_{-\infty}^\infty {dw \over (w + 1 - i\delta)^2} \left(-{1 \over w}\right)^{n_2} \Oiimm{\big(-{1 \over w} \xi_2^+,-w\xi_2^-\big)} \approx 0 \,.
\ee
These identities lead to further simplifications, and the terms in $\delta \Hext$ that survive can be identified with (\ref{op-Hext}).

Although we have argued for the simplification at the operator level, one can also understand it by working inside correlators.
We discuss this briefly in appendix \ref{appendix:contact}.

\subsubsection{Spectator operators in the same wedge\label{SameWedge}}

Now we consider what happens in a correlator when all spectator operators are inserted in the same (meaning left or right) Rindler wedge.\footnote{The arguments that follow
fail if the spectator operators are inserted in the future or past wedge, since in that case all of the singular configurations in Fig.\ \ref{fig:singularities} are possible.}  Again we'll find that the
expression for $\delta \Hext$ can be simplified, with $E$ replaced by the perturbing operator $\Oii$.  This will play an important role in our study of the
KMS condition in section \ref{sect:KMS}.

Building on the analysis in section \ref{ActionOnVacuumSection}, and using the definition of $\delta\Hext$ from eq.~\eqref{deltaHext}, along with eqs.~\eqref{rotating}, \eqref{fixed}, \eqref{eq:m-pole-generic}, and \eqref{eq:f-pole-generic}, we see that when $r = 0$, with both spectator operators $\Oi$ and $\Oiii$ inserted in the right wedge, the $w$-plane singularities only appear for $\operatorname{Re}w>0$.  Furthermore, the poles are positioned above or below the real axis depending on whether the operator sits to the left or right of $\delta\Hext$. 

As in previous cases, we must check whether the integration contour is pinched by poles in the $w$-plane. With both $\Oi$ and $\Oiii$ in the right wedge, the only possible
singularity occurs when $\xi^+_{1,3} = \xi^+_2$. This can be seen in Fig.\ \ref{fig:singularities}, where the only singular configuration that can be realized with both
spectators in the right wedge is case (i).
From section (\ref{sect:resolve}), the Wightman prescription inherited from the CFT properly resolves these potential pinchings.

\begin{figure}[h]
	\centering
	\includegraphics[width=0.4\textwidth]{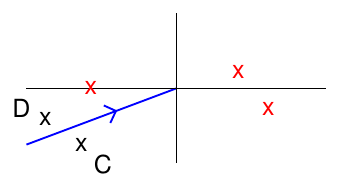}
	\includegraphics[width=0.4\textwidth]{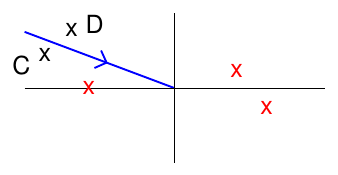}
	\caption{Contour \protect\circled{2} of $\delta H_A^{(1)}$ (left) and $\delta H_A^{(2)}$ (right).  The potential pinch singularities as $r \rightarrow \pi$ are resolved by the CFT Wightman prescription, so as $r \rightarrow \pi$ the contours can be subtracted.  This gives a contour that encircles the $w=-1$ pole, which produces the commutator in (\ref{OneSideCommutator}).	\label{fig:KMScontour_dHA}}
\end{figure}

Following the argument in section (\ref{ActionOnVacuumSection}), we conclude that when all spectator operators are in the right or left wedge, the correlator simplifies to
\begin{equation}
	\left\langle\cdots \delta \Hext\cdots \right\rangle = \left\langle\cdots \left(\delta H_A^{(1)}\big\vert_\text{contour \circled{2}} - \delta H_A^{(2)}\big\vert_\text{contour \circled{2}}\right) \cdots \right\rangle
\end{equation}
where, as shown in Fig.\ \ref{fig:KMScontour_dHA} and section (\ref{ActionOnVacuumSection}),\footnote{The overall signs come from (\ref{1minus2minus3}), $\delta H_A^{(1)} = \circled{1} - \circled{2} - \circled{3}$.}
\begin{equation}
	\delta H_A^{(1)}\big\vert_\text{contour \circled{2}} = -i \lambda \int_{-e^{i\delta}\infty}^{0}\frac{dw}{(w+1)^2}\left(-{1\over w}\right)^{n_2}\Oii\left(-e^{i\delta}{1\over w}\xi_2^+, - w\xi_2^-\right)
\end{equation}
and
\begin{equation}
	\delta H_A^{(2)}\big\vert_\text{contour \circled{2}} = -i \lambda \int_{-e^{-i\delta}\infty}^0\frac{dw}{(w+1)^2}\left(-{1\over w}\right)^{n_2}\Oii\left(-e^{-i\delta}{1\over w}\xi_2^+, - w\xi_2^-\right).
\end{equation}

Since the Wightman prescription properly resolves all singularities, there is no problem setting $r = \pi$.  Then the two integrals can be combined in to a single contour that
encircles the pole at $w = -1$.  This means that inside the correlator we can make the replacement
\be
\label{SameWedgeContour}
\delta \Hext \rightarrow - i \lambda \oint\limits_{\substack{{\rm counter-clockwise} \\ w = -1}} {dw \over (w + 1)^2} \left(-{1 \over w}\right)^{n_2} \left.\Oii\right\vert_{\big(-{1\over w} \xi_2^+,-w \xi_2^-\big)} \,.
\ee
By a slight variant of the calculation (\ref{op-Hext}), this allows us to replace $\delta \Hext$ with $i \lambda [\Hextvac, \Oii(\xi_2^+,\xi_2^-)]$ inside a correlator when all
spectator operators are in the same (left or right) wedge.
\begin{equation}
\label{OneSideCommutator}
	\left\langle\cdots \delta \Hext\cdots \right\rangle = \left\langle \cdots\, i\lambda \left[\Hextvac,\,\Oii\right]\, \cdots \right\rangle
\end{equation}

Although we have argued for the simplification at the operator level, one can also understand it by working inside correlators.
We discuss this briefly in appendix \ref{appendix:contact}.

\subsubsection{Generic (non-singular) points\label{GenericCommutator}}

Finally, we consider what happens in a correlator when the spectator operators are located at generic points, generic
meaning that the correlator with $\delta \Hext$ is non-singular.
We will see that at generic points $E$ can be replaced with the perturbing operator $\Oii$.

The argument is quite simple.  In section \ref{sect:2ptNonSing} we obtained an expression for the correlator of $\delta H_A$ with a string of spectator operators, involving
a counter-clockwise integral around the mobile poles at $w = w_m = e^{\pm ir} \xi_2^+ / \xi_k^+$.  For examples of this result with one and two spectator operators, see (\ref{FJdeltaHA4}) and (\ref{3ptContour}).

There is a similar expression for the correlator of $\delta H_{\bar{A}}$ with a string of spectator operators, however in going from
$\delta H_A$ to $\delta H_{\bar{A}}$ the mobile and fixed poles are exchanged.  That is, the correlator of $\delta H_{\bar{A}}$ with a string of spectator operators is given
by a clockwise (note the difference!) contour integral around the mobile poles at $w = w_m = e^{\pm ir} \xi_k^- / \xi_2^-$.  For examples of this with one and two spectator
operators, see (\ref{FJdeltaHAbar4}) and (\ref{3ptContour2}).

At generic points, where the contour is not pinched, there is no difficulty in sending $r \rightarrow \pi$.  At $r = \pi$ the integrands for $\delta H_A$
and $\delta H_{\bar{A}}$ are the same.  When we take the difference to get the correlator of $\delta \Hext = \delta H_A - \delta H_{\bar{A}}$, we end up with a counter-clockwise
integral that surrounds all of the poles except for the pole at $w = -1$.  This, at the cost of an overall $-$ sign, can be replaced with a counter-clockwise integral that only encircles the pole at $w = -1$.  This means that inside the correlator we can make the replacement
\be
\delta \Hext \rightarrow - i \lambda \oint\limits_{\substack{{\rm counter-clockwise} \\ w = -1}} {dw \over (w + 1)^2} \left(-{1 \over w}\right)^{n_2} \left.\Oii\right\vert_{\big(-{1\over w} \xi_2^+,-w \xi_2^-\big)} \,.
\ee
By a slight variant of the calculation (\ref{op-Hext}), this allows us to replace $\delta \Hext$ with $i \lambda [\Hextvac, \Oii(\xi_2^+,\xi_2^-)]$ inside a correlator at non-singular points.  This replacement is responsible for the simplification which we observed in the three-point function (\ref{3ptcommutator}), as can be seen with the help of (\ref{boost}).

In fact, this argument establishes a stronger result.  Even at a singular point, if the Wightman prescription inherited from the CFT
correctly resolves the way in which poles collide and pinch the contour, then there
is no problem with simply setting $r = \pi$.  The above argument goes through, and we are allowed to replace $\delta \Hext$ with $i \lambda [\Hextvac, \Oii(\xi_2^+,\xi_2^-)]$.

\subsection{Contact terms\label{sect:contact}}

In section \ref{GenericCommutator} we gave an argument which shows that, at non-singular points, $\delta\Hext$ can be replaced with
$i \lambda [\Hextvac, \Oii(\xi_2^+,\xi_2^-)]$ inside a correlator.  However the argument can fail at singular points, which motivates setting
\be
\delta \Hext = i \lambda [\Hextvac, \Oii(\xi_2^+,\xi_2^-)] + \delta \Hext_{\rm contact} \,.
\ee
We will refer to the extra contribution to $\delta \Hext$ as a contact term.  Here we make some general remarks on properties of $\delta \Hext_{\rm contact}$.
We give a more explicit discussion with some formulas in appendix \ref{appendix:contact}.

A first remark is that, since $\delta \Hext_{\rm contact}$ can only contribute to a correlation function at singular points, its correlators with strings of
spectator operators have delta-function-like support.  This motivates calling them contact terms.

A second remark is that, as we mentioned at the end of section \ref{GenericCommutator}, contact terms arise when the $i \delta$ prescription for continuing
$r \, : \, 0 \rightarrow \pi^-$ has a non-trivial effect on the correlator, that is, when it leads to a resolution of a pinching singularity which does not follow from the
Wightman prescription in the CFT.  Geometrically, this happens when spectator
operators are arranged as shown in the bottom two panels (situations (iii) and (iv)) of Figure \ref{fig:singularities}.

Finally, as shown in section \ref{ActionOnVacuumSection}, contact terms do not arise when $\delta \Hext$ acts on the unperturbed vacuum.
In other words $\delta \Hext_{\rm contact}$ annihilates the unperturbed vacuum, $\delta \Hext_{\rm contact} \vert 0 \rangle = 0$.  In this way contact terms are
reminiscent of the endpoint contributions to $\delta \Hext$ studied in \cite{Kabat:2020oic,Kabat:2021akg}, since endpoint contributions are built from light-ray operators that likewise annihilate the unperturbed vacuum.

\section{KMS condition\label{sect:KMS}}

In this section, we discuss the Kubo-Martin-Schwinger (KMS) condition. We begin by presenting the KMS condition, emphasizing all three components of the condition following the treatment in Sorce \cite{Sorce:2023gio}.  Related treatments may be found in many places including \cite{Haag:1992hx,Borchers:2000pv,Jones:2015kbb}.  We highlight the algebraic interpretation and uniqueness results
that follow. Afterward, we explicitly demonstrate that our first-order correction to the modular Hamiltonian satisfies these conditions.

\subsection{Definition and Uniqueness}

The KMS condition is central to the algebraic characterization of thermal equilibrium states and plays a fundamental role in quantum statistical mechanics and quantum field theory. Following Sorce \cite{Sorce:2023gio}, the KMS condition comprises three essential criteria that must be satisfied by the modular Hamiltonian associated with a given cyclic and separating vector $|\Omega\rangle$:

\begin{enumerate}
    \item \textbf{Modular symmetry}: The state vector $|\Omega\rangle$ must remain invariant under modular flow generated by the modular Hamiltonian $\Hext$, namely:
    \begin{equation}
        e^{-i\Hext t}|\Omega\rangle = |\Omega\rangle \quad \text{for all real } t.
    \end{equation}

    \item \textbf{Modular automorphism}: The modular Hamiltonian must generate an automorphism of the algebra of observables. That is, for every operator $a$ belonging to the algebra $\mathcal{A}$,
    \begin{equation}
        e^{i\Hext t} a e^{-i\Hext t} \in \mathcal{A} \quad \text{for all real } t.
    \end{equation}
    Here $\mathcal{A}$ could refer to the algebra of operators in the right or left wedge.

    \item \textbf{KMS analyticity condition}: Under modular evolution, correlation functions must satisfy a specific analytic continuation property. For any pair of operators $A$ and $B$ in the algebra $\mathcal{A}$, the correlation function under modular evolution obeys:
    \begin{equation}\label{usualKMS}
        \langle A_s B \rangle = \langle B A_{s+2\pi i} \rangle,
    \end{equation}
    where the modular-evolved operator $A_s$ is defined as:
    \begin{equation}
        A_s = e^{is\Hext/2\pi} A e^{-is\Hext/2\pi},
    \end{equation}
    and all expectation values $\langle \cdots \rangle$ are evaluated in the state associated with $|\Omega\rangle$.
Again $\mathcal{A}$ could refer to the algebra of operators in the right or left wedge.

Although (\ref{usualKMS}) is the standard statement of the KMS condition, it really only applies to bounded operators.\footnote{For
bounded operators one defines $G(s) = \langle A_s B \rangle$ and $F(s) = \langle B A_s \rangle$.  $G(s)$ is analytic in the strip
$-2\pi < {\rm Im} \, s < 0$, $F(s)$ is analytic in the strip $0 < {\rm Im} \, s < 2 \pi$, and KMS analyticity is the relation
$G(s) = F(s + 2 \pi i)$ \cite{Haag:1992hx}.}  We will be working with Wightman correlators of local operators, for which the
appropriate statement of KMS analyticity is
    \begin{equation}
        \langle A_s B \rangle = \langle B A_{s+2\pi i - i \delta} \rangle \,.
    \end{equation}
Here $\delta \rightarrow 0^+$ is an infinitesimal parameter which approaches zero more slowly than the Wightman $i \epsilon$'s
which are used to define the correlator.  Since this may be unfamiliar, we review this prescription for vacuum  modular flow
in appendix \ref{appendix:VacuumFlow}.
\end{enumerate}

Crucially, the modular Hamiltonian satisfying these three conditions is unique.\footnote{In addition to \cite{Sorce:2023gio}, see for example section 1.4 of \cite{Borchers:2000pv} or appendix A of \cite{Jones:2015kbb}.}  This uniqueness theorem ensures that given a cyclic and separating vector $|\Omega\rangle$, there exists exactly one modular Hamiltonian that fulfills these requirements. In subsequent sections, we verify that our perturbative construction indeed satisfies all three criteria, thereby independently establishing its validity.

Incidentally, perhaps the most surprising feature of our perturbative construction is the appearance of the contact terms discussed in section \ref{sect:contact}.  One could ask where contact terms play a role in satisfying the KMS conditions.
We will see that, although they are not required for modular symmetry or analyticity, they are crucial in satisfying the automorphism condition.

\subsection{Modular symmetry}

The modular symmetry condition is
\begin{equation}\label{ModSymm}
    e^{-i\Hext t} |\Omega\rangle = |\Omega\rangle,
\end{equation}
where $|\Omega\rangle$ is the cyclic and separating vector associated with the von Neumann algebra under consideration.  To see that this is satisfied to first order in $\lambda$, note that
\bea
\nonumber
\Hext \vert \Omega \rangle & = & \big(\Hextvac + \delta \Hext\big)\big(\identity - i \lambda \Oii\big) \vert 0 \rangle + {\cal O}(\lambda^2) \\
& = & \big( \delta \Hext - i \lambda [\Hextvac,\Oii]\big) \vert 0 \rangle + {\cal O}(\lambda^2)
\eea
where we have used $\Hextvac \vert 0 \rangle = 0$.  As shown in (\ref{op-Hext}), when acting on the vacuum $\delta\Hext$ can be replaced with $i\lambda[\Hextvac,\Oii]$.  Thus $\Hext \vert \Omega \rangle =
{\cal O}(\lambda^2)$, which means that (\ref{ModSymm}) is satisfied to first order in $\lambda$.

One can also show that the modular symmetry condition is satisfied by using (\ref{lambda_expansion}) to expand the left side of (\ref{ModSymm}) to first order in $\lambda$.

\subsection{Modular automorphism}

The condition of modular automorphism, which follows directly from Tomita's theorem, states that the modular flow generated by the modular Hamiltonian preserves the algebra of observables. That is, say for an operator ${\cal O}^{(R)}$ in the right wedge, it should be the case that $\delta {\cal O}^{(R)} = i [\Hext, {\cal O}^{(R)}]$ is also an element of the
right subalgebra.  Since the left and right subalgebras are commutants of each other (Haag duality), an equivalent statement is that the double commutator should vanish,
\begin{equation}
    [\mathcal{O}^{(L)},[\delta\Hext,\mathcal{O}^{(R)}]] = 0
\end{equation}
where ${\cal O}^{(L)}$ and ${\cal O}^{(R)}$ are arbitrary operators localized in complementary regions.

The vanishing of the double commutator is guaranteed by the structure of $\delta \Hext$ we have found.
As we saw in the operator equation (\ref{op-deltaH}), $\delta \Hext$ is expressed as a difference $\delta H_{A} - \delta H_{\bar{A}}$.  Here each term is an integral of pseudo-local operators, with the property that $\delta H_{A}$ commutes with any operator on the left, and $\delta H_{\bar{A}}$ commutes with any operator on the right.  So the double commutator
is guaranteed to vanish.

One can see this explicitly in the correlators we have computed, such as (\ref{eq:habar3pfcorrotherfixingbothDan}).
\begin{align}
	&\langle\{\mathcal{O}^{(L)}(\xi_1^+,\xi_1^-),\,\delta H_{A},\,\mathcal{O}^{(R)}(\xi_3^+,\xi_3^-)\}\rangle=-\frac{2\pi\lambda\,\xi_2^+}{(\xi_{13}^-)(\xi_{13}^+)^2}\nonumber\\
	&\Big[\frac{(\xi_1^+)^2}{(\xi_{12}^+)^2(\xi_2^+\xi_2^--\xi_1^+\xi_1^-)(\xi_2^+\xi_2^--\xi_1^+\xi_3^--i\epsilon_{23})}
	-\frac{(\xi_3^+)^2}{(\xi_{23}^+-i\epsilon_{23})^2(\xi_2^+\xi_2^--\xi_3^+\xi_3^-)(\xi_2^+\xi_2^--\xi_1^-\xi_3^++i\epsilon_{23})}\Big]
\end{align}
Notice the absence of $i\epsilon_{12}$ and $i \epsilon_{13}$ in this expression. When evaluating the outer commutator, this means the ordering between $\mathcal{O}^{(L)}(\xi_1^+,\xi_1^-)$ and $[\delta H_{A},\mathcal{O}^{(R)}(\xi_3^+,\xi_3^-)]$ is irrelevant, so $\delta H_A$ doesn't contribute to the double commutator.  One can likewise see that
$\delta H_{\bar{A}}$ doesn't contribute to the double commutator, as follows from the absence of $i \epsilon_{13}$ and $i\epsilon_{23}$ in \eqref{eq:habar3pfcorrotherfixingboth}.

Note that the contact terms discussed in section \ref{sect:contact} are essential for the modular automorphism condition.  If it weren't for $\delta \Hext_{\rm contact}$, we could replace $\delta \Hext$ with
$i \lambda [\Hextvac, \Oii(\xi_2^+,\xi_2^-)]$.
Then $\delta \Hext$ would be a local operator in the future wedge (a Lorentz boost of the perturbation, as in (\ref{boost})),
and for such an operator, the double commutator would not in general vanish.
By contrast, in checking the modular symmetry and KMS analyticity conditions, the replacement of $\delta \Hext$
with $i \lambda [\Hextvac, \Oii(\xi_2^+,\xi_2^-)]$ is allowed.\footnote{For modular symmetry, because $\delta \Hext$ acts
on the vacuum, and for KMS analyticity, because both spectator operators are in the right wedge.}  So those two conditions do not require contact terms.

\subsection{KMS analyticity condition\label{sect:KMSanalyticity}}

We will use the following form of the KMS condition:
\begin{equation}
\langle A_{s} B \rangle = \langle B A_{s+2\pi i - i \delta} \rangle
\end{equation}
where $\langle\,\cdots\rangle$ denotes expectation value in the excited state, and the excited state modular Hamiltonian $\Hext$ evolves the operator by $A_{s} = e^{is\Hext/2\pi}Ae^{-is\Hext/2\pi}$.  It's convenient to set $t = s + 2 \pi i - i \delta$
and write the KMS condition as
\begin{equation}
\langle Ae^{-is\Hext/2\pi}B\rangle = \langle Be^{i t \Hext / 2\pi}A\rangle.
\end{equation}

Since we have the modular Hamiltonian $\Hext$ to first order, we expand the exponential in the KMS condition to first order using the identity
\begin{equation}
\label{lambda_expansion}
e^{X+\lambda Y} = e^X + \lambda \int_0^1 d\alpha \, e^{(1-\alpha) X} Y e^{\alpha X} + \mathcal{O}(\lambda^2)	
\end{equation}
which implies
\begin{equation}
e^{-is\Hext/2\pi} = e^{-is\Hextvac/2\pi} - \frac{i}{2\pi} e^{-is\Hextvac/2\pi} \int_0^sd\alpha\,\delta\Hext\vert_\alpha + \mathcal{O}(\lambda^2)\,.
\end{equation}
Here $\Hext = \Hextvac + \delta \Hext$ and $\delta\Hext\vert_\alpha = e^{i\alpha\Hextvac/2\pi}\delta\Hext e^{-i\alpha\Hextvac/2\pi}$.
We suppress $\mathcal{O}(\lambda^2)$ in the future. A similar expansion is available for $e^{i t \Hext / 2 \pi}$.

Imposing that the KMS condition must be satisfied order by order, we expand and equate both sides of the KMS condition to first order in $\lambda$, where $\langle \, \cdots \rangle_{0}$ denotes an expectation value in the vacuum state: 
\bea
\nonumber
& &i\lambda\langle \Oii AB\vert_{-s} \rangle_{0} - \frac{i}{2\pi} \int_0^s d\alpha \, \langle A\vert_{s} \delta \Hext\vert_\alpha B \rangle_{0} - i\lambda\langle A\vert_{s} B \Oii \rangle_{0} \\
\label{eq:fokms}
&= &i\lambda\langle \Oii BA\vert_{t} \rangle_{0} - \frac{i}{2\pi} \int_0^{-t} d\alpha \, \langle B\vert_{-t} \delta \Hext\vert_\alpha A \rangle_{0} - i\lambda\langle B\vert_{-t} A \Oii \rangle_{0}\,.
\eea
To proceed, we must compute the integrals in (\ref{eq:fokms}).  At non-singular points this is done directly in appendix \ref{appendix:nonsing}.  Allowing for singularities, this can be done as follows.

\noindent{\em First integral} \\
The KMS analyticity condition only needs to be satisfied when $A$ and $B$ are operators in the same (either left or right) Rindler wedge.  In this case, as shown in section
\ref{SameWedge}, we can replace $\delta \Hext$ with $i\lambda[\Hextvac,\, \Oii]$.  Since $\Hextvac$ generates vacuum modular flow, we have
\be
\delta \Hext \vert_\alpha = 2 \pi \lambda {\partial \over \partial \alpha} \Oii \vert_\alpha
\ee
which means
\begin{equation}\label{eq:IntegrateDeltaHCommutator}
    \int_0^s d\alpha\, \delta \Hext \vert_\alpha = 2 \pi \lambda \left(\Oii\vert_s - \Oii\right).
\end{equation}

\noindent{\em Second integral} \\
We now turn to the second integral, focusing on the analytic continuation which is required to define it as a function of the
complex parameter $t$.

\begin{figure}[h]
	\centering
	\includegraphics[width=0.4\textwidth]{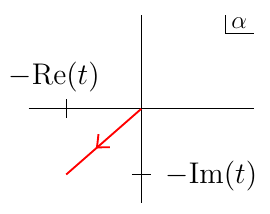} \qquad\quad
	\raisebox{6mm}{\includegraphics[width=0.45\textwidth]{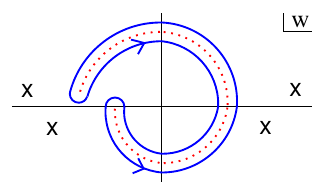}}
	\caption{
	The left panel shows the complex $\alpha$ plane.  The red curve is $\gamma_\alpha$, a choice of contour that goes from $0$ to $-t$.  The right panel shows the complex $w$ plane.
	The dotted red curve is the path traced out by the pole at $w = - e^{-\alpha}$.  As $\alpha \, : \, 0 \rightarrow -t$, the pole moves in a counter-clockwise spiral from
	$w = -1$ to $w = - e^t$.
	   The blue curve is a suitable choice of $w$ contour: it is independent of $\alpha$ and always surrounds the pole at $w = - e^{-\alpha}$,
	without encircling any other poles.}
	\label{fig:KMScontouer_dw}
\end{figure}

For real $t$ and $\alpha$, the correlator $\langle B\vert_{-t} \delta \Hext \vert_\alpha A \rangle_0$ involves $\delta \Hext$ with two
spectator operators in the right wedge.  So, as in the first integral, for real $t$ and $\alpha$ we are allowed to replace
$\delta \Hext$ with $i\lambda[\Hextvac,\, \Oii]$.  This can be represented as a contour integral around $w=-1$,
as in (\ref{SameWedgeContour}).
Thus a starting point for the analytic continuation is
\begin{equation}
	\int_0^{-t}d\alpha\, \delta \Hext\vert_\alpha = 
	- i \lambda \int_0^{-t}d\alpha\oint_{w=-1} \frac{dw}{(w + 1)^2} \left(-\frac{e^{\alpha}}{w}\right)^{n_2} \Oii\left(-\frac{1}{w} e^{\alpha}\xi_2^+,-w e^{-\alpha}\xi_2^-\right) \,.
\end{equation}
We now make the substitution $w=e^{\alpha}\tilde{w}$. Relabeling $\tilde{w}$ back to $w$, the $w$ contour acquires an $\alpha$ dependence:
\begin{equation}
	\int_0^{-t}d\alpha\, \delta \Hext\vert_\alpha = 
	- i \lambda \int_0^{-t}d\alpha\oint_{w=-e^{-\alpha}} \frac{e^\alpha dw}{(e^\alpha w + 1)^2} \left(-\frac{1}{w}\right)^{n_2} \Oii\left(-\frac{1}{w} \xi_2^+,-w \xi_2^-\right).
\end{equation}
We'd like to perform the $\alpha$ integral first, however exchanging the order of integration requires a contour for $w$ that is
independent of $\alpha$.  To achieve this, we first specify $\gamma_\alpha$, a contour in the complex $\alpha$-plane that
goes from the origin to $-t$.  The pole at $w = - e^{-\alpha}$ traces out a corresponding path $\gamma_w = - e^{-\gamma_\alpha}$ in the complex $w$ plane.  We deform the $w$ contour so that it surrounds $\gamma_w$ without encircling any other poles.
Once fixed in this way, the $w$ contour is independent of $\alpha$, and we can exchange the order of integration.  Figure~\ref{fig:KMScontouer_dw} shows an example of this procedure.

With this contour prescription, we can exchange the $w$ and $\alpha$ integrations and perform the $\alpha$ integral to obtain
\begin{equation}
\int_0^{-t}d\alpha\,\delta\Hext\vert_\alpha = i\lambda \oint {dw \over w} \left({e^t \over w + e^t} - {1 \over w+1}\right)
\left(-\frac{1}{w}\right)^{n_2}\Oii\left(-\frac{1}{w}\xi^+_2, -w\xi^-_2\right).
\end{equation}
The contour surrounds the poles at $w = -e^t$ and $w = -1$, and evaluating the residues gives
\be
\label{SecondIntegral}
\int_0^{-t}d\alpha\,\delta\Hext\vert_\alpha = 2 \pi \lambda \left(\Oii\vert_{-t} - \Oii\right)
\ee
as expected from (\ref{eq:IntegrateDeltaHCommutator}).

\noindent{\em KMS analyticity} \\
We now return to the analyticity condition (\ref{eq:fokms}).  Using (\ref{eq:IntegrateDeltaHCommutator}) and (\ref{SecondIntegral}) to evaluate the integrals, the KMS analyticity condition becomes
\bea
\nonumber
&& \langle \Oii A B\vert_{-s} \rangle_{0}
    - \langle A\vert_{s}  \Oii\vert_s B \rangle_{0}
    + \langle A\vert_{s}  \Oii B \rangle_{0}
    - \langle A|_{s} B \Oii \rangle_{0} \\[3pt]
\label{commKMS}
&&= \langle \Oii B A|_{t} \rangle_{0}
    - \langle B|_{-t} \Oii\vert_{-t} A \rangle_{0}
    + \langle B|_{-t} \Oii A \rangle_{0}
    - \langle B|_{-t} A \Oii \rangle_{0} \,.
\eea
It remains to show that the two sides are equal as distributions.

To do this, recall that $t = s + 2 \pi i - i \delta$, where $\delta \rightarrow 0^+$ is an infinitesimal parameter which approaches zero
more slowly than the Wightman $i \epsilon$'s which are used to define the correlator.  Note that for any local operator
\begin{equation}
\mathcal{O}(\xi^+,\xi^-)\vert_{t} = e^{nt} {\cal O}(e^{t}\xi^+,\, e^{-t}\xi^-) = e^{ns} \mathcal{O}(e^s\xi^+-i\delta\xi^+,\, e^{-s}\xi^-+i\delta\xi^-)
\end{equation}
and likewise
\begin{equation}
\mathcal{O}(\xi^+,\xi^-)\vert_{-t} = e^{-ns} \mathcal{O}(e^{-s}\xi^++i\delta\xi^+,\, e^{s}\xi^--i\delta\xi^-) \,.
\end{equation}
These are exactly the pseudo-local operators we encountered in section \ref{sec:exop}.  In particular, taking $A$ and $B$ to be local
operators in the right wedge (meaning $\xi^+,\,\xi^- > 0$), we have
\bea
\nonumber
&& A\vert_t = A^{-+}\vert_s \\
&& B\vert_{-t} = B^{+-}\vert_{-s} \,,
\eea
and with $\Oii$ an operator in the future wedge ($\xi^+ > 0$, $\xi^- < 0$), we have
\be
\Oii \vert_{-t} = \Oiipp\vert_{-s} \,.
\ee
However, with $A$ and $B$ in the right wedge, $\Oii$ can only be null separated from $A$ or $B$ at equal values of $\xi^+$.  So it is only the prescription for resolving singularities at equal values of $\xi^+$  that
matters, and we can equally well set
\be
\Oii \vert_{-t} = \Oiipm\vert_{-s} \,.
\ee
Making these replacements in the right side of (\ref{commKMS}), the pseudo-local operators can be moved freely past the
local operators, and we can write the right side of (\ref{commKMS}) as\footnote{In the second term, the ordering of the two
pseudo-local operators $B^{+-}$ and $\Oiipm$ is inherited from (\ref{commKMS}).  They appear in the order $B^{+-}\vert_{-s} \Oiipm\vert_{-s}$,
due to the CFT Wightman prescription.}
\be
\label{KMSrhs}
\langle A^{-+}\vert_s \Oii B \rangle_{0}
    - \langle A B^{+-}\vert_{-s} \Oiipm\vert_{-s} \rangle_{0}
    + \langle \Oii A B^{+-}\vert_{-s}\rangle_{0}
    - \langle A \Oii B^{+-}\vert_{-s}\rangle_{0} \,.
\ee
From section \ref{sec:exop}, recall that ${\cal O}^{-+}$ is the same as a local operator on the far left in a correlator, and ${\cal O}^{+-}$
is the same as a local operator on the far right.  Since the pseudo-local operators in (\ref{KMSrhs}) are already in these positions, we can drop the
subscripts, and the right side of KMS becomes
\be
\langle A\vert_s \Oii B \rangle_{0}
    - \langle A B\vert_{-s} \Oii\vert_{-s} \rangle_{0}
    + \langle \Oii A B\vert_{-s}\rangle_{0}
    - \langle A \Oii B\vert_{-s}\rangle_{0} \,.
\ee
With some help from vacuum Lorentz invariance, this exactly matches the left side of the KMS condition (\ref{commKMS}).

\section{Conclusions}

In this work we found the first-order correction to the modular Hamiltonian for a state made by perturbing the vacuum with
a local operator in the future wedge.  We developed a prescription which let us calculate correlators of $\delta \Hext$ with arbitrary strings of
spectator operators.  We went on to extract an operator expression for $\delta \Hext$, in terms of pseudo-local operators integrated over spacelike
hyperbolas.  We showed that $\delta \Hext$ could be expressed as a commutator of $\Hextvac$ with $E$, listed situations in which $E$ could be
replaced with the perturbing operator $\Oii$, and showed that the KMS conditions are satisfied.

The main technical challenge was making sense of the complex modular flow in the Sarosi-Ugajin formulas.  We did this by introducing a prescription
for continuing $r \, : \, 0 \rightarrow \pi$, with $r$ approaching $\pi$ from below.  This led to well-defined correlators that, as we showed, satisfy the KMS
conditions.  However other approaches are possible, and it would be interesting to explore them further.  In particular, the algebra of bounded operators
has a so-called Tomita subalgebra for which modular flow is entire (analytic, with no singularities at finite distance in the complex plane) \cite{Takesaki,Lashkari:2018oke}.
It should be possible to reproduce our results
by working with the Tomita subalgebra, where there is no difficulty in defining complex modular flow.

Modular Hamiltonians have diverse applications, ranging from condensed matter and statistical physics \cite{Cardy:2016fqc} to quantum field
theory \cite{Borchers:2000pv}, quantum information \cite{Witten:2018lha} and holography \cite{Faulkner:2018faa}.
There are many directions in which this work could be extended, and there are several tempting conjectures.  Here we list a few.

\noindent{\em General conformal dimensions\label{subsec:nonZ}} \\
\noindent
In this work we focused on two- and three-point functions, with operators of integer conformal dimensions.  This simplified the
analysis, since correlators only had poles.  It would be desirable to extend the analysis to generic operators and (perhaps by an
OPE argument) to higher-point correlators.  It's tempting to speculate that, although the analysis will change, the final result
(\ref{op-deltaH}) for $\delta \Hext$ will remain the same.

\noindent{\em Perturbation smeared over multiple wedges\label{subsec:multiple}} \\
\noindent
We worked with a local perturbation inserted in the future wedge.  It would be interesting to consider a more general perturbation,
obtained by smearing a local operator over an extended spacetime region.
\begin{equation}
	G=\int d\xi^+ d\xi^-\, f(\xi^+,\xi^-)\,\mathcal{O}(\xi^+,\xi^-)
\end{equation}
Such smearing is generically necessary to make the perturbed state normalizeable.\footnote{See for example the discussion
in section 5 of \cite{Bousso:2014uxa}.}

If one restricts to a smearing $f(\xi^+,\xi^-)$ with support in the future wedge, this is quite straightforward.  One can simply integrate operator expressions
such as (\ref{op-deltaH}) or correlators such as (\ref{eq:habar3pfcorrotherfixingbothDan}), (\ref{eq:habar3pfcorrotherfixingboth}) against $f(\xi_2^+,\xi_2^-)$,
and no significant new features arise.

The question becomes more interesting when the perturbation is smeared across multiple wedges, including the Rindler horizons and the entangling
surface.  For perturbations that are smeared on a spacelike slice through the entangling surface, $\delta \Hext$ picks up endpoint contributions
\cite{Kabat:2020oic,Kabat:2021akg}.  These are due to additional singularities at $w = 0$ and $w = \infty$, which can arise when the state is perturbed at $\xi_2^+ = 0$ or
$\xi_2^- = 0$,
but which are absent for perturbations in the
future wedge.  For a general smearing across multiple wedges, it would be interesting to understand how the various contributions
to $\delta \Hext$ assemble into a unique operator that satisfies the KMS conditions.

\noindent{\em Modular flow near the horizon} \\
\noindent
It's widely believed that in any reasonable state, close to the entangling surface, modular flow approaches a Lorentz boost.\footnote{For a discussion of the
conjecture see section 4.1 in \cite{Jensen:2023yxy} and for rigorous bounds see \cite{Fredenhagen1985}.}  Since the vacuum modular Hamiltonian
already generates a Lorentz boost, an equivalent statement is that the commutator $[\delta \Hext, {\cal O}(\xi^+,\xi^-)]$
of $\delta \Hext$ with any local operator will vanish (rapidly enough, in an appropriate sense) as the local operator approaches the origin.

We are in a position to test this conjecture using our explicit results for $\delta \Hext$.  However, for a local perturbation in the future wedge, there isn't
much to test.  For a local perturbation in the future wedge, there is no difficulty in placing a spectator operator at the origin.  The origin is a generic position,
at which the correlator is non-singular.  This can be seen explicitly in (\ref{eq:deltaHa3pfiep}), which is non-singular when $\Oi$ is placed at the origin, or in (\ref{eq:habar3pfcorrotherfixing}), which is non-singular when $\Oiii$ is placed at the origin.  Since the correlator is non-singular, factors of $\epsilon_{2k}$
can be dropped, which means operator order doesn't matter.  Thus $\delta \Hext$ commutes with an operator at the origin.

Smearing the perturbation against a function $f(\xi^+,\xi^-)$ with support in the future wedge doesn't change this conclusion.  However it's not obvious
what happens when the perturbation is smeared across multiple wedges.  In this case it would be interesting to establish explicitly that
the commutator vanishes sufficiently rapidly near the origin.  In the context of AdS/CFT, it would be interesting to develop an understanding of
bulk modular flow \cite{Jafferis:2015del}, and to study its behavior in the vicinity of the HRT surface.

\noindent{\em An all-orders speculation} \\
\noindent
Suppose for a moment that the perturbation $G$ is localized in the $R$ (or $L$) wedge.  In this case it is straightforward to show that the excited state modular Hamiltonian can be obtained from the vacuum modular Hamilton by conjugation \cite{Kabat:2020oic}.  That is,
for a state
\be
\vert \psi \rangle = U \vert 0 \rangle \quad {\rm with} \quad U = e^{-i \lambda G}, \quad \hbox{\rm $G$ localized in $R$ or $L$ wedge}
\ee
the extended modular Hamiltonian is
\be
\label{commutator_all-orders}
\Hext = U \Hextvac U^\dagger \,.
\ee
At first order in $\lambda$ this implies
\be
\label{commutator_first-order}
\delta \Hext = -i \lambda \big[ G, \Hextvac \big] \,.
\ee
If we began by obtaining (\ref{commutator_first-order}) from the Sarosi--Ugajin formula, we would find that at higher orders in $\lambda$ the perturbation series
exponentiates, and we could recover (\ref{commutator_all-orders}) by re-summing the perturbation series.

It is tempting to speculate that, for a perturbation localized in $F$, the perturbation series also exponentiates.  Given the
first-order correction (\ref{commutator_deltaHext4}), this would mean that for a state
\be
\vert \psi \rangle = U \vert 0 \rangle \quad {\rm with} \quad U = e^{-i \lambda G}, \quad \hbox{\rm $G$ localized in $F$ wedge}
\ee
the extended modular Hamiltonian is
\be
\Hext = V \Hextvac V^\dagger \quad {\rm with} \quad V = e^{-i \lambda E} \,.
\ee
That is, the speculation is that the excited state modular Hamiltonian is still given by conjugating the vacuum modular Hamiltonian, but with an exponential involving $E$ rather than $G$.  It would be interesting to test this conjecture beyond first order in $\lambda$.  If correct, we would have a construction of the modular operator $\Delta = e^{-\Hext}$ associated with the state $\vert \psi \rangle$,
in addition to the modular operator $\Delta_0 = e^{-\Hextvac}$ associated with the state $\vert 0 \rangle$.  We could then construct a unitary flow
\be
u(t) = \Delta^{it} \Delta_0^{-it}
\ee
which satisfies the properties of a Radon--Nikodym cocycle.\footnote{Let $\Delta_\Omega$ be the modular operator for a state $\vert \Omega \rangle$, and denote modular flow by $\sigma^\Omega_t(A) = \Delta_\Omega^{it} A \Delta_\Omega^{-it}$.  Given two modular operators, let $u_{\Omega_1\Omega_2}(t) = \Delta_{\Omega_1}^{it} \Delta_{\Omega_2}^{-it}$.  From the definitions, it is straightforward to check
\bea
\nonumber
&& \hbox{\rm composition law (cocycle identity)} \, : \quad u_{\Omega_1\Omega_2}(t_1+t_2) = u_{\Omega_1\Omega_2}(t_1)\sigma^{\Omega_2}_{t_1}\big(u_{\Omega_1\Omega_2}(t_2)\big) \\
\nonumber
&& \hbox{\rm chain rule} \, : \quad u_{\Omega_1\Omega_2}(t) u_{\Omega_2\Omega_3}(t) = u_{\Omega_1\Omega_3}(t) \\
&& \hbox{\rm intertwining property} \, : \quad u_{\Omega_1\Omega_2}(t)\sigma_t^{\Omega_2}(A) = \sigma_t^{\Omega_1}(A) u_{\Omega_1\Omega_2}(t) \\
\nonumber
&& \hbox{\rm initial condition} \, : \quad u_{\Omega_1\Omega_2}(0) = \identity
\eea
See for example section V.2.3 in \cite{Haag:1992hx}.}

\bigskip
\goodbreak
\centerline{\bf Acknowledgements}
\noindent
We are grateful to Nima Lashkari for valuable discussions, and to Gilad Lifschytz and Phuc Nguyen for discussions and preliminary investigations on these topics.  The work of XJ, DK and AM was supported by U.S.\ National Science Foundation grant PHY-2412480.  The work of DS is supported by the DST-FIST grant number SR/FST/PSI-225/2016 and SERB CRG grant CRG/2023/000904. 

\appendix
\section{More on contact terms}\label{appendix:contact}

In section \ref{sect:simplifying} we discussed situations in which $\delta \Hext$ can be replaced with $i \lambda [\Hextvac, \Oii(\xi_2^+,\xi_2^-)]$, however this replacement is not valid in general.  In section \ref{sect:contact} we wrote
\be
\delta \Hext = i \lambda [\Hextvac, \Oii(\xi_2^+,\xi_2^-)] + \delta \Hext_{\rm contact}
\ee
and made some general remarks on properties of the contact contribution.  Here we analyze the contact contribution more explicitly, using the prescription for resolving singularities
inside correlators developed in sections \ref{sect:resolve} and \ref{subsec:iepdeltahabar}.
For concreteness, we focus on operators with $\Delta = 2$ and $n = 0$.

Let us first make some remarks on how contact terms arise.  From the Sarosi-Ugajin formulas (\ref{operatordeltaHA}), (\ref{eq:deltahabarSU}) we can schematically write
\begin{equation}
\label{appendix3pt1}
	\langle\big\lbrace\mathcal{O}(\xi_1^+,\xi_1^-),\,\delta H_{A},\,\mathcal{O}(\xi_3^+,\xi_3^-)\big\rbrace\rangle=\frac{i\lambda}{\xi_{13}^+\xi_{13}^-}\int_0^\infty \frac{dw}{(w+1)^2}\left(\text{Term 1}-\text{Term 2}\right)\,,
\end{equation}
and 
\begin{equation}
\label{appendix3pt2}
	\langle\big\lbrace\mathcal{O}(\xi_1^+,\xi_1^-),\,\delta H_{\bar{A}},\,\mathcal{O}(\xi_3^+,\xi_3^-)\big\rbrace\rangle=\frac{i\lambda}{\xi_{13}^+\xi_{13}^-}\int_0^\infty \frac{dw}{(w+1)^2}\left(\text{Term 3}-\text{Term 4}\right)\,.
\end{equation} 
If we ignored the Wightman $i \epsilon$'s, and naively set $r = \pi$, then the four terms would look identical.
\begin{equation}
	\text{Term $i$}=\frac{1}{(\xi_1^++\frac{1}{w}\xi_2^+)(-\frac{1}{w}\xi_2^+-\xi_3^+)(\xi_1^-+w\xi_2^-)(-w\xi_2^--\xi_3^-)}
\end{equation}
However, if we had followed the analytic continuation $r \, : \, 0 \rightarrow \pi$, the four terms would look different. For simplicity we focus on the $\delta H_A$ correlator,
although a similar discussion applies to $\delta H_{\bar{A}}$.  In the $\delta H_A$ correlator there are mobile poles at $w_i^\pm$ where
\begin{eqnarray}
\label{term1poles}
&& \text{term 1:} \quad w_1^- = e^{- ir}\left({\xi_2^+ \over \xi_1^+} + i \epsilon_{12}\right) \quad {\rm and} \quad w_3^-=e^{- ir}\left(\frac{\xi_2^+}{\xi_3^+} - i \epsilon_{23}\right) \\
\label{term2poles}
&& \text{term 2:} \quad w_1^+ = e^{+ ir}\left({\xi_2^+ \over \xi_1^+} + i \epsilon_{12}\right) \quad {\rm and} \quad w_3^+ = e^{+ ir}\left(\frac{\xi_2^+}{\xi_3^+} - i \epsilon_{23}\right)
\end{eqnarray}
There are also fixed poles in both term 1 and term 2, at
\be
\label{term12fixed}
w^f_1 = - {\xi_1^- \over \xi_2^-} + i \epsilon_{12} \quad {\rm and} \quad w^f_3 = - {\xi_3^- \over \xi_2^-} - i \epsilon_{23}\,,
\ee
besides the pole at $w = -1$ from the integration measure.

As mentioned in section \ref{GenericCommutator}, contact terms arise when the Wightman prescription inherited from the CFT fails to unambiguously resolve a pole collision.
In these situations the analytic continuation $r \, : \, 0 \rightarrow \pi$ is essential.  When does this happen?

From section \ref{SameWedge} we know that if the spectator operators $\Oi$ and $\Oiii$ are in the same (left or right) wedge, then $\Hext_{\rm contact}$ will not contribute.
From section \ref{ActionOnVacuumSection} we know that if $\delta \Hext$ is on the far left or far right in a correlator, again $\Hext_{\rm contact}$ will not contribute.
Therefore, to pick up a contact contribution, we focus on correlators of the form $\langle \OiL \, \delta \Hext \, \OiiiR \rangle$, with spectator operators in the left and right wedges and $\delta \Hext$ in the middle of the correlator.

Given the operator locations, we have
\bea
\nonumber
&& \xi_1^+,\xi_1^- < 0 \\
&& \xi_2^+ > 0, \, \xi_2^- < 0 \\
\nonumber
&& \xi_3^+,\xi_3^- > 0
\eea
and given the desired operator ordering, we take
\be
\epsilon_{12},\,\epsilon_{23} > 0\,.
\ee
Then there are two possible pole collisions in which the Wightman prescription inherited from the CFT is inadequate, and the $i \delta$ prescription for continuing
$r \, : \, 0 \rightarrow \pi$ plays an essential role.\footnote{There are higher-order singularities, where multiple poles simultaneously collide and pinch a contour,
which we will not attempt to resolve.}
\begin{itemize}
\item
$w_1^-$ begins just above the negative real axis and rotates clockwise, to pinch the contour against $w^f_3$ which is located just below the positive real axis.  If one simply
set $r = \pi$, then both poles would be in the fourth quadrant, and collision of poles would be ambiguous (it would depend on whether $\epsilon_{12}$ or $\epsilon_{23}$ was
larger).  The $i \delta$ prescription resolves the ambiguity, by showing that $w_1^-$ hits $w^f_3$ from above.
\item
$w_3^+$ begins just below the positive real axis and rotates counter-clockwise, dragging the contour and pinching against $w_1^f$ which is located just above the negative real
axis.  Again, if one simply set $r = \pi$, the collision would be ambiguous, since both poles would be just above the negative real axis.  The $i \delta$ prescription resolves the
singularity by showing that $w_3^+$ hits $w_1^f$ from above.
\end{itemize}
These potentially ambiguous singularities are possible because the operators $\Oi$ and $\Oiii$ are in different (left and right) wedges, and $\delta H_A$ is in the middle
of the correlator.  One can likewise list the potentially ambiguous pole collisions for correlators involving $\delta H_{\bar{A}}$.

Having clarified the origin of the contact terms, we will now give an explicit expression for the correlator $\langle 0 \vert \OiL \delta \Hext_{\rm contact} \OiiiR \vert 0 \rangle$.  The strategy is to evaluate the correlator for $\delta \Hext$, then subtract the correlator for $i \lambda [\Hextvac, \Oii]$.
The relevant correlators with $\delta H_A$ and $\delta H_{\bar{A}}$ are given in \eqref{eq:habar3pfcorrotherfixingbothDan} and \eqref{eq:habar3pfcorrotherfixingboth}.  Taking the
difference gives a rather lengthy expression for
\be
\label{O1deltaHO3}
\langle 0 \vert \OiL \delta\Hext \OiiiR \vert 0 \rangle = \langle 0 \vert \OiL \delta H_A \OiiiR \vert 0 \rangle - \langle 0 \vert \OiL \delta H_{\bar{A}} \OiiiR \vert 0 \rangle
\ee
which we will not bother to write.  The correlator with $i \lambda [\Hextvac, \Oii(\xi_2^+,\xi_2^-)]$ at non-singular points is given in (\ref{3ptcommutator}).
To resolve singularities we apply the CFT Wightman prescription $t_i \rightarrow t_i - i \epsilon_i$ (or just calculate directly) to obtain
\be
\label{3ptcommutator2}
\langle \OiL \, i \lambda [\Hextvac, \Oii] \, \OiiiR \rangle = 2 \pi \lambda \Big(\xi_2^+ {\partial \over \partial \xi_2^+} - \xi_2^- {\partial \over \partial \xi_2^-}\Big) \,
{1 \over \xi_{12}^+ \, \xi_{13}^+ \, (\xi_{23}^+ - i \epsilon_{23}) \, (\xi_{12}^- + i \epsilon_{12}) \, \xi_{13}^- \, \xi_{23}^-} \,.
\ee
It is straightforward but tedious to check that at non-singular points, where $\epsilon_{12}$ and $\epsilon_{23}$ can be neglected, the two expressions (\ref{O1deltaHO3}) and
(\ref{3ptcommutator2}) are equal.  Both expressions have singularities when one of the spectator operators is null separated from the perturbation.  However one can check
that the null singularities in the two expressions are identical, resolved in the same way by the CFT Wightman prescription.  For example, whether one expands (\ref{O1deltaHO3}) or (\ref{3ptcommutator2})
about $\xi_2^+ = \xi_3^+$, one obtains
\be
- {2 \pi \lambda \over (\xi_{13}^+)^2 \xi_{13}^-} \left({\xi_3^+ \over \xi_{12}^- \xi_{23}^- (\xi_{23}^+ - i \epsilon_{23})^2} +
{(\xi_2^-)^2 - \xi_1^- \xi_3^- \over (\xi_{12}^-)^2 (\xi_{23}^-)^2 (\xi_{23}^+ - i \epsilon_{23})} + {\rm regular}\right) \,.
\ee
Contact terms come from the singularities in \eqref{O1deltaHO3} which are absent from (\ref{3ptcommutator2}).
To isolate these singularities, we use
\be
{1 \over x + i \epsilon} = {\rm PV} {1 \over x} - i \pi \delta(x)
\ee
to replace the singular factors
\be
{1 \over \xi_2^+ \xi_2^- - \xi_1^+ \xi_3^- \pm i \epsilon} \quad {\rm and} \quad {1 \over \xi_2^+ \xi_2^- - \xi_1^- \xi_3^+ \pm i \epsilon}
\ee
with principle values and delta functions.  In computing a correlator involving $\delta \Hext_{\rm contact}$, the principle value pieces cancel between (\ref{O1deltaHO3}) and
(\ref{3ptcommutator2}).  The delta functions survive and give
\bea
\nonumber
\langle \OiL \delta \Hext_{\rm contact} \OiiiR \rangle & = & {i 2 \pi^2 \lambda \over (\xi_{13}^+)^2 (\xi_{13}^-)^2} \Bigg[\left({\xi_1^+ \xi_2^+ \over (\xi_{12}^+)^2}
+ {\xi_2^- \xi_3^- \over (\xi_{23}^-)^2}\right)\delta\big(\xi_2^+\xi_2^- - \xi_1^+\xi_3^-\big) \\
& & \hspace{1.9cm} - \left({\xi_1^- \xi_2^- \over (\xi_{12}^-)^2} + {\xi_2^+\xi_3^+ \over (\xi_{23}^+)^2}\right) \delta\big(\xi_2^+ \xi_2^- - \xi_1^- \xi_3^+\big)\Bigg] \,.
\eea

\noindent{\em When are contact terms absent?} \\
\noindent
Our focus has been on working inside correlators, to clarify the origin of the contact terms and to give explicit expressions
for them.  However the same techniques can be used to show when contact terms are absent.  This leads to the simplifications which
we discussed in sections \ref{ActionOnVacuumSection} and \ref{SameWedge} from an operator perspective.  Here we show how the simplifications arise inside correlators.

First we consider a situation in which $\delta \Hext$ is inserted on the far left or far right in a correlator, which means the infinitesimal
quantities $\epsilon_{2j}$ all have the same sign (either all positive or all negative).  For concreteness, let's consider the correlator
(\ref{appendix3pt1}), with $\delta H_A$ on the far left so that $\epsilon_{21},\,\epsilon_{23} > 0$.  There are two possible collisions, in which one of the mobile poles (\ref{term1poles}), (\ref{term2poles}) rotates and pinches the integration contour
against one of the fixed poles (\ref{term12fixed}).
\begin{itemize}
\item
$w_1^-$ begins just below the negative real axis, rotates clockwise, and pinches the contour against $w_3^f$ which is located just
below the positive real axis.  This gives a singularity at $\xi_2^+ \xi_2^- = \xi_1^+ \xi_3^-$ that can be seen in (\ref{eq:habar3pfcorrotherfixingbothDan}).  Note that the way in which poles collide, with $w_1^-$ hitting $w_3^f$ from above, is fixed
by the CFT Wightman prescription.  There is no need to be careful about continuing in $r$, and one can simply set $r = \pi$.
\item
$w_3^+$ begins just below the positive real axis and rotates counter-clockwise.  It drags the integration contour with it and pinches
against $w_1^f$ which is located just below the negative real axis.  This gives a singularity at $\xi_2^+ \xi_2^- = \xi_3^+ \xi_1^-$ which can
be seen in (\ref{eq:habar3pfcorrotherfixingbothDan}).  Note that the way in which poles collide, with $w_3^+$ hitting $w_1^f$ from above, is fixed
by the CFT Wightman prescription.  There is no need to be careful about continuing in $r$, and one can simply set $r = \pi$.
\end{itemize}
One can study the pole collisions in (\ref{appendix3pt2}) under the same conditions, and one finds that again the CFT Wightman
prescription is sufficient to properly decide the way in which the integration contour gets pinched.\footnote{Contrast this to the previous situation, where $\delta \Hext$ was in the middle of the correlator.  In that case the CFT Wightman prescription left the pole collision ambiguous and the $i \delta$ prescription was necessary to resolve the ambiguity.}  What does this mean for a correlator involving
$\delta \Hext$?  Since there is no difficulty in setting $r = \pi$, we are in the situation considered in section \ref{GenericCommutator}:
the contour can be deformed to only surround the pole at $w = -1$, which means we can replace $\delta \Hext$ with
$i \lambda [\Hextvac, \Oii]$.

Finally, we consider the situation in which all spectator operators are inserted in the same (left or right) wedge.  In a sense, this situation
is simpler.  If both spectator operators are in (say) the right wedge, then all of the mobile and fixed poles in (\ref{term1poles}), (\ref{term2poles}), (\ref{term12fixed}) begin in the right half plane.  The mobile poles rotate, but there is nothing for them to pinch
against, so there are no singularities to resolve.  Again we can set $r = \pi$ and replace $\delta \Hext$ with
$i \lambda [\Hextvac, \Oii]$.

\section{KMS and Wightman}\label{appendix:VacuumFlow}
In this appendix we summarize how vacuum modular flow and the KMS analyticity condition apply to a CFT 2-point correlator.
For a related discussion see \cite{Papadodimas:2012aq}.

\begin{figure}
\begin{center}
\includegraphics[height=5cm]{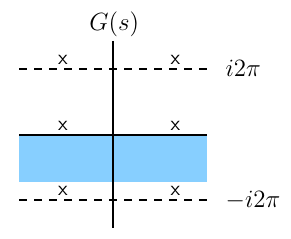} \hspace{1cm} \includegraphics[height=5cm]{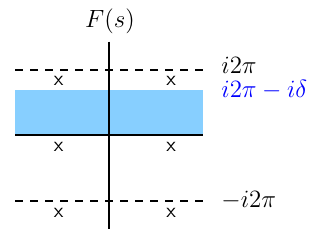}
\end{center}
\caption{Poles in $G(s)$ (left panel) and $F(s)$ (right panel).  The functions are analytic in the strips indicated in blue.
For real $s$, the distributions are related by $G(s) = F(s + i 2 \pi - i \delta)$.\label{fig:VacuumPoles}}
\end{figure}

We consider Wightman correlators of modular-flowed operators
\bea
\label{VacuumFlowG}
G(s) = \langle 0 \vert {\cal O}(\xi_1^+,\xi_1^-) \big\vert_s {\cal O}(\xi_2^+,\xi_2^-) \vert 0 \rangle = e^{n s}
{1 \over \left(e^s \xi_1^+ - \xi_2^+ - i \epsilon \right)^{\Delta + n} \left(e^{-s} \xi_1^- - \xi_2^- + i \epsilon \right)^{\Delta - n}} \\
F(s) = \langle 0 \vert {\cal O}(\xi_2^+,\xi_2^-) {\cal O}(\xi_1^+,\xi_1^-) \big\vert_s \vert 0 \rangle = e^{n s}
{1 \over \left(e^s \xi_1^+ - \xi_2^+ + i \epsilon \right)^{\Delta + n} \left(e^{-s} \xi_1^- - \xi_2^- - i \epsilon \right)^{\Delta - n}}
\eea
We take both operators to be in the right wedge, so that $\xi_i^+,\,\xi_i^- > 0$.  Assuming that the left- and right-moving
conformal dimensions are integers, the functions $G(s)$ and $F(s)$ have poles at
\bea
\nonumber
&& G(s) \, : \quad s = \log {\xi_2^+ \over \xi_1^+} + i 2 \pi {\mathbb Z} + i \epsilon \quad {\rm and} \quad s = - \log {\xi_2^- \over \xi_1^-} + i 2 \pi {\mathbb Z} + i \epsilon \\
&& F(s) \, : \quad s = \log {\xi_2^+ \over \xi_1^+} + i 2 \pi {\mathbb Z} - i \epsilon \quad {\rm and} \quad s = - \log {\xi_2^- \over \xi_1^-} + i 2 \pi {\mathbb Z} - i \epsilon
\eea
The poles are illustrated in Fig.\ \ref{fig:VacuumPoles}.  Note that $G(s)$ is analytic in the strip $-2\pi + \delta < {\rm Im} \, s < 0$, where $\delta$
is an infinitesimal parameter that approaches $0^+$ more slowly than the quantity $\epsilon$ which is used to define the Wightman
correlator.  Likewise $F(s)$ is analytic in the strip $0 < {\rm Im} \, s < 2\pi - \delta$.

To see how $F(s)$ and $G(s)$ are related, which is the statement of the KMS analyticity condition, note that
\bea
\nonumber
F(s + i 2 \pi - i \delta) & = & e^{ns} {1 \over \left(e^s \xi_1^+ e^{-i \delta} - \xi_2^+ + i \epsilon \right)^{\Delta + n}
\left(e^{-s} \xi_1^- e^{i \delta} - \xi_2^- - i \epsilon \right)^{\Delta - n}} \\
\label{override}
& = & e^{ns} {1 \over \left(e^s \xi_1^+ - \xi_2^+ - i \delta \right)^{\Delta + n}
\left(e^{-s} \xi_1^- - \xi_2^- + i \delta \right)^{\Delta - n}} \\
\nonumber
& = & G(s) \,.
\eea
In the second line we used the fact that $\delta$ approaches zero more slowly than $\epsilon$, and in the last line we used the
fact that as $\delta \rightarrow 0$ the resulting distribution is identical to (\ref{VacuumFlowG}).  This leads us to the statement
of KMS analyticity for Wightman correlators used in section \ref{sect:KMS}, namely $G(s) = F(s + i 2 \pi - i \delta)$.

Incidentally, in (\ref{override}), note that $i \delta$ overrides the Wightman $i \epsilon$ prescription inherited from the CFT, which
changes the way in which the singularity is resolved.  This change is crucial for KMS analyticity.  A similar phenomenon played an important role in resolving singularities in section \ref{sect:resolve}.

\section{KMS analyticity for non-singular operator configurations\label{appendix:nonsing}}

In this section, we check the validity of the KMS analyticity condition \eqref{eq:fokms} for operators with $\Delta = 2$ and $n = 0$ by directly computing the three-point functions that appear there. Unlike section \ref{sect:KMSanalyticity}, we restrict our attention to non-singular
operator configurations.

We first rewrite \eqref{eq:fokms} for convenience.
\begin{eqnarray}\label{eq:fokmsnew}
		-i\lambda\left(\langle G A B_{-s}\rangle-\langle A_s B G\rangle-\langle G B A_{s+2\pi i}\rangle+\langle B_{-s-2\pi i} A G\rangle\right)\nonumber\\
		=-\frac{i}{2\pi}\left(\int_{0}^{s}d\alpha\langle A_s\delta\Hext_\alpha B\rangle-\int_{0}^{-s-2\pi i}d\alpha\langle B_{-s-2\pi i}\delta\Hext_\alpha A\rangle\right)
\end{eqnarray}
We take $G=\mathcal{O}(\xi_2^+,\xi_2^-)$ and $B$ and $A$ to be $\mathcal{O}(\xi_1^+,\xi_1^-)$ and $\mathcal{O}(\xi_3^+,\xi_3^-)$ respectively. In this section we focus on non-singular operator configurations.  Aside from this, we suppress any explicit discussion of the locations of the $A$ and $B$ operators. They should both be in the same wedge, according to the statement of KMS, however at non-singular points this property will not play a role.

Using the three-point function (\ref{eq:3pfarbitloc}), we obtain the LHS of \eqref{eq:fokmsnew} to be
\begin{align}\label{eq:KMSLHS}
	\frac{ 2 i\lambda\, e^{2s} (e^s - 1)\, (X+Y)}{(\xi_{12}^-)
  (\xi_{12}^+)
  (\xi_{23}^-)
  (\xi_{23}^+)
  \left(e^s \xi_1^- - \xi_2^-\right)
  \left(e^s \xi_1^- - \xi_3^-\right)
  \left(e^s \xi_2^+ - \xi_1^+\right)
  \left(e^s \xi_3^+ - \xi_1^+\right)
  \left(e^s \xi_2^- - \xi_3^-\right)
  \left(e^s \xi_3^+ - \xi_2^+\right)}\,,
  \end{align}
where 
\begin{equation}
	X=\xi_1^+ \left(
     -\xi_2^-\, \xi_3^+ (e^s \xi_1^- + \xi_3^-)
        + (\xi_2^-)^2 (e^s \xi_3^+ - \xi_2^+ + \xi_3^+)
        + \xi_1^-\, \xi_2^+\, \xi_3^-
      \right)
\end{equation}
and
\begin{equation}
	Y=\xi_2^+ \left(
        e^s \left(
          \xi_1^-\, \xi_2^-\, \xi_2^+
          + \xi_1^-\, \xi_3^- (\xi_3^+ - \xi_2^+)
          - (\xi_2^-)^2 \xi_3^+
        \right)
        + \xi_2^+\, \xi_3^- (\xi_2^- - \xi_1^-)
      \right)\,.
\end{equation}

To evaluate the RHS of \eqref{eq:fokmsnew}, we note that 
\begin{equation}
	\delta H_A|_{\alpha}=i\lambda\int_0^\infty\frac{dw}{(1+w)^2}\left[\mathcal{O}\left(\frac{e^{-ir+\alpha}\,\xi_2^+}{w},-we^{-\alpha}\xi_2^-\right)-\mathcal{O}\left(\frac{e^{ir+\alpha}\,\xi_2^+}{w},-we^{-\alpha}\xi_2^-\right)\right]
\end{equation}
and
\begin{equation}
	\delta H_{\bar{A}}|_{\alpha}=i\lambda\int_0^\infty\frac{dw}{(1+w)^2}\left[\mathcal{O}\left(-\frac{e^{\alpha}\,\xi_2^+}{w},we^{-ir-\alpha}\xi_2^-\right)-\mathcal{O}\left(-\frac{e^{\alpha}\,\xi_2^+}{w},we^{ir-\alpha}\xi_2^-\right)\right]\,.
\end{equation}
We will therefore have to compute correlators like (the notation $(\dots)|_{\alpha}$ below implies vacuum modular flow)
\begin{equation}\label{eq:alpha3p}
	\langle\mathcal{O}\left(e^{-(s+2\pi i)}\xi_1^+,e^{(s+2\pi i)}\xi_1^-\right)\,\delta H_{i}|_{\alpha}\,\mathcal{O}\left(\xi_3^+,\xi_3^-\right)\rangle \quad\text{and}\quad \langle\mathcal{O}\left(e^{s}\xi_3^+,e^{-s}\xi_3^-\right)\,\delta H_{i}|_{\alpha}\,\mathcal{O}\left(\xi_1^+,\xi_1^-\right)\rangle
\end{equation}
with $i=A,\bar{A}$, and then evaluate the $w$ and $\alpha$ integrals. The mobile poles that appear in the three-point functions of \eqref{eq:alpha3p} are located at
(apologies to the reader -- our notation for the poles here is different than in appendix A)
\begin{align}\label{eq:KMSpoles}
	\tilde{w}^{\mp}_1&=e^{\mp ir+\alpha}\left(\frac{e^s\xi_2^+}{\xi_1^+}+i\epsilon_{12}\right)\,,\nonumber\\
	\tilde{w}^{\mp}_2&=e^{\mp ir+\alpha}\left(\frac{\xi_2^+}{\xi_3^+}-i\epsilon_{23}\right)\,,\nonumber\\
	\tilde{w}^{\pm}_3&=e^{\pm ir+\alpha}\left(\frac{e^s\xi_1^-}{\xi_2^-}-i\epsilon_{12}\right)\,,\nonumber\\
	\tilde{w}^{\pm}_4&=e^{\pm ir+\alpha}\left(\frac{\xi_3^-}{\xi_2^-}+i\epsilon_{23}\right)\,.
\end{align}
On the other hand, the set of fixed poles are located at $\tilde{w}_{f}=\tilde{w}^{\pm}_{f}|_{r=\pi}$ (for $f=1,2,3$ and 4). Computing the residues of the $w$ integrals and then performing the $\alpha$ integrals, we finally obtain (in what follows, $a=e^{-s}$)
\begin{align}\label{eq:t1}
&\frac{i}{2\pi}\int_{0}^{-s-2\pi i} d\alpha\,\langle\mathcal{O}\left(e^{-(s+2\pi i)}\xi_1^+,e^{(s+2\pi i)}\xi_1^-\right)\,\delta H_{A}|_{\alpha}\,\mathcal{O}\left(\xi_3^+,\xi_3^-\right)\rangle=	-\frac{i (a-1) a^2 \lambda \xi_2^+}
{(\xi_1^- - a \xi_3^-)(\xi_3^+ - a \xi_1^+)^2}\nonumber\\
&\left(
  \frac{(\xi_1^+)^2}{
    (\xi_{12}^+)(a \xi_1^+ - \xi_2^+)
    (\xi_1^- \xi_1^+ - \xi_2^- \xi_2^+)
    (a \xi_1^+ \xi_3^- - \xi_2^- \xi_2^+)
  }
  +
  \frac{(\xi_3^+)^2}{
    (\xi_{23}^+)(a \xi_2^+ - \xi_3^+)
    (\xi_3^- \xi_3^+ - \xi_2^- \xi_2^+)
    (a \xi_2^- \xi_2^+ - \xi_1^- \xi_3^+)
  }
\right)\,,
\end{align}

\begin{align}
&\frac{i}{2\pi}\int_{0}^{s} d\alpha\,\langle\mathcal{O}\left(e^{s}\xi_3^+,e^{-s}\xi_3^-\right)\,\delta H_{A}|_{\alpha}\,\mathcal{O}\left(\xi_1^+,\xi_1^-\right)\rangle=	-\frac{i (a-1) a^2 \lambda \xi_2^+}
{(\xi_3^+ - a \xi_1^+)^2(a \xi_3^- - \xi_1^-)}\nonumber\\
&\left(
  \frac{(\xi_1^+)^2}{
    (\xi_{12}^+)(a \xi_1^+ - \xi_2^+)
    (\xi_1^- \xi_1^+ - \xi_2^- \xi_2^+)(a \xi_1^+ \xi_3^- - \xi_2^- \xi_2^+)}
  -
  \frac{(\xi_3^+)^2}{(\xi_{23}^+)(a \xi_2^+ - \xi_3^+)
    (\xi_3^- \xi_3^+ - \xi_2^- \xi_2^+)(\xi_1^- \xi_3^+ - a \xi_2^- \xi_2^+)}
\right)\,,
\end{align}

\begin{align}
&\frac{i}{2\pi}\int_{0}^{-s-2\pi i} d\alpha\,\langle\mathcal{O}\left(e^{-(s+2\pi i)}\xi_1^+,e^{(s+2\pi i)}\xi_1^-\right)\,\delta H_{\bar{A}}|_{\alpha}\,\mathcal{O}\left(\xi_3^+,\xi_3^-\right)\rangle=	\frac{i (a-1) a^2 \lambda \xi_2^-}
{(\xi_1^- - a \xi_3^-)^2 (a \xi_1^+ - \xi_3^+)}\nonumber\\
&\left(\frac{(\xi_1^-)^2}{(\xi_{12}^-)(\xi_1^- - a \xi_2^-)(\xi_1^- \xi_1^+ - \xi_2^- \xi_2^+)(a \xi_2^- \xi_2^+ - \xi_1^- \xi_3^+)}
  +
  \frac{(\xi_3^-)^2}{(\xi_{23}^-)(\xi_2^- - a \xi_3^-)(\xi_2^- \xi_2^+ - \xi_3^- \xi_3^+)(\xi_2^- \xi_2^+ - a \xi_1^+ \xi_3^-)}
\right)\,,
\end{align}

\begin{align}\label{eq:t4}
&\frac{i}{2\pi}\int_{0}^{s} d\alpha\,\langle\mathcal{O}\left(e^{s}\xi_3^+,e^{-s}\xi_3^-\right)\,\delta H_{\bar{A}}|_{\alpha}\,\mathcal{O}\left(\xi_1^+,\xi_1^-\right)\rangle=\frac{i (a-1) a^2 \lambda \xi_2^-}
{(\xi_1^- - a \xi_3^-)^2 (a \xi_1^+ - \xi_3^+)}\nonumber\\
& \left(
  \frac{(\xi_1^-)^2}{(\xi_{12}^-)(\xi_1^- - a \xi_2^-)(\xi_1^- \xi_1^+ - \xi_2^- \xi_2^+)(\xi_1^- \xi_3^+ - a \xi_2^- \xi_2^+)}
  +
  \frac{(\xi_3^-)^2}{
    (\xi_{23}^-)(a \xi_3^- - \xi_2^-)(\xi_2^- \xi_2^+ - \xi_3^- \xi_3^+)(\xi_2^- \xi_2^+ - a \xi_1^+ \xi_3^-)}
\right)\,.
\end{align}
One can combine \eqref{eq:t1} through \eqref{eq:t4} as per the RHS of \eqref{eq:fokmsnew}. By doing so, one can explicitly confirm that the combination precisely matches with \eqref{eq:KMSLHS}. Hence, for non-singular points, the KMS analyticity condition is satisfied.


\begin{thebibliography}{10}

\bibitem{Haag:1992hx}
R.~Haag, {\em {Local quantum physics: Fields, particles, algebras}}.
\newblock Springer Verlag, 1992.

\bibitem{Borchers:2000pv}
H.~J. Borchers, ``{On revolutionizing quantum field theory with Tomita's
  modular theory},'' \href{http://dx.doi.org/10.1063/1.533323}{{\em J. Math.
  Phys.} {\bfseries 41} (2000) 3604--3673}.

\bibitem{Jones:2015kbb}
V.~F.~R. Jones, ``{Von Neumann algebras}''.
  \url{https://math.berkeley.edu/~vfr/VonNeumann2009.pdf}.

\bibitem{Witten:2018lha}
E.~Witten, ``{APS medal for exceptional achievement in research: Invited
  article on entanglement properties of quantum field theory},''
  \href{http://dx.doi.org/10.1103/RevModPhys.90.045003}{{\em Rev. Mod. Phys.}
  {\bfseries 90} no.~4, (2018) 045003},
\href{http://arxiv.org/abs/1803.04993}{{\ttfamily arXiv:1803.04993 [hep-th]}}.

\bibitem{Sorce:2023gio}
J.~Sorce, ``{An intuitive construction of modular flow},''
  \href{http://dx.doi.org/10.1007/JHEP12(2023)079}{{\em JHEP} {\bfseries 12}
  (2023) 079}, \href{http://arxiv.org/abs/2309.16766}{{\ttfamily
  arXiv:2309.16766 [hep-th]}}.

\bibitem{Papadodimas:2013jku}
K.~Papadodimas and S.~Raju, ``{State-dependent bulk-boundary maps and black
  hole complementarity},''
  \href{http://dx.doi.org/10.1103/PhysRevD.89.086010}{{\em Phys. Rev.}
  {\bfseries D89} no.~8, (2014) 086010},
\href{http://arxiv.org/abs/1310.6335}{{\ttfamily arXiv:1310.6335 [hep-th]}}.

\bibitem{Jafferis:2015del}
D.~L. Jafferis, A.~Lewkowycz, J.~Maldacena, and S.~J. Suh, ``{Relative entropy
  equals bulk relative entropy},''
  \href{http://dx.doi.org/10.1007/JHEP06(2016)004}{{\em JHEP} {\bfseries 06}
  (2016) 004},
\href{http://arxiv.org/abs/1512.06431}{{\ttfamily arXiv:1512.06431 [hep-th]}}.

\bibitem{Bisognano:1976za}
J.~J. Bisognano and E.~H. Wichmann, ``{On the duality condition for quantum
  fields},''
\href{http://dx.doi.org/10.1063/1.522898}{{\em J. Math. Phys.} {\bfseries 17}
  (1976) 303--321}.

\bibitem{Hislop:1981uh}
P.~D. Hislop and R.~Longo, ``{Modular structure of the local algebras
  associated with the free massless scalar field theory},''
  \href{http://dx.doi.org/10.1007/BF01208372}{{\em Commun.\ Math.\ Phys.}
  {\bfseries 84} (1982) 71}.

\bibitem{Faulkner:2016mzt}
T.~Faulkner, R.~G. Leigh, O.~Parrikar, and H.~Wang, ``{Modular Hamiltonians for
  deformed half-spaces and the averaged null energy condition},''
  \href{http://dx.doi.org/10.1007/JHEP09(2016)038}{{\em JHEP} {\bfseries 09}
  (2016) 038}, \href{http://arxiv.org/abs/1605.08072}{{\ttfamily
  arXiv:1605.08072 [hep-th]}}.

\bibitem{Sarosi:2017rsq}
G.~S\'arosi and T.~Ugajin, ``{Modular Hamiltonians of excited states, OPE
  blocks and emergent bulk fields},''
  \href{http://dx.doi.org/10.1007/JHEP01(2018)012}{{\em JHEP} {\bfseries 01}
  (2018) 012},
\href{http://arxiv.org/abs/1705.01486}{{\ttfamily arXiv:1705.01486 [hep-th]}}.

\bibitem{Eisler:2022rnp}
V.~Eisler, E.~Tonni, and I.~Peschel, ``{Local and non-local properties of the
  entanglement Hamiltonian for two disjoint intervals},''
  \href{http://dx.doi.org/10.1088/1742-5468/ac8151}{{\em J. Stat. Mech.}
  {\bfseries 2208} no.~8, (2022) 083101},
  \href{http://arxiv.org/abs/2204.03966}{{\ttfamily arXiv:2204.03966
  [cond-mat.stat-mech]}}.

\bibitem{Cardy:2016fqc}
J.~Cardy and E.~Tonni, ``{Entanglement Hamiltonians in two-dimensional
  conformal field theory},''
  \href{http://dx.doi.org/10.1088/1742-5468/2016/12/123103}{{\em J. Stat.
  Mech.} {\bfseries 1612} no.~12, (2016) 123103},
  \href{http://arxiv.org/abs/1608.01283}{{\ttfamily arXiv:1608.01283
  [cond-mat.stat-mech]}}.
  
\bibitem{Bratteli:1996xq}
O.~Bratteli and D.~W. Robinson,
  \href{http://dx.doi.org/10.1007/978-3-662-03444-6}{{\em {Operator algebras
  and quantum statistical mechanics. Vol. 2: Equilibrium states. Models in
  quantum statistical mechanics}}}.
\newblock Springer Verlag, 1996.
\newblock See proposition 5.3.29.

\bibitem{Lashkari:2018tjh}
N.~Lashkari, H.~Liu, and S.~Rajagopal, ``{Perturbation theory for the logarithm
  of a positive operator},''
  \href{http://dx.doi.org/10.1007/JHEP11(2023)097}{{\em JHEP} {\bfseries 11}
  (2023) 097}, \href{http://arxiv.org/abs/1811.05619}{{\ttfamily
  arXiv:1811.05619 [hep-th]}}.

\bibitem{Rosenhaus:2014woa}
V.~Rosenhaus and M.~Smolkin, ``{Entanglement entropy: A perturbative
  calculation},'' \href{http://dx.doi.org/10.1007/JHEP12(2014)179}{{\em JHEP}
  {\bfseries 12} (2014) 179}, \href{http://arxiv.org/abs/1403.3733}{{\ttfamily
  arXiv:1403.3733 [hep-th]}}.

\bibitem{Kabat:2020oic}
D.~Kabat, G.~Lifschytz, P.~Nguyen, and D.~Sarkar, ``{Endpoint contributions to
  excited-state modular Hamiltonians},''
  \href{http://dx.doi.org/10.1007/JHEP12(2020)128}{{\em JHEP} {\bfseries 12}
  (2020) 128}, \href{http://arxiv.org/abs/2006.13317}{{\ttfamily
  arXiv:2006.13317 [hep-th]}}.

\bibitem{Kabat:2021akg}
D.~Kabat, G.~Lifschytz, P.~Nguyen, and D.~Sarkar, ``{Light-ray moments as
  endpoint contributions to modular Hamiltonians},''
  \href{http://dx.doi.org/10.1007/JHEP09(2021)074}{{\em JHEP} {\bfseries 09}
  (2021) 074}, \href{http://arxiv.org/abs/2103.08636}{{\ttfamily
  arXiv:2103.08636 [hep-th]}}.
  
\bibitem{Kudler-Flam:2023hkl}
J.~Kudler-Flam, S.~Leutheusser, A.~A. Rahman, G.~Satishchandran, and A.~J.
  Speranza, ``{Covariant regulator for entanglement entropy: Proofs of the
  Bekenstein bound and the quantum null energy condition},''
  \href{http://dx.doi.org/10.1103/PhysRevD.111.105001}{{\em Phys. Rev. D}
  {\bfseries 111} no.~10, (2025) 105001},
  \href{http://arxiv.org/abs/2312.07646}{{\ttfamily arXiv:2312.07646
  [hep-th]}}.

\bibitem{Kabat:2018pbj}
D.~Kabat and G.~Lifschytz, ``{Does boundary quantum mechanics imply quantum
  mechanics in the bulk?},''
  \href{http://dx.doi.org/10.1007/JHEP03(2018)151}{{\em JHEP} {\bfseries 03}
  (2018) 151}, \href{http://arxiv.org/abs/1801.08101}{{\ttfamily
  arXiv:1801.08101 [hep-th]}}.

\bibitem{Takesaki}
M.~Takesaki, \href{http://dx.doi.org/10.1007/978-3-662-10451-4}{{\em {Theory of
  Operator Algebras II}}}.
\newblock Springer Verlag, 2003.

\bibitem{Lashkari:2018oke}
N.~Lashkari, H.~Liu, and S.~Rajagopal, ``{Modular flow of excited states},''
  \href{http://arxiv.org/abs/1811.05052}{{\ttfamily arXiv:1811.05052
  [hep-th]}}.

\bibitem{Faulkner:2018faa}
T.~Faulkner, M.~Li, and H.~Wang, ``{A modular toolkit for bulk
  reconstruction},'' \href{http://dx.doi.org/10.1007/JHEP04(2019)119}{{\em
  JHEP} {\bfseries 04} (2019) 119},
  \href{http://arxiv.org/abs/1806.10560}{{\ttfamily arXiv:1806.10560
  [hep-th]}}.

\bibitem{Bousso:2014uxa}
R.~Bousso, H.~Casini, Z.~Fisher, and J.~Maldacena, ``{Entropy on a null surface
  for interacting quantum field theories and the Bousso bound},''
  \href{http://dx.doi.org/10.1103/PhysRevD.91.084030}{{\em Phys. Rev. D}
  {\bfseries 91} no.~8, (2015) 084030},
  \href{http://arxiv.org/abs/1406.4545}{{\ttfamily arXiv:1406.4545 [hep-th]}}.

\bibitem{Jensen:2023yxy}
K.~Jensen, J.~Sorce, and A.~J. Speranza, ``{Generalized entropy for general
  subregions in quantum gravity},''
  \href{http://dx.doi.org/10.1007/JHEP12(2023)020}{{\em JHEP} {\bfseries 12}
  (2023) 020}, \href{http://arxiv.org/abs/2306.01837}{{\ttfamily
  arXiv:2306.01837 [hep-th]}}.

\bibitem{Fredenhagen1985}
K.~Fredenhagen, ``On the modular structure of local algebras of observables,''
  \href{http://dx.doi.org/10.1007/BF01206179}{{\em Communications in
  Mathematical Physics} {\bfseries 97} no.~1, (1985) 79--89}.

\bibitem{Papadodimas:2012aq}
K.~Papadodimas and S.~Raju, ``{An infalling observer in AdS/CFT},''
  \href{http://dx.doi.org/10.1007/JHEP10(2013)212}{{\em JHEP} {\bfseries 10}
  (2013) 212},
\href{http://arxiv.org/abs/1211.6767}{{\ttfamily arXiv:1211.6767 [hep-th]}}.

\end{thebibliography}
\providecommand{\href}[2]{#2}\begingroup\raggedright\endgroup

\end{document}